\documentclass[aps,prb,twocolumn,english,showpacs,superscriptaddress,amssymb,amsfonts]{revtex4}
\usepackage[T1]{fontenc}
\usepackage[latin9]{inputenc}
\usepackage{amsmath}
\usepackage{dsfont}
\usepackage{amstext}

\usepackage{tocvsec2}

\usepackage{amssymb}
\usepackage{amsbsy}
\usepackage{amsthm}
\usepackage{epsfig}
\usepackage{graphicx}
\usepackage{bbm}
\usepackage{hyperref}
\usepackage{color}

\newcommand{\Ref}[1]{Ref.~\onlinecite{#1}}

\newcommand{\bst}{{\boldsymbol{T}}}
\newcommand{\bse}{{\boldsymbol{e}}}
\newcommand{\bsg}{{\boldsymbol{g}}}
\newcommand{\fnp}{{\boldsymbol{P}_f}}
\newcommand{\ie}{{\emph{i.e.~}}}
\makeatletter
\newcommand{\rmnum}[1]{\romannumeral #1}
\newcommand{\Rmnum}[1]{\expandafter\@slowromancap\romannumeral #1@}
\makeatother
\newcommand{\imth}{\hspace{1pt}\mathrm{i}\hspace{1pt}}

\newcommand{\eg}{{\emph{e.g.~}}}
\newcommand{\etc}{{\emph{etc.~}}}

\newcommand{\mbz}{{\mathbb{Z}}}
\newcommand{\bpm}{\begin{pmatrix}}
\newcommand{\epm}{\end{pmatrix}}
\newcommand{\bea}{\begin{eqnarray}}
\newcommand{\eea}{\end{eqnarray}}

\makeatother

\usepackage{babel}
\begin{document}
\title{Theory and classification of interacting `integer' topological phases in two dimensions: A Chern-Simons approach}

\author{Yuan-Ming Lu}
\affiliation{Department of Physics, University of California, Berkeley, CA 94720, USA}
\affiliation{Materials Science Division, Lawrence Berkeley National Laboratories, Berkeley, CA 94720}

\author{Ashvin Vishwanath}
\affiliation{Department of Physics, University of California, Berkeley, CA 94720, USA}
\affiliation{Materials Science Division, Lawrence Berkeley National Laboratories, Berkeley, CA 94720}
\begin{abstract}
We study topological phases of interacting systems in two spatial dimensions in the absence of topological order (i.e. with a unique ground state on closed manifolds and no fractional excitations). These are the closest interacting analogs of integer Quantum Hall states, topological insulators and superconductors. We adapt the well-known Chern-Simons {K}-matrix description of Quantum Hall states to classify such `integer' topological phases. Our main result is a general formalism that incorporates symmetries into the {{K}}-matrix description. Remarkably, this simple analysis yields the same list of topological phases as a recent group cohomology classification, and in addition provides field theories and explicit edge theories for all these phases. The bosonic topological phases, which only appear in the presence of interactions and which remain well defined in the presence of disorder include (i) bosonic insulators with a Hall conductance quantized to even integers (ii) a bosonic analog of quantum spin Hall insulators and (iii) a bosonic analog of a chiral topological superconductor, whose K matrix is the Cartan matrix of Lie group E$_8$. We also discuss interacting fermion systems where symmetries are realized in a projective fashion, where we find the present formalism can handle a wider range of symmetries than a recent group super-cohomology classification. Lastly we construct microscopic models of these phases from coupled one-dimensional systems.
\end{abstract}


\maketitle


{\small \setcounter{tocdepth}{2} \tableofcontents}

\section{Introduction}
The recent discovery of topological insulators\cite{Hasan2010,Qi2011,Hasan2011,zhang2007, hseih2008, hseih2009, hseih2009_2, xia2009, chen2009, roushan2009,kane2005, zhang2006, fu2007, moore2007, roy2009, zhang2008} has lead to a renewed interest in phases of matter that are not described within the usual Landau paradigm of symmetry breaking and order parameters\cite{Wen2004}. Topological insulators and topological superconductors, like integer quantum Hall states, are gapped in the bulk but differ from trivial phases in the topology of their electronic states. They are characterized by gapless edge excitations that reflect the bulk topology. A new aspect of $\mbz_2$ spin-orbit topological insulators is the role of symmetry (time reversal in this case) in protecting the topological distinction. In the absence of symmetry, the topological properties, such as gapless edge states, are generally destroyed. As with the integer Hall effect, topological insulators and superconductors can be described in terms of non-interacting particles. A complete classification of such free fermion topological phases that are stable to disorder, in all spatial dimensions, has been achieved\cite{Schnyder2008,Kitaev2009}.  The remaining outstanding questions for fundamental theory have to do with interacting systems.

Interacting topological phases have been studied at two levels. First, the stability of the non-interacting classification to interactions has been examined\cite{Fidkowski2010,Qi2012,Yao2012,Ryu2012}. In some cases interactions {\em reduce} the number of topological phases\cite{Fidkowski2010,Turner2011,Gu2012,Qi2012,Yao2012,Ryu2012}, i.e. two topologically distinct phases of free fermions be continuously connected, via intermediate states that involve interactions. The second possibility, of interactions leading to {\em new} phases, not possible within non-interacting particles has also been studied. Largely, these have attempted to extend the analogy between integer and fractional quantum Hall states, to topological insulators. Thus, fractional topological insulators have been theoretically discussed\cite{Ran2008a,Ruegg2012,Levin2009,Maciejko2010,Swingle2011,Neupert2011a,Santos2011}, along with lattice realizations of fractional quantum Hall insulators\cite{Tang2011,Sun2011,Neupert2011,Sheng2011,Wang2011,Parameswaran2011,Vaezi2011,Lu2012,McGreevy2012,Regnault2011}. These phases are topologically ordered - in that they involve fractional excitations in the bulk  and feature ground state degeneracies that depend on the topology of the space on which they are defined. They are also characterized by a finite topological entanglement entropy (TEE) in the ground state. In contrast, integer quantum Hall and topological insulators (and superconductors), despite being topologically distinct, are not topologically ordered. Bulk excitations are essentially like electrons or groups of electrons, and the ground state is unique when defined on a manifold without boundaries. The TEE vanishes for these phases. Henceforth we shall refer to gapped phases without topological order as being {\em short range entangled} ({\bf SRE}) states.\footnote{This terminology differs slightly from that of Chen-Gu-Wen\cite{Chen2011}, who require a state to also be non-chiral to be short range entangled.}. It seems appropriate to define interacting ``integer" topological phases, as new topological phases without topological order, but which only appear in the presence of interactions.  Do such phases exist? And if so, how can they be studied?

In one dimension, topological order is absent, and all topological phases found are ``integer" (or SRE) phases. They include examples like the Haldane (or AKLT) state of gapped spin-1 chains\cite{Auerbach1994}.  Using the matrix product representation of gapped states\cite{Fidkowski2011,Turner2011,Chen2011a,Cirac} they are argued to be classified by projective representations of the symmetry group ($G$) or equivalently by the second group cohomology ${\mathcal H}^2(G,{\mathcal C})$ of symmetry group $G$. In higher dimensions, such rigorous results are not available. Nevertheless, new work indicates that these are also amenable to theoretical study. Recently, Chen, Gu and Wen\cite{Chen2011} have proposed that higher dimensional group cohomology describes $d=2,3$ dimensional interacting topological states without topological order. For example, bosonic systems were studied, where there are no topological phases in the absence of interactions. With interactions, topological phases were predicted in two (and three) dimensions, without topological order. While Chen et al. \cite{Chen2011} restrict attention to the non-chiral subset of these states (\ie ones that do not have a net imbalance of left and right movers at the edge of a two dimensional system) protected by symmetry, Kitaev\footnote{Alexei Kitaev, unpublished. See \url{http://online.kitp.ucsb.edu/online/topomat11/kitaev/}} has also considered chiral states. Explicit examples of such phases in special cases has been given\cite{Chen2011b,Levin2012}. However predictions in the general case rely on writing Wess-Zumino-Witten terms for generalized sigma models. While this is a powerful approach, the physical meaning of the phases that are predicted are obscure. For example, the nature of edge excitations in these phases is not apparent. Moreover, a knowledge of the Borel-group-cohomolgy machinery is required, which is mathematically sophisticated even by the standards of the field.

{\bf $K$-matrix formulation:} Here we take a completely different and simpler approach to the problem, focusing on the case of two spatial dimensions. We rely on the $K$ matrix formulation of quantum Hall states, a symmetric integer matrix that appears in the Chern-Simons action:
($\hbar=1$, and summation is implied over repeated indices $\mu,\nu,\lambda=0,1,2$):
\begin{eqnarray}\label{cs action}
&4\pi\mathcal{S}_{CS}=\int d^2xdt~\sum_{I,J}{\epsilon_{\mu\nu\lambda}}a^I_\mu [{\bf K}]_{I,J}\partial_\nu a_\lambda^J
\end{eqnarray}
While this has been utilized to discuss quantum Hall states with Abelian topological order, here we show that it is also a powerful tool to discuss topological phases in the {\em absence} of topological order. The latter requires $|\det{\bf K}|=1$ (\ie $\bf K$ is a {\em unimodular} matrix). The bulk action also determines topological properties of the edge states. For example, the signature of the $K$ matrix (number of positive minus negative eigenvalues) is the chirality of edge states - the imbalance between number of right and left moving edge modes. Maximally chiral states have all edge modes moving in the same direction. {Physically the fluxes $\epsilon^{\mu\nu\lambda} \partial_\mu a_\nu$ are related to densities and currents of bosons of different flavors}.
\begin{figure}
 \includegraphics[width=0.45\textwidth]{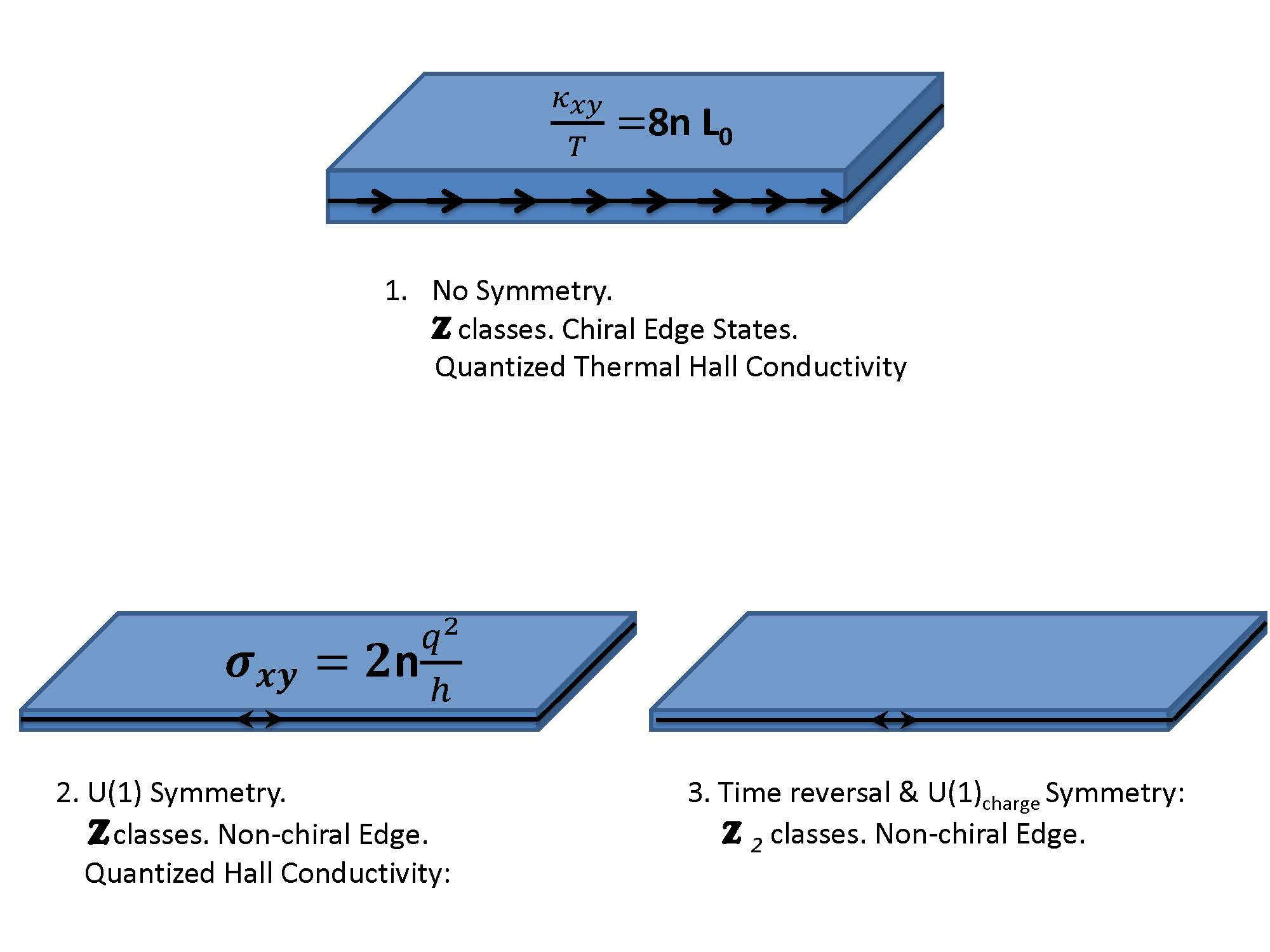}
\caption{Summary of some simple `integer' bosonic topological phases. 1) A chiral phase of bosons (no symmetry required). An integer multiple of eight chiral bosons at the edge is needed to evade topological order, leading to a quantized thermal Hall conductance $\kappa_{xy}/T=8nL_0$ in units of the the universal thermal conductance $L_0=\frac{\pi^2k_B^2}{3h}$. These are bosonic analogs of chiral superconductors. (2) A non-chiral phase of bosons protected by $U(1)$ symmetry (eg. charge conservation). Distinct phases can be labeled by the quantized Hall conductance $\sigma_{xy}=2n\sigma_0$, which are even integer multiples of the universal conductance $\sigma_0=q^2/h$ for particles with charge $q$. These are bosonic analogs of the integer quantum Hall phases. (3) a non-chiral phase stabilized in the presence of time reversal and $U(1)$ charge conservation symmetries, the same symmetries used to define quantum spin Hall (topological) insulators. A $Z_2$ topological classification is obtained, although bosonic time reversal that squares to $+1$ is involved.}\label{fig:summary}
\end{figure}

{\bf Strategy and results:} Let us briefly review our strategy and results. Although we mainly focus on non-chiral states, we begin by looking for maximally chiral states of bosons without topological order. These are bosonic analogs of the integer quantum Hall effect or chiral superconductors of fermions. It is readily shown that the smallest dimension of bosonic $K$ matrix that yields a maximally chiral state is 8. This is consistent with the prediction of Kitaev, derived from topological field theory. Here we explicitly construct a candidate $K$ matrix for this state, corresponding to the Cartan matrix of the group $E_8$.

\begin{table*}
\centering
\begin{ruledtabular}
\begin{tabular}{|l|c|l|}
  \hline
  {\em Symmetry} & \em Topological Classification & {\em Comments} \\
  \hline
   No symmetry (chiral) & $\mbz$ & $E_8$ state and derivatives with chiral central charge\cite{Kitaev2006} $c_-=8n$. \\
   \hline
  $Z_2^T$ & $\mbz_1$ & Time reversal symmetry \\
  $U(1)$ & $\mbz$ & Charge conserved. Quantized Hall conductance $\sigma_{xy}=2nq^2/h$ with $n\in\mbz$ \\
  $U(1)\rtimes Z_2^T$ & $\mbz_2$ & Bosonic Quantum Spin Hall with charge $U(1)$ and time reversal ${\mathcal T}^2=+1$.  \\
  $U(1)\times Z_2^T$ & $\mbz_1$ &  $U(1)$ spin conservation and time reversal.\\
  $Z_n$ & $\mbz_n$ &  $U(1)$ broken down to a discrete subgroup\\
  $Z_n\rtimes Z_2^T$ & $\mbz^2_{(n,2)}$ & $(a,b)\equiv$~greatest common divisor of $a$ and $b$. \\
  $Z_n\times Z_2^T$ & $\mbz^2_{(n,2)}$ &  \\
   $U(1)\times Z_2$ & $\mbz\times \mbz^2_{2}$ &  \\
  \hline
\end{tabular}
\caption{Topological classification of gapped D=2+1 dimensional phases of bosons with short range entanglement (no topological order).}
\label{table1}
\end{ruledtabular}
\end{table*}

We then consider non-chiral states of bosons, with equal number of left and right moving edge modes. In the absence of symmetry we argue that there are no nontrivial topological phases with $|\det{\bf K}|=1$. However, the presence of a symmetry can lead to new topological phases. The main result of this work is a scheme to classify topological phases with $|\det{\bf K}|=1$, that are protected by symmetry.

Given a particular symmetry (\eg time reversal, charge conservation etc.) we study distinct ways in which the symmetry can act on the elementary quasiparticles. The symmetries are realized by a set of symmetry transformations on elementary quasiparticles, which form a (faithful) representation of the symmetry group $G$. Distinct realizations are potentially different phases - analogous to the space group classification of crystals. However, an additional requirement to realize a non-trivial topological phase is the existence of symmetry protected edge states, \ie {\em either the edge is gapless, or if it is gapped, it must spontaneously break the symmetry}. Note, an internal symmetry can never provide such protection to a purely one-dimensional system - hence the edge states enjoy these special properties by virtue of their connection to the bulk topological phase. We will call such phases Symmetry Protected Topological ({\bf SPT}) phases following the terminology of \Ref{Chen2011}. To access these states, we supplement the Chern-Simons action with insertion/removal of ``local" quasiparticles that are bosonic and have trivial mutual statistics with any other excitation. Symmetry imposes additional, and crucial, constraints on the possible terms. The set of these symmetry-allowed perturbations can be used to analyze if stable edge states exist or not. For the most part we restrict our attention to $2\times 2$ $K$ matrices.

Remarkably, this simple analysis yields the same set of interacting topological phases as the group cohomology classification of Chen \emph{et al} for a large set of symmetry groups $G$ (see table \ref{table1}). For example, bosonic phases with $G=U(1)$ are classified by an integer which is just the quantized Hall conductance in units of $2Q^2/h$ where $Q$ is the unit of boson charge. When $U(1)$ is broken to a discrete subgroup $\mbz_n$, the set of topological phases is also reduced to $\mbz_n$. Similarly both schemes find a $\mbz_2$ classification of bosonic insulators with conserved charge and time reversal symmetry ($G=U(1) \rtimes Z_2^T$, the semidirect product ensures this is the usual relation between charge and time reversal), the analog of fermionic quantized spin Hall insulators, despite the fact that the time reversal operation is ``bosonic" and squares to $+1$. An advantage of the present formulation is that the edge states of these phases are explicit - typically being non-chiral $c=1$ conformal field theory (CFT) when gapless. Moreover, being cast in the familiar Abelian Chern-Simons form, it is amenable to further investigation using standard field theory methods.
We focus on symmetries (such as time reversal) that are realized locally. Spatial symmetries such as translation invariance, inversion \etc will be left for future work. Since we do not make any assumption about spatial uniformity - the topological phases we find are well defined in the presence of disorder.

A disadvantage of our method is that it is less suited to discuss non-Abelian Lie group symmetries, and we are currently restricted to two spatial dimensions, neither of which is a restriction for group cohomology theory\cite{Chen2011}. Also, our method does not automatically produce a group structure for the set of topological states. On general grounds, one expects the set of topological phases protected by a particular symmetry to form an Abelian group, which is automatically satisfied in the group cohomology classification and in the classification of free fermion topological phases. We handle this by defining a group structure on pairs of phases described within the $K$ matrix formulation. With this refinement the group structure of the resulting sets of phases is readily determined.

{\bf Topological Phases of Interacting Fermions:} We extend our discussion to classifying topological phases of interacting fermion, in the absence of topological order.  A key difference from the bosonic case is that symmetries are realized projectively on the fermion fields. We compare our results to a recent super-cohomology classification of interacting fermion phases\cite{Gu2012}. In addition to the relative simplicity of our method, an advantage over super-cohomology classification is that we are able to handle Kramers time reversal symmetry ${\mathcal T^2=(-1)^{\hat{N}_f}}$ ($\hat{N}_f$ is the total fermion number operator). A disadvantage, shared by the super-cohomology classification, is that we are not able to capture chiral or nonchiral states with odd numbers of Majorana edge modes. As expected, we recover the $\mbz_2$ classification of time reversal symmetric quantum spin Hall insulators, from this interacting formalism as well. We also compare our results with the recent work\cite{Qi2012,Yao2012,Ryu2012} on topological phases of interacting fermions with $\mbz_2\times \mbz_2^T$ symmetry.

{\bf Microscopic Quasi-1D Realization:} Finally, to give a deeper insight into the obtained topological phases we utilize a quasi-one-dimensional (coupled wire \cite{Kane2002,Teo2011}) approach to construct a candidate state consistent with the edge content that emerges from the classification. The {$K$} matrix approach naturally suggests such a construction. This sheds light on various paradoxical results such as the fact that there is a bosonic analog of the quantum spin Hall state although time reversal acts only on bosons with $T^2=1$.

Some aspects of this work are similar in spirit to a number of previous works that have discussed the role of symmetry and stability of edge states in various specific contexts\cite{Haldane1995,Kane2005a,Wu2006,Xu2006,Levin2009,Kou2008,Cho2011,Levin2012,Levin2012a}. For example, \Ref{Levin2009,Levin2012a} discussed the stability of edge states in fractional topological insulators. However, the general machinery presented here to generate symmetry protected topological states has not previously been discussed.

\section{$K$ matrix formulation of $2+1$-D topological phases}

It is believed\cite{Read1990,Blok1990a,Wen1992} that $K$ matrix provides a complete classification of all Abelian fraction quantum Hall (FQH) states in $2+1$ dimensions. In this section we briefly review the $K$ matrix formulation of Abelian FQH states. We then discuss how it can be applied to study states without topological order. In particular we point out that in the absence of symmetry, fields that have trivial (or bosonic) self and mutual statistics will be `Higgsed', and the stability of the edge is examined in the presence of these terms. This is then applied to study chiral topological phases of bosons that lack topological order.

\subsection{A brief review of the $K$ matrix formulation}\label{K MAT FQH}

The low-energy effective theory of an Abelian quantum Hall state is captured by:($\hbar=1$)
\begin{eqnarray}\label{cs action}
&\mathcal{L}_{CS}=\frac{\epsilon^{\mu\nu\lambda}}{4\pi}a^I_\mu K_{I,J}\partial_\nu a_\lambda^J-a^I_\mu j_I^\mu+\cdots
\end{eqnarray}
(summation over repeated indices is assumed).  The $a_\mu^I,~I=1,2,\cdots,N$ are internal gauge fields coupled to quasiparticles currents $j_I^\mu$, and $\bf K$ is a symmetric matrix with integer entries. For states built entirely out of underlying bosons, the diagonal elements of ${\bf K}$ are all \emph{even integers}, while for those built from underlying fermions (electrons), at least one diagonal entry is an odd integer.

The topological order is also characterized by the ${K}$ matrix. The ground state degeneracy (GSD) on a torus can be calculated by quantizing the Chern-Simons theory (\ref{cs action}) and is given by\cite{Keski-Vakkuri1993}
\begin{eqnarray}\label{gsd}
\text{GSD on a torus}=|\det{\bf K}|
\end{eqnarray}
We will mainly be interested in states without topological order \ie with $|\det{\bf K}|=1$.

Quasiparticles are characterized by integer vector ${\bf l}$, and couple minimally to the combination $\sum_I l_I a_{\mu}^{I}$. The self (exchange) statistics $\theta$ of a quasiparticle is obtained by integrating out the gauge fields:
\begin{eqnarray}\label{qp statistics}
\theta=\pi{\bf l}^T{\bf K}^{-1}{\bf l}
\end{eqnarray}
while the mutual (braiding) statistics on taking quasiparticle ${\bf l_1}$ around quasiparticle ${\bf l_2}$ is:
\begin{eqnarray}\label{qp statistics_mutual}
\theta_{12}=2\pi{\bf l_1}^T{\bf K}^{-1}{\bf l_2}
\end{eqnarray}

The effective action describing the gapless edge excitations of a FQH state characterized by ${\bf K}$ can also be derived\cite{Wen1995} from gauge invariance of Lagrangian (\ref{cs action}) on a manifold with boundary:
\begin{eqnarray}\label{edge action:FQH}
\mathcal{S}^0_{edge}=\int\frac{\text{d}t\text{d}x}{4\pi}\sum_{I,J}\big(K_{I,J}\partial_t\phi_I\partial_x\phi_J-V_{I,J}\partial_x\phi_I\partial_x\phi_J\big)
\end{eqnarray}
Here $V_{I,J}$ is a positive definite constant matrix, that is non-universal. However, the commutation relations between fields is fixed by the first term that is simply the $K$ matrix in the bulk. The number of right movers $n_+$ and left movers $n_-$ are given by the signature of symmetric matrix ${\bf K}$, \ie the matrix ${\bf K}$ has $n_+$ positive eigenvalues and $n_-$ negative eigenvalues.



One important question is: are different FQH states characterized by different ${K}$ matrices fundamentally different? In other words, can two different ${K}$ matrices represent the same phase? This means that two FQH states have exactly the same set of quasiparticles but these quasiparticles are labeled in two different ways. It turns out\cite{Read1990,Frohlich1990,Wen1992} that a generic change of label (or change of basis) for the same set of quasiparticles is realized by the following $GL(N,\mbz)$ transformation:
\begin{eqnarray}\label{GL(N,Z) guage field}
a_\mu^I\rightarrow\sum_{J}W_{I,J}a^J_\mu,~~~{\bf W}\in GL(N,\mbz).
\end{eqnarray}
Here $GL(N,\mbz)$ denotes all $N\times N$ integer matrix with determinant $\pm1$. After this relabeling of quasiparticles the ${\bf K}$ matrix and currents $j_\mu^I$ transform as
\begin{eqnarray}\label{gl(n,z)}
&{\bf K}\rightarrow {\bf W}^T{\bf K}{\bf W},\\
&\notag j_\mu^I\rightarrow\sum_{J}W_{J,I}j_\mu^J.
\end{eqnarray}
Any two ${K}$ matrices related by such a $GL(N,\mbz)$ transformation represent the same state (in the absence of any global symmetry). It's straightforward to see that physical properties such as the determinant and the signature of a ${K}$ matrix is invariant under such a $GL(N,\mbz)$ transformation.

When there is a $U(1)$ symmetry associated with charge conservation, one couples an external $U(1)$ gauge field $A_\mu$ to the conserved $U(1)$ current with charge $q$ via an integer vector ${\bf t}\equiv(t_1,\cdots,t_N)^T$ called the \emph{charge vector}\cite{Wen1992}. This is incorporated by adding the following term to the Lagrangian (\ref{cs action}) above: $2\pi {\mathcal L}_{\rm charge}=-{q\epsilon_{\mu\nu\lambda}}t_I A_\mu\partial_\nu a_\lambda^I$. By integrating out internal gauge fields $\{a_\mu^I\}$ one can obtains the quantized Hall conductance:
\begin{eqnarray}
\sigma_{xy}=\frac{q^2}{2\pi}{\bf t}^T{\bf K}^{-1}{\bf t}
\end{eqnarray}
and the $U(1)$ charge of a quasiparticle with integer vector $\bf l$ is given by $Q=q{\bf t}^T{\bf K}^{-1}{\bf l}$.

The many-body wavefunction of a multi-layer FQH state described by effective theory (\ref{cs action}) is given by\cite{Halperin1983,Keski-Vakkuri1993}
\begin{eqnarray}\label{multilayer wavefunction}
\Psi_{\bf K}=\prod_{i<j,I,J}\big(z_i^{(I)}-z_j^{(J)}\big)^{{\bf K}_{I,J}}e^{-\sum_{i,I}|z_i^{(I)}|^2/4}.
\end{eqnarray}
in a disc geometry. Here $z_i^{(I)}\equiv x_i^{(I)}+\imth y_i^{(I)}$ denotes the two-dimensional coordinates of the $i$-th particle in the $I$-th layer. Multi-particle pseudopotentials can be constructed\cite{Barkeshli2010a,Ardonne2011,Davenport2012} as ideal Hamiltonians, whose zero-energy ground states are the above multi-layer FQH states.

\subsection{$K$ matrix + Higgs formulation}
\label{K+HIGGS}

The Lagrangian (\ref{cs action})
seems to have $U(1)^N$ symmetry (or $N$ conserved currents) due to the existence of $N$ internal gauge fields $\{a^I_\mu\}$. When these correspond to bosonic excitations (featured by trivial self and mutual statistics with other quasiparticles), and in the absence of any symmetry, one generically does not expect them to be conserved. This can be implemented by adding  terms to the action (\ref{cs action}) that create and destroy these bosonic particles, which we (in the absence of a better phrase) call Higgs terms. To be precise, denote the annihilation operator for a quasiparticle of $I$-th type as $b_I$ and the associated creation operator as $b_I^{-1}\equiv b_I^\dagger$. If an integer vector ${\bf l}=(l_1,\cdots,l_N)^T$ characterizes a boson, then we demand:
\begin{eqnarray}\label{boson}
\pi{\bf l}^T{\bf K}^{-1}{\bf l}=0\mod 2\pi
\end{eqnarray}
and
\begin{eqnarray}\label{boson_mutual}
2\pi{\bf l}^T{\bf K}^{-1}{\bf l'}=0\mod 2\pi
\end{eqnarray}
for trivial mutual (braiding) statistics with all other quasiparticles ${\bf l'}$. Then, in the absence of symmetry we can add a Higgs term
\begin{eqnarray}{\label{Higgs}}
&\delta\mathcal{L}_{CS}=C_{\bf l}\prod_{I}b_I^{l_I}+~h.c.,~~~C_{\bf l}=\text{const.}
\end{eqnarray}
to the Lagrangian $\mathcal{L}_{CS}$ that condenses the boson with vector ${\bf l}$. Note, since this particle has trivial statistics we can dispense with the gauge field in this expression, whose only role here is to keep track of statistics. One can show that the topological properties of the corresponding state, such as the quasiparticle statistics (\ref{qp statistics}) and ground state degeneracy (\ref{gsd}) are not influenced by these Higgs terms. Therefore the generic Lagrangian describing a $2+1$-D gapped Abelian phase is the following
\begin{eqnarray}\label{cs higgs}
\mathcal{L}_{2+1}=\mathcal{L}_{CS}+\sum_{\{{\bf l}={\,\rm bosonic}\}}\big(C_{\bf l}\prod_{I}b_I^{l_I}+~h.c.\big)
\end{eqnarray}
Taken at face value, the Chern-Simons theory which attaches flux to particles would require monopole terms to account for a change in particle number. We have argued above why this may be un-necessary\footnote{In fact, these ``Higgs" terms can be imposed by local perturbations that tunnel between different layers added to the ideal Hamiltonians for multi-layer FQH states (\ref{multilayer wavefunction}). Since there is an energy gap separating the ground states and excited states of ideal Hamiltonian (pseudopotentials), any weak perturbations that is local will not affect the topological properties of the phase.} In any event, as we will see below, the only action of the Higgs terms we will need is their effect on the boundary - which does not suffer from these problems.

{\bf Stability of Edge States:} Now, the action of edge excitations corresponding to bulk Lagrangian (\ref{cs higgs}) is
\begin{eqnarray}\label{edge action}
&\mathcal{S}_{edge}=\mathcal{S}^0_{edge}+\mathcal{S}^1_{edge},\\
&\notag\mathcal{S}^1_{edge}=\sum_{\{{\bf l}={\, \rm bosonic}\}}\tilde C_{\bf l}\int\text{d}t\text{d}x\cos(\sum_I l_I\phi_I+\alpha_{\bf l}).
\end{eqnarray}
where $\mathcal{S}^0_{edge}$ is given in (\ref{edge action:FQH}). The bare action $\mathcal{S}^0_{edge}$ indicates the following Kac-Moody algebra:
\begin{eqnarray}\label{KM algebra}
&[\partial_x\phi_I(x),\partial_y\phi_J(y)]=2\pi\imth{\bf K}^{-1}_{I,J}\partial_x\delta(x-y)
\end{eqnarray}
Notice that each allowed Higgs term (\ref{Higgs}) in the bulk has a one-to-one correspondence with those on the edge in $\mathcal{S}^1_{edge}$, \ie
\begin{equation}
\left [ C_{\bf l}\prod_{I}b_I^{l_I}+~h.c. \right ] \rightarrow \left [ \tilde C_{\bf l}\cos(\sum_I l_I\phi_I+\alpha_{\bf l})\right ]
\end{equation}
While all these perturbations are present at the edge a more stringent requirement needs to be placed if they are to gap out the edge modes. For example, we expect a maximally chiral edge, where all modes move in the same direction, to be stable even in the absence of any symmetry. The required condition can be deduced by studying the commutation relation implied by the Kac-Moody algebra above for the field ${\bf l}^T\cdot {\boldsymbol \phi}=\sum_I l_I\phi_I$:
\begin{eqnarray}
[{\bf l}^T\cdot\partial_x{\bf \phi}(x),{\bf l}^T\cdot \partial_y{\bf\phi}(y)]=2\pi\imth({\bf l}^T{\bf K}^{-1}{\bf l})\partial_x\delta(x-y)
\end{eqnarray}
thus, in order to be able to localize this field at a classical value, and gap out an edge mode, we require that the commutator vanishes \ie
\begin{eqnarray}\label{condition:local}
{\bf l}^T{\bf K}^{-1}{\bf l}=0
\end{eqnarray}
For a maximally chiral state where ${\bf K}^{-1}$ is a positive definite matrix, no non-vanishing vector satisfies this condition. Hence the edge states cannot be gapped.
Similarly, when there are an imbalanced number of right and left moving modes, $n_+\neq n_-$ the system has a net number of chiral modes and we call it a $2+1$-D chiral phase. In a $2+1$-D chiral phase even in the absence of any symmetry there will be gapless edge excitations\cite{Haldane1995}.

To completely gap out an edge one requires equal number of counter-propagating modes, \ie a non-chiral edge. Then, dimension of $\bf K$ matrix $N$ is even and $N/2=n_+= n_-$. Let us call \ie $\cos({\bf l_1}^T{\bf\phi}+\alpha_{\bf l_1})$ and $\cos({\bf l_2}^T{\bf\phi}+\alpha_{\bf l_2})$ \emph{independent Higgs terms} if and only if:
\begin{eqnarray}\label{condition:commute}
{\bf l_1}^T{\bf K}^{-1}{\bf l_1}={\bf l_2}^T{\bf K}^{-1}{\bf l_2}= {\bf l_1}^T{\bf K}^{-1}{\bf l_2}=0
\end{eqnarray}
In this case they form a pair of commuting variables according to the Kac-Moody algebra. According to Heisenberg's uncertainty principle, these mutually commutating fields $\{{\bf l_n}^T{\bf\phi}|n=1,2,\cdots\}$ can be pinned at certain classical values simultaneously, and consequently their associated edge excitations will be gapped. Then, to completely gap out the edge one needs a set of $N/2$ independent Higgs terms that are pairwise commuting. In the absence of any symmetry, this is typically possible. However, in the next section we will see that symmetry can forbid some Higgs terms leading to SPT phases with non-trivial edge structure. Now let us first consider a chiral state of bosons without topological order.

\subsection{A chiral bosonic phase without topological order: The $E_8$ state}

A phase without topological order is characterized by a symmetric ${K}$ matrix with $|\det{\bf K}|=1$. A chiral state in $2+1$-D requires the signature $(n_+,n_-)$ of its ${K}$ matrix to satisfy that $n_+\neq n_-$. Such a state has gapless edge excitations and a non-zero quantized thermal Hall conductance\cite{Kane1996}. There are many such examples for a fermionic system: \eg an integer quantum Hall state whose ${K}$ matrix is the unit matrix of size $N$. On the other hand, in a bosonic system without topological order, the existence of such states is less obvious.

We therefore seek a ${K}$ matrix with the following properties (\rmnum{1}) $|\det{\bf K}|=1$ (\rmnum{2}) the diagonal elements $K_{I,I}$ are all even integers and (\rmnum{3}) a maximally chiral phases, where all the edge states propagate in a single direction. Then, all eigenvalues of $K$ must have the same sign (say positive), so ${\bf K}$ is a positive definite symmetric unimodular matrix.

It is helpful to map the problem of finding such a $\bf K$ to the following crystallographic problem. Diagonalizing ${\bf K}$ and multiplying each normalized eigenvector by the square root of its eigenvalue one obtains a set of primitive lattice vectors ${\rm \bf e}_I$  such that $K_{IJ}={\rm \bf e}_I\cdot{\rm \bf e}_J$. The inner product of a pair of vectors $l_I{\rm \bf e}_I$ and $l'_I{\rm \bf e}_I$ are given by $l'_IK_{IJ}l_J$, while the volume of the unit cell is given by $\left [ {\rm Det}K\right ]^{1/2}$. The latter can be seen by writing the components of the vectors as a square matrix: $[{\bf k}]_{aI}=[e_I]_a$. Then ${\rm Det} k$ is the volume of the unit cell. However, $K_{IJ}=\sum_a k_{aI}k_{aJ}=({\rm \bf k^Tk})_{IJ}$. Thus ${\rm Det}{\bf K} =[{\rm Det}{\bf k}]^2$.

Thus, for a phase without topological order, we require the volume of the lattice unit cell to be unity $[{\rm Det}k]=1$ (unimodular lattice). Furthermore, for a bosonic state, we need that all lattice vectors have even length $l_IK_{IJ}l_J ={\rm even~integer}$, since the $K$ matrix has even diagonal entries (even lattice). It is known that the minimum dimension this can occur in is eight\cite{Serre1973}. In fact, the root lattice of the exceptional Lie group $E_8$ is the smallest dimensional unimodular, even lattice\footnote{Wikipedia entry for $E_8$ root lattice (Gosset lattice): \url{http://en.wikipedia.org/wiki/E8\_(mathematics)\#E8\_root\_system}.}. Such lattices only occur in dimensions that are a multiple of $8$.

A specific form of the $K$ matrix is:
\begin{equation}
K=\left(
    \begin{array}{cccccccc}
      2 & -1 & 0 & 0 & 0 & 0 & 0 & 0\\
      -1 & 2 & -1 & 0 & 0 & 0 & 0 & 0 \\
      0 & -1 & 2 & -1 & 0 & 0 & 0 & -1 \\
      0 & 0 & -1 & 2 & -1 & 0 & 0 & 0 \\
      0 & 0 & 0 & -1 & 2 & -1 & 0 & 0 \\
      0 & 0 & 0 & 0 & -1 & 2 & -1 & 0 \\
      0 & 0 & 0 & 0 & 0 & -1 & 2 & 0 \\
      0 & 0 & -1 & 0 & 0 & 0 & 0 & 2 \\
    \end{array}
  \right)
\end{equation}
 This matrix has unit determinant and all eigenvalues are positive. It defines a topological phase of bosons without topological order, with eight chiral bosons at the edge. Note $K^{-1}$ can be related to $K$ by a $GL(8,\mbz)$ transformation $S^{\rm T} K^{-1} S$ if we take $S=K$, and so they are physically identical.  Thus, even without computing the inverse, all particles are seen to have trivial statistics ($\pi l'^{T}K^{-1}l=2\pi m$). This ${\bf K}$ is the Cartan matrix for $E_8$, hence the name of the state.

This result was previously pointed out by Kitaev\cite{Kitaev2006,Frohlich1988}, utilizing the fact that the central charge of the edge states ($c_-=c-\bar{c}$) of a chiral topological phase are determined by the statistics of emergent excitations only modulo 8. Thus, phases with trivial statistics are allowed whenever $c_-=0\mod8$.

Combining these phases leads to an integer classification of chiral topological states, which are characterized by a quantized thermal hall conductivity \cite{Kane1996} which are integer multiples of the universal thermal conductivity: $\lim_{T\rightarrow 0} \frac{\kappa_{xy}}{T}=8\frac{\pi^2k_B^2}{3h}$.
\section{Incorporating symmetries in $K$ matrix formulation}
\label{SYMMETRY}
In this section we will be interested in incorporating global symmetries into the $K$ matrix + Higgs formulation. This will lead to new symmetry protected topological phases. We only consider internal symmetries, spatial symmetries like inversion, translation \etc will not be discussed.


Now let us restrict ourselves to $2+1$-D {\em non-chiral} phases with equal numbers of counter propagating modes $n_+=n_-$ for signature $(n_+,n_-)$ of matrix ${\bf K}$.  In the absence of symmetry any edge Higgs term that satisfies (\ref{Higgs}) can be added. In such a phase there will be no gapless edge excitations in the absence of any symmetry, \ie all edge modes will be gapped out by the Higgs terms $\mathcal{S}^1_{edge}$.

However this is not true any more when there are symmetries in the system. In the presence of symmetry, only those bosonic quasiparticles which transform trivially under the symmetry operation can condense. This means certain Higgs terms which transform nontrivially under the symmetry operation are not allowed and cannot be added to effective theory (\ref{cs higgs}) or (\ref{edge action}).

How do quasiparticles transform under a symmetry operation?
In general the Lagrangian (\ref{cs action}) and (\ref{cs higgs}) should be invariant under the symmetry transformation on quasiparticle currents $\{j^\mu_I\}$. Notice that when a $K$ matrix is acted on by a $GL(N,\mbz)$ transformation (\ref{gl(n,z)}), it describes the same physical state. Only the labels of different quasiparticles are changed. Given a state described by a certain ${\bf K}$ matrix, the allowed $GL(N,\mbz)$ transformations ${\bf W}$ which transform the quasiparticles under symmetry must leave the ${K}$ matrix invariant \ie ${\bf K}={\bf W}^T{\bf K}{\bf W}$. Besides any global $U(1)$ phase transformation on the quasiparticle annihilation operator $b_I\rightarrow e^{\imth\delta\phi_I}b_I$ also keeps the Lagrangian (\ref{cs action}) invariant. Notice that such a phase shift $\delta\phi_I$ is defined modulo $2\pi$ due to the quantization of quasiparticle number in a compact theory.

Therefore a generic realization of unitary symmetry $g$ on the $\{\phi_I\}$ fields at the edge (we will not require the transformation law in the bulk in the following) is given by
\begin{eqnarray}
&\phi_I\rightarrow\sum_J W^g_{I,J}\phi_I+\delta\phi_I^g.\label{unitary}
\end{eqnarray}
where $\delta\phi_I^g\in[0,2\pi)$ are constants and matrix $W^g\in GL(N,\mbz)$ satisfies
\begin{eqnarray}\label{unitary:gl(n,z)}
{\bf K}=(W^g)^T{\bf K}W^g.
\end{eqnarray}

For an anti-unitary symmetry $h$ (such as time reversal symmetry $Z_2^T$), in general it is realized in the following way:
\begin{eqnarray}
&\phi_I\rightarrow-\sum_J W^h_{I,J}\phi_J+\delta\phi_I^h.\label{antiunitary}
\end{eqnarray}
where $\delta\phi_I^h\in[0,2\pi)$ are constants and matrix $W^h\in GL(N,\mbz)$ satisfies
\begin{eqnarray}\label{antiunitary:gl(n,z)}
{\bf K}=-(W^h)^T{\bf K}W^h.
\end{eqnarray}
The anti-unitary symmetry operation $h$ is realized by the above transformations followed by complex conjugation $\mathcal{C}$. Notice that ${K}$ matrix changes sign under the above $GL(N,\mbz)$ transformation since a Chern-Simons term $\epsilon^{\mu\nu\lambda}a^I_\mu\partial_\nu a^J_\lambda$ always changes sign under time reversal. 

{\bf Group compatibility conditions:} It may appear we have wide latitude in determining how the generators of a symmetry group act on the quasiparticles in our theory. However, there is an important constraint. For an arbitrary symmetry group $G$, the multiplication rule of its group elements is completely determined by certain algebraic relations of the group generators $\{g_1,g_2,\cdots\}$:
\begin{eqnarray}\label{algebra}
\mathcal{A}_{\{n_a\}}\equiv\prod_ag_a^{n_a}=\bse
\end{eqnarray}
where $\bse$ is the identity element of group $G$ and $\{n_a\}$ are all integers. A bosonic quasiparticle (which satisfied Eqn. \ref{boson}) is a physical excitation and must transform trivially under the identity element.
Thus all boson insertion operators satisfying (\ref{boson}) and (\ref{boson_mutual}) should be invariant under the symmetry operation $\mathcal{A}_{\{n_a\}}$:
\begin{eqnarray}\label{condition:algebra}
&\mathcal{A}_{\{n_a\}}:~~~\sum_{I}l_I\phi_I\rightarrow\sum_{I}l_I\phi_I\mod2\pi,\\
&\notag\forall~{\bf l}~\text{satisfying}~{\bf l}^T{\bf K}^{-1}{\bf l}=0\mod 2.
\end{eqnarray}
These algebraic requirements serve as constraints to the possible $GL(N,\mbz)$ transformations $W^{g_a}$ and $U(1)$ phase rotations $\{\delta\phi_I^{g_a}\}$ and we shall call them \emph{group compatibility conditions} for the edge states described by effective theory (\ref{edge action}) and their associated bulk topological phases.
In the case of bosonic phases without topological order, ${\bf K}$ is a symmetric unimodular matrix whose diagonal elements are all even integers. Then, any integer vector ${\bf l}$ satisfies the conditions (\ref{boson}) and (\ref{boson_mutual}), \ie all quasiparticles of a bosonic SRE phase are bosons. Therefore the group compatibility conditions (\ref{condition:algebra}) for symmetry transformations are simplified as
\begin{eqnarray}\label{condition:algebra:bosonic SRE}
&\text{Under}~\mathcal{A}_{\{n_a\}}=\bse:~~~\phi_I\rightarrow\phi_I\mod2\pi,\\
&\notag~~~I=1,2,\cdots,N.
\end{eqnarray}
By solving these algebraic equations we can find out all sets of inequivalent symmetry transformations $\{W^{g_a},\delta\phi_I^{g_a}\}$ for generators $\{g_a\}$ of group $G$.

As an aside we note that for phases with topological order, or in the presence of fermionic quasiparticles, symmetries are realized projectively\cite{Wen2002}. Then, even the identity elements (\ref{algebra}) can induce a nontrivial transformation on quasiparticles.

{\bf Gauge Equivalence:} A question naturally arises: do different symmetry transformations represent different SPT phases? We answer this question in two parts. First we comment on the equivalency or inequivalency between two sets of symmetry transformations $\{W^{g_a},\delta\phi_I^{g_a}\}$. Notice that one can always change the label of quasiparticles by a $GL(N,\mbz)$ transformation $X$ (as long as $X^T{\bf K}X={\bf K}$), or perform a global $U(1)$ gauge transformation $\phi_I\rightarrow\phi_I+\Delta\phi_I$ which keeps Lagrangian (\ref{cs action}) invariant. Under such a ``gauge" transformation the symmetry operations $\{W^{g_a},\delta\phi_I^{g_a}\}$ will transform as
\begin{eqnarray}\label{gauge transformation on symmetry}
&W^g\rightarrow X^{-1}W^gX,\\
&\notag\delta\phi^g_I\rightarrow X^{-1}\Big[\delta\phi^g_I-\Delta\phi_I+\eta\sum_JW^g_{I,J}\Delta\phi_J\Big],\\
&\notag \text{if}~~~X\in GL(N,\mbz),~X^T{\bf K}X={\bf K}.
\end{eqnarray}
where $\eta=\pm1$ if $g$ is a unitary (anti-unitary) symmetry. If two sets of symmetry operations $\{W^{g_a},\delta\phi_I^{g_a}\}$ associated with group $G$ generated by $\{g_a\}$ are related by the above gauge transformation (\ref{gauge transformation on symmetry}), then these two sets of symmetry operations are essentially identical.

{\bf Edge Stability and Criteria for SPT Phases:} With a set of symmetry transformations $\{W^{g_a},\delta\phi_I^{g_a}\}$ one can determine what Higgs terms in (\ref{cs higgs}) and (\ref{edge action}) that are allowed in the presence of the symmetry group $G=\{g\}$. In general only those Higgs terms transform trivially under the symmetry transformations $\{W^{g_a},\delta\phi_I^{g_a}\}$ are allowed, which induce certain allowed set of edge perturbations. To determine the fate of possible gapless modes at the edge, one considers terms that commute with each other and with themselves, \ie terms involving variables $\sum_I l_I\phi_I$ that satisfy  conditions (\ref{condition:local}) and (\ref{condition:commute}). Then one can simultaneously minimize these terms like classical variables. Note, scaling dimensions of these edge terms are immaterial to this discussion.


{\em\underline{Criterion for Trivial Phase}:} If there is a set of independent Higgs terms $\{C_{a}\cos({\bf l}^T_a\phi+\alpha_{a})\}$ allowed by symmetry, so that any other variables ${\bf l}^T\phi$ on the edge is either a linear combination of these bosonic variables $\{{\bf l}^T_a\phi\}$ or doesn't commute with every condensed bosonic quasparticles in $\{{\bf l}^T_a\phi\}$, then the edge of the system will be completely gapped in the presence of independent Higgs terms $\{C_{a}\cos({\bf l}^T_a\phi+\alpha_{a})\}$. When the independent Higgs terms $\{C_{a}\cos({\bf l}^T_a\phi+\alpha_{a})\}$ are simultaneously minimized on the edge, the \emph{elementary bosonic variables} $\{{\bf v}_a^T\phi\}$
\begin{eqnarray}\label{elementary independent bosons}
{\bf v}_a=(v_{a,1},\cdots,v_{a,N})\equiv\frac{{\bf l}_a}{\text{gcd}(l_{a,1},l_{a,2},\cdots,l_{a,N})},~\forall~a.
\end{eqnarray}
will all condense and be localized at various classical values $\langle{\bf v}_a^T\phi\rangle=B_a$ ($\text{gcd}$ is short for greatest common divisor). Notice that if the set of independent elementary bosonic variables $\{{\bf v}^T_a\phi\}$ are invariant under any symmetry transformation \ie
\begin{eqnarray}
\text{Under}~\forall~g\in G:~~~\{\sum_Iv_{a,I}\phi_I\}\rightarrow\{\sum_Iv_{a,I}\phi_I\}.
\end{eqnarray}
then the edge states can be all gapped out without breaking the symmetry $G$ at all. We call such a non-chiral SRE phase a \emph{trivial} phase since in general it doesn't support gapless edge states in the presence of symmetry $G$. These principles will be illustrated by examples in detail in the following section. In comparison to the aforementioned trivial phase, \emph{a nontrivial SPT phase has a gapless edge structure which cannot be gapped without breaking the symmetry $G$}. The only two possible situations for the edge structure of such a phase are:

(\rmnum{1}) {\em Gapless:} The maximal set of independent Higgs terms $\{C_{a}\cos({\bf l}^T_a\phi+\alpha_{a})\}$ allowed by symmetry cannot gap out all the edge states. In other words, on the edge there exist at least one variable ${\bf l}\phi$ that commutes with all the condensed bosonic variables $\{{\bf l}^T_a\phi\}$. Hence this degree of freedom ${\bf l}\phi$ remains gapless even in the presence of all the symmetry-allowed independent Higgs terms $\{C_{a}\cos({\bf l}^T_a\phi+\alpha_{a})\}$. An example is the bosonic SPT phases protected by $U(1)$ symmetry in $2+1$-D, as will be discussed in detail later.

OR

(\rmnum{2}) {\em Spontaneous Symmetry Breaking on the Edge:} Although all the edge states will be gapped in the presence of independent Higgs terms $\{C_{a}\cos({\bf l}^T_a\phi+\alpha_{a})\}$ allowed by symmetry, not all of the associated elementary bosonic variables $\{{\bf v}^T_a\phi\}$ in (\ref{elementary independent bosons}) are invariant under symmetry transformations $\{\eta_gW^g,\delta\phi^g\},~\forall g\in G$ where $G$ is the symmetry group. This means at least one elementary bosonic variable ${\bf v}^T_a\phi$ in (\ref{elementary independent bosons}) would transform nontrivially under symmetry group $G$. Therefore in order to gap out the edge by condensing all the independent elementary bosons, one has to spontaneously break the symmetry on the edge. An example is the bosonic SPT phase protected by $U(1)\rtimes Z_2^T$ symmetry (or by $Z_2$ symmetry) in $2+1$-D, as will be shown later.

{\bf Group structure of phases protected by symmetry group $G$:}  In general, the set of different {\em phases} that appear in a topological classification are expected to form an Abelian group, as proved for non-interacting fermions\cite{Kitaev2009,Wen2012} and conjectured for interacting bosonic systems\cite{Chen2011} (since group cohomology classification leads to an Abelian group). How does this group structure appear within our ${K}$ matrix formulation? Let $\{\Psi_G[{\bf K},\{W^{g_a},\delta\phi^{g_a}\}]\}$ to denote the set of trivial and SPT phases in the presence of symmetry $G$. Here $\Psi_G[{\bf K},\{W^{g_a},\delta\phi^{g_a}\}]$ represents a phase with matrix ${\bf K}$ in action (\ref{edge action}) and transformation rules $\{W^{g_a},\delta\phi^{g_a}\}$ for symmetry group  $G=\{g_a\}$. We would like to attach to this set a group product.

A natural Abelian product rule $\oplus$ of two phases $\Psi_G[{\bf K},\{W^{g_a},\delta\phi^{g_a}\}]$ and $\Psi_G[\tilde{\bf K},\{\tilde{W}^{g_a},\tilde{\delta\phi}^{g_a}\}]$ is to take their matrix direct sum:
\begin{eqnarray}
&\notag\Psi_G[{\bf K},\{W^{g_a},\delta\phi^{g_a}\}]\oplus\Psi_G[\tilde{\bf K},\{\tilde{W}^{g_a},\tilde{\delta\phi}^{g_a}\}]\\
&=\Psi_G[{\bf K}\oplus\tilde{\bf K},\{W^{g_a}\oplus\tilde{W}^{g_a},\delta\phi^{g_a}\oplus\tilde{\delta\phi}^{g_a}\}].\label{mupliplication rule:group of phases}
\end{eqnarray}
This seems to suggest that one cannot obtain a full classification of all different SPT phases when restricted to a ${K}$ matrix with a fixed dimension. However notice that two phases $\Psi_G^1$ and $\Psi_G^2$ can be identified as the same one if $\Psi_G^1\oplus\bse_G=\Psi_G^2$, \ie adding a trivial phase (denoted by $\bse_G$) to $\Psi_G^1$ yields the phase $\Psi_G^2$. We can use this fact to reduce the dimensions of ${K}$ matrix by throwing away the ``trivial" parts of the edge structure, which can be gapped without breaking any symmetry.

The identity element in the group $\{\Psi_G\}$ corresponds to the trivial phase $\bse_G\equiv\Psi_G[{\bf K}_0,W^{g_a}\equiv I_{N\times N},\delta\phi^{g_a}\equiv0]$, where ${\bf K}_0$ can be any $N\times N$ unimodular symmetric matrix corresponding to a non-chiral SRE phase in $2+1$-D. It edge states can be gapped out without breaking the symmetry of group $G$.

We can also define the ``inverse" of a phase $\Psi_G[{\bf K},\{W^{g_a},\delta\phi^{g_a}\}]$ in the group to be
\begin{eqnarray}\label{inverse:group of phase}
\Psi_G[{\bf K},\{W^{g_a},\delta\phi^{g_a}\}]^{-1}=\Psi_G[-{\bf K},\{W^{g_a},\delta\phi^{g_a}\}].
\end{eqnarray}
\ie by changing the sign of its ${K}$ matrix we obtain the inverse of a phase. This is simply because we can always gap out the edge of
\begin{eqnarray}
&\notag\Psi_G[{\bf K},\{W^{g_a},\delta\phi^{g_a}\}]\oplus\Psi_G[-{\bf K},\{W^{g_a},\delta\phi^{g_a}\}]=\bse_G
\end{eqnarray}
without breaking the symmetry. Consider the variables $\{\phi_I\}$ of phase $\Psi_G^{-1}[{\bf K},\{W^{g_a},\delta\phi^{g_a}\}]$ and $\{\tilde\phi_I\}$ of phase $\Psi_G[-{\bf K},\{W^{g_a},\delta\phi^{g_a}\}]$ on the edge. Then edge perturbations such as: $\{C_a\cos\big[{\bf l}^T_a(\phi-\tilde\phi)+\alpha_a\big]\}$ will not be affected by the phase factors $\delta\phi^{g_a}$. We can then condense a set of independent elementary bosonic variables $\{{\bf l}^T_a(\phi-\tilde\phi)\}$ satisfying (\ref{condition:local}) and ($\ref{condition:commute}$). This can be readily shown when either $W=I$ (arbitrary $\bf K$) or when $ \bf K$ is a $2\times 2$ matrix (arbitrary $W$). This is simply because $\{\phi_I\}$ and $\{\tilde\phi_I\}$ satisfies the Kac-Moody algebra with opposite ${K}$ matrices. No symmetry will be broken by condensing these bosons, when a proper set of vectors $\{{\bf l}_a\}$ are chosen. To check if two putative SPT states are the same phase or are different phases, we use the above group multiplication rules to combine one state with the inverse of the other and check if a trivial phase results. If so, these two are the same SPT phase.

Now that the identity element, the inverse of an element and the multiplication rules are defined, we can identify the group structure. We will perform this analysis below to clarify the connection between phases generated by our formalism.

{\bf Miscellaneous Considerations:}
We focus on the cases where $\{W^{g_a},\delta\phi_I^{g_a}\}$ form a faithful representation of symmetry group $G$, where $G$ is the symmetry group of the system's Hamiltonian. The case when it forms an unfaithful representation, \ie when more than one group element acts like the identity on all quasiparticles (this set of group elements form an invariant subgroup $H$), actually corresponds to the faithful representation of the quotient group ($G/H$). Thus, studying faithful representations suffices. This is discussed in more detail in Appendix \ref{Appendix:faithful}. In some cases we will find that a solution for symmetry transformation may not be realizable in a theory with local fields (for example one that exchanges a pair of fields that are canonical conjugate of each other in a $c=1$ edge), which will then not be included in the minimal set of SPT phases. For phases that are reported in Table \ref{table1} we have checked that they have symmetries that can be realized starting from microscopic degrees of freedom (such as from the coupled wire construction).

In the following we illustrate the above principles, first by classifying bosonic SPT phases in $2+1$-D, whose topology is protected by a certain symmetry group $G$.

\section{$K$-matrix classification of bosonic SPT phases}\label{BOSONIC SPT}

In this section we will focus on bosonic non-chiral SRE states in the presence of certain symmetry, \ie bosonic non-chiral SPT phases in $2+1$-D. They are described by a symmetric, unimodular ${K}$ matrix, whose diagonal elements are even integers and with the same number of positive eigenvalues and negative eigenvalues ($n_+=n_-$). Therefore the dimension of matrix ${\bf K}$ must be \emph{even} and:
\begin{eqnarray}
\det{\bf K}=(-1)^{\text{dim}({\bf K})/2}~\text{for a non-chiral SRE phase}.\notag
\end{eqnarray}
In this section we restrict ourselves to a $2\times2$ ${K}$ matrix. SPT phases that are necessarily described by ${K}$ matrices of a larger size, may be missed by this restriction. However our results are internally consistent and also capture at least all the topological states of the group cohomology classification of bosonic SPT phases in \Ref{Chen2011}. It appears that a ${K}$ matrix of dimension 2 is sufficient to represent and classify SPT phases in $2+1$-D in many cases. The reason behind this unexpected success is the following: although we focus on SPT phases described by a $K$ matrix of size $2\times2$, when analyzing the group structure formed by SPT phases with symmetry $G$ we need to multiply two phases together by a direct sum of their $K$ matrices. Consequently we are in fact considering $K$ matrices of size $2n\times 2n$ obtained from direct sums of original $2\times2$ $K$ matrices. Therefore it's actually not surprising that many of the bosonic SPT phases in $2+1$-D can be described and classified by a $2\times2$ $K$ matrix and associated symmetry transformations.

As proved in Appendix \ref{app:K2x2} a $2\times2~{K}$ matrix with determinant $-1$ for a bosonic system (see $n=1$ in theorem (\ref{K2x2:standard form})) is always equivalent to the standard form $\sigma_x=\begin{pmatrix}0&1\\1&0\end{pmatrix}$ by certain $GL(N,\mbz)$ transformations ($\sigma_\alpha,~\alpha=x,y,z$ are Pauli matrices). In the following we always choose the $2\times2$ matrix ${\bf K}=\sigma_x$  to represent a generic bosonic non-chiral SRE state. In the following we use general principles discussed earlier in section (\ref{SYMMETRY}) to study non-chiral bosonic SPT phases with different symmetries. Note that the $GL(2,\mbz)$ transformations ${\bf X}$ that keeps ${\bf K}=\sigma_x$ invariant under (\ref{gl(n,z)}) are ${\bf X}=\pm I_{2\times2},~\pm\sigma_x$. For such a non-chiral bosonic SRE phase, the unperturbed edge theory is:
\begin{eqnarray}\label{bare:boson:edge}
&\mathcal{S}_{edge}^0=\frac1{4\pi}\int\text{d}t\text{d}x\big(\partial_t\phi_1\partial_x\phi_2+\partial_t\phi_2\partial_x\phi_1\\
&\notag-\sum_{I,J}V_{I,J}\partial_x\phi_I\partial_x\phi_J\big)
\end{eqnarray}
where $V_{I,J}$ is a positive definite constant matrix, as discussed in section \ref{K+HIGGS}. This implies the following commutation relations (Kac-Moody algebra) for the edge fields $\{\phi_1,\phi_2\}$:
\begin{eqnarray}
&\notag[\partial_x\phi_1(x),\partial_y\phi_1(y)]=[\partial_x\phi_2(x),\partial_y\phi_2(y)]=0,\\
&[\partial_x\phi_1(x),\partial_y\phi_2(y)]=2\pi\imth\partial_x\delta(x-y).\label{KM algebra:boson non-chiral SRE}
\end{eqnarray}
In the absence of any symmetry, the edge states in (\ref{bare:boson:edge}) can always be gapped, since a set of independent Higgs terms satisfying (\ref{condition:local}) and (\ref{condition:commute}) on the edge can be choose as either $\{C_l\cos(l\phi_1+\alpha_l),~l\in\mbz\}$ or $\{C_l\cos(l\phi_2+\alpha_l),~l\in\mbz\}$. These added terms destroy the edge states. All degrees of freedom on the edge are gapped when variable $\phi_1$ (or $\phi_2$) is localized at a classical value.

Now, let us consider various symmetries.


\subsection{$Z_2^T$ symmetry: $\mbz_1$ class}\label{BOSON SPT:Z2T}

$Z_2^T$ symmetry (time reversal) is generated by $\bst$. The algebra (\ref{algebra}) which defines $Z_2^T$ symmetry group is
\begin{eqnarray}\label{symmetry:T}
\bst^2=\bse
\end{eqnarray}
where $\bse$ is the identity operation. Time reversal symmetry $\bst$ is implemented (following rules (\ref{antiunitary})) by a matrix $W^\bst\in GL(2,\mbz)$ and a vector of $U(1)$ phase changes ${\bf\delta\phi}_\bst$ (defined modulo $2\pi$) satisfying the constraint (\ref{condition:algebra:bosonic SRE}) and (\ref{antiunitary:gl(n,z)}) for a bosonic SRE system
\begin{eqnarray}
&\label{z2t:constraint:W} (W^\bst)^2=I_{2\times2},~~~(W^\bst)^T{\bf K}W^\bst=-{\bf K};\\
&\label{z2t:constraint:dphi} \delta\phi_I^\bst-\sum_JW^\bst_{I,J}\delta\phi_J^\bst=0\mod2\pi,~I=1,2.
\end{eqnarray}
The only $GL(2,\mbz)$ matrix solutions to (\ref{z2t:constraint:W}) are $W^\bst=\pm\sigma_z$. Notice that $W^\bst=\sigma_z$ and $W^\bst=-\sigma_z$ are related a gauge transformation (\ref{gauge transformation on symmetry}) $X=\sigma_x$. Therefore one can always choose
\begin{eqnarray}
&\notag W^\bst=+\sigma_z.
\end{eqnarray}
by a proper gauge fixing. Then the constraint (\ref{z2t:constraint:dphi}) becomes $(I_{2\times2}-\sigma_z)\delta\phi^\bst=0\mod2\pi$ and it leads to
\begin{eqnarray}\label{sym trans:z2t}
&\notag\delta\phi^\bst_2=n_2\pi\mod2\pi,~~~n_2=0,1.
\end{eqnarray}
Under a gauge transformation $\Delta\phi_I$ in (\ref{gauge transformation on symmetry}) the compact $U(1)$ phase shift $\delta\phi^\bst$ transforms to $\delta\phi^\bst-(I_{2\times2}+\sigma_z)\Delta\phi$. As a result we can always choose a gauge so that
\begin{eqnarray}
&\notag\delta\phi^\bst_1=0\mod2\pi.
\end{eqnarray}
So a generic bosonic non-chiral SPT phase in the presence of time reversal symmetry $\bst$ has symmetry transformation
\begin{eqnarray}
\{W^\bst,\delta\phi^\bst\}=\{\sigma_z,\begin{pmatrix}0\\n_2\pi\end{pmatrix}\}~~~n_2=0,1.
\end{eqnarray}
Since each bulk Higgs term has a one-to-one correspondence with that on the edge, hereafter we'll only write down those Higgs terms $\tilde C_{\bf l}\cos(\sum_I l_I\phi_I+\alpha_{\bf l})$ on the edge. In the case of $Z_2^T$ symmetry the allowed Higgs terms are
\begin{eqnarray}
&\notag\mathcal{S}^1_{edge}=\sum_{l_1\geq0,l_2}C_{\bf l}\int\text{d}x\text{d}t\Big[\cos(l_1\phi_1+l_2\phi_2+\alpha_{\bf l})\\
&\notag+\cos(-l_1\phi_1+l_2\phi_2+n_2l_2\pi+\alpha_{\bf l})\Big].
\end{eqnarray}
For $n_2=0,1$ the allowed Higgs terms are different, \eg $\cos(\phi_2)$ terms are allowed for $n_2=0$ but not allowed for $n_2=1$. Thus there is a distinction between these states. However, this is not a topological distinction as argued below. For both  $n_2=0$ and $n_2=1$ cases we can write the same set of symmetry-allowed \emph{independent} Higgs terms formed by mutually commuting operators (\ref{KM algebra:boson non-chiral SRE}): \ie
\begin{eqnarray}
&\mathcal{S}^1_{edge}=\sum_{l_1}C_{l_1,0}\int\text{d}x\text{d}t\cos\Big(l_1\phi_1(x,t)\Big).
\end{eqnarray}
 Thus, in both cases the edge can be gapped, therefore they belong to the same trivial phase.If the variable $\phi_1$ is localized at expectation value eg. $\langle\phi_1\rangle=0$ by the Higgs terms, all excitations on the edge would be gapped but the time reversal symmetry ($\phi_1\rightarrow-\phi_1$) is not broken by this expectation value.

\subsection{$U(1)$ symmetry: $\mbz$ classes}

The elements of $U(1)$ group can be labeled as $U_\theta$ where $\theta\in[0,2\pi)$ and the identity element is $U_0$. The multiplication rule is given by
\begin{eqnarray}\label{symmetry:U(1)}
U_{\theta_1}U_{\theta_2}=U_{(\theta_1+\theta_2\mod2\pi)}
\end{eqnarray}
and therefore
A generic form of symmetry transformations $\{W^{g_a},\delta\phi_I^{g_a}\}$ satisfying constraint (\ref{condition:algebra:bosonic SRE}) for $U(1)$ group is
\begin{eqnarray}
&\{W^{U_\theta}=I_{2\times2},\delta\phi^{U_\theta}=\theta{\bf t}=\theta\begin{pmatrix}t_1\\t_2\end{pmatrix}\},~~~(t_1,t_2)=1.\notag
\end{eqnarray}
where $(t_1,t_2)$ denotes the \emph{greatest common divisor} of integers $t_1$ and $t_2$. Notice that only when $(t_1,t_2)=1$ the above symmetry transformations $\{W^{U_\theta},\phi^{U_\theta}\}$ form a faithful representation of symmetry group $U(1)$, which is what we assume here.\footnote{In general if $(t_1,t_2)=t>1$ symmetry transformations $\{W^{U_\theta},\phi^{U_\theta}\}$ will form an unfaithful representation of symmetry group $U(1)$. Then the symmetry group $G_\Psi$ of ground state will be broken from $G=U(1)$ (symmetry group of Hamiltonian) down to $\mbz_t$ if the edge is gapped by condensing bosons in the presence of symmetry-breaking independent Higgs terms.}

Here ${\bf t}^\prime\equiv{\bf K}{\bf t}=\begin{pmatrix}t_2\\t_1\end{pmatrix}$ is nothing but the charge vector defined in the context of ${K}$ matrix formulation of a FQH state (see section \ref{K MAT FQH}). As proved in Appendix \ref{app:K2x2:boson:U(1)} for a bosonic SRE phase with ${\bf K}=\sigma_x$, an arbitrary charge vector ${\bf t}$ with $(t_1,t_2)=1$ is equivalent to the standard form
\begin{eqnarray}
&{\bf t}=\begin{pmatrix}t_1\\t_2\end{pmatrix}\simeq\begin{pmatrix}q\\1\end{pmatrix}\simeq\begin{pmatrix}1\\q\end{pmatrix},~q\in\mbz.\notag
\end{eqnarray}
by certain $GL(N,\mbz)$ gauge transformations. Therefore the inequivalent symmetry transformations under constraint (\ref{condition:algebra:bosonic SRE}) for a bosonic non-chiral SRE phase with $U(1)$ symmetry are
\begin{eqnarray}
&\{W^{U_\theta}=I_{2\times2},\delta\phi^{U_\theta}=\theta\begin{pmatrix}1\\q\end{pmatrix}\},~~~q\in\mbz.\notag
\end{eqnarray}
The associated symmetry-allowed Higgs terms are
\begin{eqnarray}
&\mathcal{S}^1_{edge}=\sum_{l\in\mbz}C_{l}\int\text{d}x\text{d}t\cos\Big[l~\big(q\phi_1-\phi_2\big)\Big].
\end{eqnarray}
Due to Kac-Moody algebra (\ref{KM algebra:boson non-chiral SRE}), they do not form a set of independent Higgs terms satisfying (\ref{condition:local})-(\ref{condition:commute}) unless $q=0$!  When $q=0$ the Higgs term can localize the variables $\phi_2$ to expectation value $\langle\phi_2\rangle=\text{const.}$, which gaps the edge excitations without breaking $U(1)$ symmetry ($\phi_1\rightarrow\phi_1+\theta$).

Now we'll determine the group structure formed by these phases. Let's label a phase by
\begin{eqnarray}
&[q]\equiv\Psi_{U(1)}[{\bf K}=\sigma_x,\{W^{U_\theta}=I_{2\times2},\delta\phi^{U_\theta}=\theta\begin{pmatrix}1\\q\end{pmatrix}\}]\notag
\end{eqnarray}
where $\bse_{U(1)}=[0]$ is the trivial phase. Consider two states $[q_1]$ with edge variables $\{\phi_1,\phi_2\}$ and $[q_2]$ with edge variables $\{\phi^\prime_1,\phi^\prime_2\}$. When they are put together the following independent edge terms are symmetry allowed: $\sum_{l\in\mbz}C_l\cos\big[l(\phi_1-\phi_1^\prime)+\alpha_l\big]$ and their associated elementary bosonic variable is $\phi_1-\phi_1^\prime$. If $q_2=-q_1$, independent Higgs terms $\sum_{l\in\mbz}C_l^\prime\cos\big[l(\phi_2+\phi_2^\prime)+\alpha_l^\prime\big]$ are also allowed by $U(1)$ symmetry. Therefore the edge states will be fully gapped without breaking the $G=U(1)$ symmetry, by condensing elementary independent bosons $\{\phi_1-\phi_1^\prime,\phi_2+\phi_2^\prime\}$. Therefore we have
\begin{eqnarray}
[q]^{-1}=[-q],~~~\forall~q\in\mbz.
\end{eqnarray}
On the other hand if $q_1\neq-q_2$ the bosonic variable $\{\phi_2+\phi_2^\prime\}$ cannot be gapped since $\cos\big[l(\phi_2+\phi_2^\prime)+\alpha_l^\prime\big]$ terms are not allowed by symmetry. Now the new variables describing the gapless edge structure can be chosen as $\{\tilde\phi_1=\phi_1,\tilde\phi_2\equiv\phi_2+\phi_2^\prime\}$ satisfying Kac-Moody algebra (\ref{KM algebra:boson non-chiral SRE}). Notice that under $U(1)$ they transform as $\tilde{W}^{U_\theta}=I_{2\times2}$ and
\begin{eqnarray}
\notag\begin{pmatrix}\delta\tilde\phi_1^{U_\theta}\\ \delta\tilde\phi_2^{U_\theta}\end{pmatrix}=\theta\begin{pmatrix}1\\q_1+q_2\end{pmatrix}
\end{eqnarray}
Therefore we have the multiplication rule of the group formed by phases $[q]$ with $U(1)$ symmetry:
\begin{eqnarray}
[q_1]\oplus[q_2]=[q_1+q_2]
\end{eqnarray}
This means phases $[q]$ labeled by different integer $q$'s are different phases in the presence of $U(1)$ symmetry, and they form nothing but the integer group $\mbz$!
Any phase $[q]$ with $q\neq0$ corresponds to a nontrivial SPT phase, whose gapless edge states cannot be gapped without breaking $U(1)$ symmetry.

There is a simple physical reason underlying these observations. The Hall conductance is a physical invariant that distinguishes these different phases:
\begin{eqnarray}
&\sigma_{xy}=\big({\bf t}^\prime\big)^T{\bf K}^{-1}{\bf t}^\prime=\begin{pmatrix}1\\q\end{pmatrix}^T{\bf K}\begin{pmatrix}1\\q\end{pmatrix}=2q.
\end{eqnarray}

\subsection{$U(1)\rtimes Z_2^T$ symmetry: $\mbz_2$ classes}

In the presence of both ``charge" $U(1)$ (group elements $U_\theta$) and time reversal $Z_2^T$ symmetry (generator $\bst$), the extra algebraic relation in addition to (\ref{symmetry:T}) and (\ref{symmetry:U(1)}) is given by
\begin{eqnarray}
&U_{-\theta}\bst=\bst U_{\theta}\notag
\end{eqnarray}
since the charge $U(1)$ symmetry doesn't commute with time reversal symmetry.  The algebraic relations for $U(1)\rtimes Z_2^T$ are
\begin{eqnarray}\label{symmetry:U(1):T_semi}
&\bst^2=\bst U_\theta\bst U_\theta=\bse.
\end{eqnarray}
in addition to (\ref{symmetry:U(1)}). The corresponding constraints (\ref{condition:algebra:bosonic SRE}) for symmetry transformations $\{W^\bst,\delta\phi^\bst\}$ and $\{W^{U_\theta}=I_{2\times2},\delta\phi^{U_\theta}=\theta{\bf t}\}$ are
\begin{eqnarray}
&\label{u1xT_semi:constaint:dphi}(I_{2\times2}-W^\bst)({\bf\delta\phi}^\bst+\theta{\bf t})=\begin{pmatrix}0\\0\end{pmatrix}\mod2\pi,~~~\forall~\theta.
\end{eqnarray}
and (\ref{z2t:constraint:W}-\ref{z2t:constraint:dphi}). Just like in the case of $Z_2^T$ symmetry, again using gauge transformation $X=\sigma_x$ in (\ref{gauge transformation on symmetry}) one can fix $W^\bst=\sigma_z$ to satisfy (\ref{z2t:constraint:W}). Solving (\ref{z2t:constraint:dphi}) and (\ref{u1xT_semi:constaint:dphi}) we have $t_2=0$ and $\delta\phi^\bst_2=n\pi,~n=0,1$. Using gauge transformation $\Delta\phi$ in (\ref{gauge transformation on symmetry}) one can fix $\delta\phi^\bst_1=0$. Hence the inequivalent symmetry transformations for $U(1)\rtimes Z_2^T$ symmetry is
\begin{eqnarray}
&W^{U_\theta}=I_{2\times2},~~~\delta\phi^{U_\theta}=\theta\begin{pmatrix}1\\0\end{pmatrix}\\
&W^\bst=\sigma_z,~~~\delta\phi^\bst=\begin{pmatrix}0\\n\pi\end{pmatrix},~~~n=0,1.\label{sym trans:U(1)rtimes Z2T}
\end{eqnarray}
The symmetry-allowed independent Higgs terms are
\begin{eqnarray}\label{u1xT:higgs}
\mathcal{S}^1_{edge}=\sum_{l\in\mbz}C_{l}\int\text{d}x\text{d}t\Big[\cos(l\phi_2)+\cos(l\phi_2+nl\pi)\Big]\notag\\
\end{eqnarray}
Apparently they consist of only the $\phi_2$ variable and hence are independent of each other. When $n=0$ all $\cos(l\phi_2),~l\in\mbz$ terms are allowed in (\ref{u1xT:higgs}) and it corresponds to the trivial phase. Variable $\phi_2$ can be localized at value $\langle\phi_2\rangle=0$ and the edge excitations will be gapped without breaking any symmetry. On the other hand when $n=1$ only $\cos(l\phi_2),~l=$even are allowed in (\ref{u1xT:higgs}) and it corresponds to the nontrivial SPT phase. If $\phi_2$ is localized at any value the edge becomes gapped. However time reversal symmetry $\bst$ will be broken since under $\bst$ we have $\phi_2\rightarrow\phi_2+\pi$.

Now let's analyze the group structure formed by these phases. We denote the two phases with ${\bf K}=\sigma_x$ and symmetry transformations (\ref{sym trans:U(1)rtimes Z2T}) where $n=0,1$ as $[0]$ and $[1]$. $[0]=\bse_{U(1)\rtimes Z_2^T}$ is the trivial phase. Consider two copies of nontrivial SPT phases $[1]$ put together: they have edge variables $\{\phi_1,\phi_2\}$ and $\{\phi_1^\prime,\phi_2^\prime\}$. Apparently one can always condensed the pair of elementary independent bosons $\{\phi_1-\phi_1^\prime,\phi_2+\phi_2^\prime\}$ on the edge, and the edge states will be fully gapped without breaking any symmetry. Therefore we have
\begin{eqnarray}
[1]\oplus[1]=[0].
\end{eqnarray}
Clearly $[0]$ and $[1]$ form a $\mbz_2$ group. As a result $n=0$ and $n=1$ label the $\mbz_2$ classes of bosonic non-chiral SPT phases for $U(1)\rtimes Z_2^T$ symmetry.

Note, the generator of charge U(1) $U_\theta=e^{i\hat{n}\theta}$ does not commute with time reversal which involves complex conjugation which sends $i\rightarrow-i$ in the exponential. Thus $U_{-\theta}\bst=\bst U_{\theta}$, which implies time reversal and charge conjugation are combined via the semi-direct product $U(1)\rtimes Z_2^T$. However, if the $U(1)$ was associated with spin rotation about $S_z$ for example, of an integer spin system, the relation with time reversal would be that of a direct product $U(1)\times Z_2^T$, since now $U_{\theta}\bst=\bst U_{\theta}$. This completely changes the topological classification and leads to no nontrivial phases as shown in Appendix\ref{Appendix:U(1)xZ_2^T}.

\subsection{$Z_N$ symmetry: $\mbz_N$ classes}

Denoting the generator of $Z_N$ group as $\bsg$, the algebraic structure (\ref{algebra}) of $Z_N$ group is given by
\begin{eqnarray}
&\label{symmetry:Zn}\bsg^N=\bse.
\end{eqnarray}
The corresponding constraints (\ref{condition:algebra:bosonic SRE}) for symmetry transformations $\{W^\bsg,\delta\phi^\bsg\}$ are
\begin{eqnarray}
&\label{constraint:zn:W}(W^\bsg)^N=I_{2\times2},~~~(W^\bsg)^T\sigma_xW^\bsg=\sigma_x,\\
&\label{constraint:zn:dphi}\sum_{a=1}^N(W^\bsg)^{a-1}\delta\phi^\bsg=\begin{pmatrix}0\\0\end{pmatrix}\mod2\pi.
\end{eqnarray}
In the following we discuss the cases of $N$ being an odd and even integer respectively.

\subsubsection{$N=$~odd integer:~~~$\mbz_N$ classes}

It's straightforward to check that the only solution to (\ref{constraint:zn:W}) is $W^\bsg=I_{2\times2}$. So the solutions to (\ref{constraint:zn:dphi}) have the general form of $\delta\phi^\bsg=\frac{2\pi k}{N}{\bf t}$ where ${\bf t}=\begin{pmatrix}t_1\\t_2\end{pmatrix}$ and $(t_1,t_2)=(k,N)=1$ for $k,t_1,t_2=0,1,\cdots,N-1$. Here we require $(t_1,t_2)=(k,N)=1$ so that the transformations (\ref{sym trans:zn:odd}) form a faithful representation of symmetry group $G=Z_N$. Making use of theorem (\ref{K2x2:U(1):standard form}), we can always reduce an arbitrary ``charge vector" ${\bf t}$ with $(t_1,t_2)=1$ to its standard form $\begin{pmatrix}1\\q\end{pmatrix}$ and hence the inequivalent symmetry transformations $\{W^\bsg,\delta\phi^\bsg\}$ for $Z_N,~N=odd$ symmetry are
\begin{eqnarray}
&\notag W^\bsg=I_{2\times2},~~~\delta\phi^\bsg=\frac{2\pi k}{N}\begin{pmatrix}1\\q\end{pmatrix},\\
&(k,N)=1,~~~q=0,1,\cdots,N-1.\label{sym trans:zn:odd}
\end{eqnarray}
It's easy to show that the Higgs terms allowed by symmetry don't depend on $k$ and they are
\begin{eqnarray}\label{higgs:zn:odd}
&\mathcal{S}^1_{edge}=\sum_{l_1+ql_2=0\mod N}C_{\bf l}\int\text{d}x\text{d}t\cos({\bf l}^T\phi+\alpha_{\bf l})
\end{eqnarray}
Notice that a Higgs term labeled by vector ${\bf l}$ is allowed only if $l_1+ql_2=0\mod N$. Apparently when $q=0$ this is a trivial phase with a set of independent Higgs terms being $\sum_{l\in\mbz}C_{l}\int\text{d}x\text{d}t\cos(l\phi_2+\alpha_{l})$, and the variable $\phi_2$ can be localized at any value without breaking the $Z_N$ symmetry. Since for different $k$ values in transformations (\ref{sym trans:zn:odd}) the symmetry-allowed Higgs terms are exactly the same, we believe different $k$ values correspond to the same phase and we will assume a representative $k=1$ in (\ref{sym trans:zn:odd}) hereafter.

To analyze the group structure of these states, let's denote the various phases with symmetry transformations (\ref{sym trans:zn:odd}) under $Z_N$ symmetry by
\begin{eqnarray}
[q]\equiv\Psi_{Z_N}[\sigma_x,\{W^\bsg=I_{2\times2},\delta\phi^\bsg=\frac{2\pi}{N}\begin{pmatrix}1\\q\end{pmatrix}\}].
\end{eqnarray}
where $[0]=[N]=\bse_{Z_N}$ is the trivial phase. Again consider two states $[q_1]$ with edge variables $\{\phi_1,\phi_2\}$ and $[q_2]$ with edge variables $\{\phi^\prime_1,\phi^\prime_2\}$. Completely in parallel with the discussions for $U(1)$ symmetry, it's straightforward to show that
\begin{eqnarray}
[q]^{-1}=[N-q],~~~[q_1]\oplus[q_2]=[q_1+q_2\mod N].
\end{eqnarray}
Therefore different phases $[q]$ with $q=0,1,\cdots,N-1$ form a $\mbz_N$ group. There are $\mbz_N$ classes of different phases labeled by $q=0,1,\cdots,N-1$ in the presence of $Z_N$ symmetry, when $N=$~odd.

\subsubsection{$N=$~even integer:~~~$\mbz_N$ classes}

Now the inequivalent solutions to (\ref{constraint:zn:W}) are $W^\bsg=\pm I_{2\times2},~\pm\sigma_x$.

(\rmnum{1}) For $W^\bsg=I_{2\times2}$ we have exactly the same solutions as (\ref{sym trans:zn:odd}) in $N=$~odd case, and hence $\mbz_N$ different classes of bosonic non-chiral SPT phases. All these $\mbz_N$ phases can be realized by coupled wire construction, as will be discussed in section \ref{COUPLED WIRE CONSTRUCTION:BOSON SPT}.

(\rmnum{2}) For $W^\bsg=-I_{2\times2}$ one can always choose a gauge $\Delta\phi$ so that $\delta\phi^\bsg=0$. The symmetry-allowed Higgs terms are $\sum_{\forall~{\bf l}}\cos({\bf l}^T\phi)$ and this describes nothing but the trivial phase as $[q=0]$ in (\ref{sym trans:zn:odd}). Its edge states can be gapped out without breaking any symmetry.

\subsubsection{$N=$~even integer:~~~other solutions}
\label{unphysical}
We discuss below additional representations of the symmetry group that appear for this particular case, which we believe are unphysical for a SRE phase with no ground state degeneracy on a torus\footnote{These representations however are known to be realized in topologically ordered phases with nontrivial ground state degeneracy on a torus, such as $Z_2$ spin liquids with translational symmetry as discussed in \Ref{Kou2008}.}. These require interchanging the two edge fields $\phi_1,\phi_2$, which have very different character when they describe fundamental bosons (one is like the phase field, and is compact, while the other is related to the integrated density). Therefore we believe it is unphysical to exchange them. Also, unlike for the other symmetry transformations, a microscopic model with this realization of symmetries was not found. Finally these additional phases are not naturally accommodated into a group structure. These points taken together lead us to drop them from the final list of topological phases with this symmetry.

(\rmnum{3}) For $W^\bsg=\sigma_x$, the gauge inequivalent solutions to (\ref{constraint:zn:dphi}) are $\delta\phi^\bsg=\frac{2\pi k}{N}\begin{pmatrix}1\\1\end{pmatrix}$ with $k=0,1,\cdots,N-1$. We require $(k,N/2)=1$ so that these transformations $\{W^\bsg,\delta\phi^\bsg\}$ form a faithful representation of symmetry group $Z_N,~N=$even. The symmetry-allowed Higgs terms are
\begin{eqnarray}\label{higgs:zn:even:+sigma_x}
&\mathcal{S}^1_{edge}=\sum_{2k(l_1+l_2)=0\mod N}C_{\bf l}\int\text{d}x\text{d}t\cdot\\
&\notag\Big[\cos({\bf l}^T\phi+\alpha_{\bf l})+\cos(l_1\phi_2+l_2\phi_1+\frac{2\pi k(l_1+l_2)}{N}+\alpha_{\bf l})\Big]
\end{eqnarray}
One can verify that they all corresponds to nontrivial SPT phases, whose edge states cannot be gapped without breaking the symmetry. More precisely, variables $\phi_1$ and $\phi_2$ cannot be localized simultaneously since they do not commute, and if only one variable (say $\phi_1$) is localized the symmetry $\bsg$ will be broken since $\phi_1\leftrightarrow\phi_2+\frac{2\pi k}{N}$. Their symmetry transformations are summarized as
\begin{eqnarray}\label{sym trans:zn:even:sigma_x}
&W^\bsg=\sigma_x,~\delta\phi^\bsg=\frac{2\pi k}{N}\begin{pmatrix}1\\1\end{pmatrix},~(k,\frac{N}2)=1.
\end{eqnarray}
But it's not clear how to realize these phases in a microscopic model or what group structure they form. We label these phases as $[\sigma_x,k]\equiv\Psi_{Z_{N}}[\sigma_x,\{W^\bsg=\sigma_x,\delta\phi^\bsg=\frac{2\pi k}{N}\begin{pmatrix}1\\1\end{pmatrix}\}]$.

(\rmnum{4}) For $W^\bsg=-\sigma_x$, the gauge inequivalent ``faithful" solutions to (\ref{constraint:zn:dphi}) are $\delta\phi^\bsg=\frac{2\pi k}{N}\begin{pmatrix}1\\-1\end{pmatrix}$ with $k=0,1,\cdots,N-1$ and $(k,N/2)=1$. The symmetry-allowed Higgs terms are
\begin{eqnarray}\label{higgs:zn:even:-sigma_x}
&\mathcal{S}^1_{edge}=\sum_{2k(l_1-l_2)=0\mod N}C_{\bf l}\int\text{d}x\text{d}t\cdot\\
&\notag\Big[\cos({\bf l}^T\phi+\alpha_{\bf l})+\cos(-l_1\phi_2-l_2\phi_1+\frac{2\pi k(l_1-l_2)}{N}+\alpha_{\bf l})\Big]
\end{eqnarray}
If we label these phases as $[-\sigma_x,k]\equiv\Psi_{Z_{N}}[\sigma_x,\{W^\bsg=-\sigma_x,\delta\phi^\bsg=\frac{2\pi k}{N}\begin{pmatrix}1\\-1\end{pmatrix}\}]$, it's easy to see that
\begin{eqnarray}
[\sigma_x,k]^{-1}=[-\sigma_x,k]
\end{eqnarray}
This is because once a $[\sigma_x,k]$ state with edge variable $\{\phi_1,\phi_2\}$ and a $[-\sigma_x,k]$ state with edge variable $\{\phi_1^\prime,\phi_2^\prime\}$ are put together, one can always condense bosons $\{\phi_1+\phi_1^\prime,\phi_2-\phi_2^\prime\}$, and the edge will be gapped without breaking $Z_N$ symmetry.\\

To summarize, no matter $N=$~odd or $N=$~even, there are $\mbz_N$ different bosonic non-chiral phases in the presence of $Z_N$ symmetry. They are characterized by different symmetry operations (\ref{sym trans:zn:odd}) associated with $Z_N$ generator $\bsg$ and symmetry-allowed Higgs terms (\ref{higgs:zn:odd}). All these $\mbz_N$ phases can be realized in coupled wire construction as will be shown in section \ref{COUPLED WIRE CONSTRUCTION:BOSON SPT}.

Besides, when $N=$~even there are extra solutions (\ref{sym trans:zn:even:sigma_x}) to constraint (\ref{constraint:zn:dphi}) for symmetry transformations associated with $Z_N$ symmetry. However the physical realization of these states and their group structure are not clear.

\subsection{$Z_N\rtimes Z_2^T$ symmetry}

The generators of $Z_N\rtimes Z_2^T$ symmetry group are $\bsg$ for $Z_N$ and $\bst$ for $Z_2^T$ satisfying the following algebra
\begin{eqnarray}
&\bsg^N=\bst^2=\bst\bsg\bst\bsg=\bse.
\end{eqnarray}
The associated constraints on symmetry operations are
\begin{eqnarray}\label{constraint:zn:z2t}
&W^\bsg W^\bst W^\bsg W^\bst=I_{2\times2},\\
&(I_{2\times2}-W^\bsg W^\bst)(\delta\phi^\bsg+W^\bsg\delta\phi^\bst)=\begin{pmatrix}0\\0\end{pmatrix}\mod2\pi.\notag
\end{eqnarray}
in addition to (\ref{z2t:constraint:W}-\ref{z2t:constraint:dphi}) and (\ref{constraint:zn:W}-\ref{constraint:zn:dphi}).

\subsubsection{$N=$~odd integer:~~~$\mbz_1$ class}

The gauge inequivalent solutions to these constraint equations are (\ref{sym trans:z2t}) and
\begin{eqnarray}\label{sym trans:zn:odd:z2t}
&W^\bsg=I_{2\times2},~~\frac{2\pi k}{N}\begin{pmatrix}1\\0\end{pmatrix},~(k,N)=1.
\end{eqnarray}
Let's label these phases by $[n_2,k]$ where $n_2$ is defined in (\ref{sym trans:z2t}). Notice that when $n_2=0$ in (\ref{sym trans:z2t}) one can always destroy the gapless edge excitations by localizing variable $\phi_2$ without breaking any symmetry (under $\bst$ we have $\phi_2\rightarrow\phi_2+n_2\pi$). Similarly when $k=0$ in (\ref{sym trans:zn:odd:z2t}) the edge can be gapped out by localizing bosonic variable $\phi_1$. So $n_2=0$ or $k=0$ both correspond to the trivial phase. On the other hand when $n_2=1$, the symmetry-allowed Higgs terms are
\begin{eqnarray}\label{higgs:zn:z2t:odd}
&\mathcal{S}^1_{edge}=\sum_{{\bf l}}C_{\bf l}\int\text{d}x\text{d}t\Big[\cos(Nl_1\phi_1+l_2\phi_2+\alpha_{\bf l})\\
&\notag+\cos(-Nl_1\phi_1+l_2\phi_2+\alpha_{\bf l}+l_2n_2\pi)\Big]
\end{eqnarray}
At first sight it seems the edge states cannot be gapped without breaking the symmetry, \ie neither $\phi_1$ nor $\phi_2$ can be localized due to symmetry. However when a state $[n_2=1,k\neq0]$ with edge variable $\{\phi_1,\phi_2\}$ is put together with a trivial state $[1,0]=\bse_{Z_N\rtimes Z_2^T}$ with edge variable $\{\phi_1^\prime,\phi_2^\prime\}$, the edge can be fully gapped by condensing bosons $\{N\phi_1+\phi_1^\prime,\phi_2-N\phi_2^\prime\}$ without breaking the $Z_N$ symmetry ($N=$odd). Therefore $[1,k]\oplus[1,0]=\bse_{Z_N\rtimes Z_2^T}$ and $[1,k]$ are all trivial phases, which in general doesn't have gapless edge states. As a result there is no nontrivial SPT phases in the presence of symmetry $Z_N\rtimes Z_2^T,~N=$odd.

\subsubsection{$N=$~even integer:~~Minimal set:~~$\mbz_2^2$ classes}

When $N=$~even we always have $W^\bsg=\pm I_{2\times2}$.

(\rmnum{1}) For $W^\bsg=I_{2\times2}$, the gauge inequivalent faithful solutions to the constraint equations are (\ref{sym trans:z2t}) and
\begin{eqnarray}\label{sym trans:zn:even:z2t:+I}
&\delta\phi^\bsg=\pi\begin{pmatrix}2k/N\\n\end{pmatrix},~~(k,N/2)=1,~~n=0,1.
\end{eqnarray}
Let's label various phases with symmetry transformations (\ref{sym trans:z2t}) and (\ref{sym trans:zn:even:z2t:+I}) as $[k,n_2,n]$ where $n_2=0,1$ is defined in (\ref{sym trans:z2t}). When $k=0$ variable $\phi_1$ can be localized without breaking any symmetry and it is the trivial SPT phase. When $n_2=n=0$ the variable $\phi_2$ can be localized and it is the trivial phase again. Therefore
\begin{eqnarray}
\bse_{Z_{N}\rtimes Z_2^T}=[0,n_2,n]=[k,0,0].
\end{eqnarray}
for $N=$~even. In the following we analyze the group structure formed by these states.

Following discussions in section \ref{SYMMETRY} we can obtain the inverse of a phase by merely changing the sign of its ${K}$ matrix. Now let's put together a state $[k,n_2,n]$ with edge variable $\{\phi_1,\phi_2\}$ is put together with a state $[k^\prime,n_2^\prime,n^\prime]^{-1}$ with edge variable $\{\phi_1^\prime,\phi_2^\prime\}$, we can condense the following independent bosonic variables $\{k^\prime\phi_1-k\phi_1^\prime,k\phi_2-k^\prime\phi_2^\prime\}$ and destroy the gapless edge states if $(k,k^\prime)=1$. The associated Higgs terms will not break the $Z_{N}\rtimes Z_2^T$ symmetry if $k^\prime n-kn^\prime=0\mod2$ and $kn_2-k^\prime n^\prime_2=0\mod2$. As a result $[k,n_2,n]\oplus[k^\prime,n_2^\prime,n^\prime]^{-1}=\bse_{Z_{N}\rtimes Z_2^T}$
\begin{eqnarray}\notag
&\notag\text{For}~(k,k^\prime)=1:~~~[k,n_2,n]=[k^\prime,n_2^\prime,n^\prime],\\
&\notag\text{if}~k^\prime n-kn^\prime=0\mod2,~~kn_2-k^\prime n^\prime_2=0\mod2.
\end{eqnarray}
Therefore we have $[2k+1,n_2,n]=[1,n_2,n]$. For $k=$even on the other hand, we know that $N/2$ must be odd since $(k,N/2)=1$ for a faithful representation. Then we can choose $k^\prime=0$ and condense independent bosons $\{\frac N2\phi_1-\phi_1^\prime,\phi_2-\frac N2\phi_2^\prime\}$ to destroy all edge state. No symmetry will be broken by doing so. Hence we showed $[2k,n_2,n]=[0,n_2,n]=\bse_{Z_{N}\rtimes Z_2^T}$. Consequently the only three nontrivial SPT phases are $[1,1,0]$, $[1,0,1]$ and $[1,1,1]$.

Similarly by putting together a state $[1,n_2,n]$ with edge variable $\{\phi_1,\phi_2\}$ is put together with a state $[1,n_2^\prime,n^\prime]$ with edge variable $\{\phi_1^\prime,\phi_2^\prime\}$, we can always localize bosonic variable $\phi_1-\phi_1^\prime$ and gap out part of the edge. What is left on the edge are described by variables $\{\tilde\phi_1=\phi_1,\tilde\phi_2=\phi_2+\phi_2^\prime\}$. They obey Kac-Moody algebra (\ref{KM algebra:boson non-chiral SRE}) and transform as a $[1,n_2+n_2^\prime,n+n^\prime]$ state. Hence we've shown that
\begin{eqnarray}
[1,n_2,n]\oplus[1,n_2^\prime,n^\prime]=[1,n_2+n_2^\prime,n+n^\prime].
\end{eqnarray}
Since $n,n_2=0,1$ are both $\mbz_2$ integers, so clearly all different 4 states $[1,n_2,n]$ form a $\mbz_2^2$ group. Consequently there are $3$ nontrivial SPT phases labeled by $k=1$ and $[n_2,n]=[0,1],~[1,0]$ or $[1,1]$ in (\ref{sym trans:z2t}) and (\ref{sym trans:zn:even:z2t:+I}).

(\rmnum{2}) For $W^\bsg=-I_{2\times2}$ we can always choose a gauge in (\ref{gauge transformation on symmetry}) so that $\delta\phi^\bsg=\begin{pmatrix}0\\0\end{pmatrix}$. From constraint equations one can derive $W^\bst=\sigma_z$ and $(I_{2\times2}\pm W^\bst)\delta\phi^\bst=\begin{pmatrix}0\\0\end{pmatrix}$ and we have
\begin{eqnarray}
\delta\phi^\bst=\begin{pmatrix}n_1\\n_2\end{pmatrix}\pi,~~~n_1,n_2=0,1.
\end{eqnarray}
Apparently if $n_1=0$ variable $\phi_1$ can be localized without breaking any symmetry, and similarly if $n_2=0$ variable $\phi_2$ can be localized without breaking the symmetry. For the nontrivial SPT phase with $n_1=n_2=1$, the edge states cannot be destroyed without breaking any symmetry. If we label this SPT phase by $[n_1=1,n_2=1]$, one can show that the group structure formed by states with symmetry transformations $W^\bsg=-I$ is the integer group $\mbz$ \ie $\{[1,1]^n,~n\in\mbz\}$. However the above symmetry transformations $\{W^\bsg=-I_{2\times2},\delta\phi^\bsg=\begin{pmatrix}0\\0\end{pmatrix}\}$ do not correspond to a faithful representation of $Z_N\rtimes Z_2^T$ group for $N=$even unless $N=2$. And it is not clear whether the states with symmetry transformations $W^\bsg=-I$ can be realized in a physical bosonic system. Therefore we won't include the states with symmetry transformations $W^\bsg=-I$ in the minimal set of topological phases with $Z_N\rtimes Z_2^T$ symmetry.\\

%

%

To summarize, there are $\mbz_2^2$ classes of different non-chiral bosonic SRE phases (including one trivial phase and three SPT phases) with $Z_N\rtimes Z_2^T$ symmetry when $N=~$even. To compare, the classification and analysis of bosonic SPT phases with $Z_N\times Z_2^T$ symmetry (the \emph{direct} product of $Z_N$ and $Z_2^T$ in contrast to the semi-direct product discussed here) is shown in Appendix \ref{Appendix:Z_NxZ_2^T}.

\section{$K$-matrix classification of fermionic SPT phases}\label{FERMION SPT}

\begin{table*}
\centering
\begin{ruledtabular}
\begin{tabular}{|l|c|l|}
  \hline
  {\em Symmetry} & \em Minimal Topological Classification & {\em Comments} \\
  \hline
 $Z_2^f$ (no symmetry)& $\mbz_1$ & No symmetry (fermion parity always
conserved) non-chiral phase\\
  $Z_2^T\times Z_2^f$ & $\mbz_1$ & Time reversal symmetric superconductor \\
    $U(1)\rtimes Z_2^T$,~$\bst^2=1$ & $\mbz_2$ & Bosonic quantum spin Hall insulator of Cooper pairs\\
  $U(1)\rtimes Z_2^T$,~$\bst^2=\fnp$ & $\mbz_2$ & Fermionic quantum spin Hall
insulator\\
  $U(1)\times Z_2^T\times Z_2^f$ & $\mbz_1$ &  $U(1)$ spin conservation and
time reversal.\\
$Z_2\times Z_2^f$ & $\mbz_4$ & Superconductor with Ising-type symmetry\\
  $Z_4$ & $\mbz_2$ &  Bosonic $Z_2$-symmetric SPT phase of Cooper pairs \\

  $(Z_2 \rtimes Z_2^T)\times Z_2^f$ & $\mbz_4$ & Discussed in
\Ref{Qi2012,Yao2012,Ryu2012} with $\mbz_8$ classification\\
  $Z_4 \times Z_2^T$ & $\mbz_2^2$ & Superconductor with $Z_4$ spin symmetry\\
  $Z_4 \rtimes Z_2^T,~\bst^2=\fnp$ & $\mbz_2^2$ & Time-reversal-symmetric charge-$4e$ superconductor\\
  \hline
\end{tabular}
\caption{Topological classification of gapped D=2+1 dimensional {\em non-chiral} phases of
fermions with short range entanglement (no topological order). Here $\fnp$ denotes fermion parity, which is always conserved. Note, states with an odd number of right(left)-moving Majorana edge modes (such as class $DIII$ topological superconductor) are not captured in this formalism.}
\label{table1}
\end{ruledtabular}
\end{table*}

According to theorem (\ref{K2x2:standard form}) in Appendix \ref{app:K2x2} a $2\times2~{K}$ matrix with determinant $-1$ for a fermionic system is always equivalent to the standard form $\begin{pmatrix}0&1\\1&1\end{pmatrix}\simeq\sigma_z$ by certain $GL(2,\mbz)$ transformations. In the following we always choose the $2\times2$ matrix ${\bf K}=\sigma_z$ to represent a generic fermionic non-chiral SRE state. In the following we use general principles discussed in section \ref{SYMMETRY} to study non-chiral fermionic SPT phases with different symmetries. Note that the only $GL(2,\mbz)$ transformations ${\bf X}$ that keeps ${\bf K}=\sigma_z$ invariant under (\ref{gl(n,z)}) are ${\bf X}=\pm I_{2\times2},~\pm\sigma_z$. For such a non-chiral fermionic SRE phase, its ``bare" Chern-Simons effective theory with no Higgs terms added is
\begin{eqnarray}\label{bare:fermion:bulk}
\mathcal{L}_{{\bf K}}=\frac1{4\pi}\epsilon^{\mu\nu\lambda}(a^1_\mu\partial_\nu a^1_\lambda-a^2_\mu\partial_\nu a^2_\lambda)-\sum_{I=1}^2a_\mu^Ij_I^\mu
\end{eqnarray}
in the bulk and
\begin{eqnarray}\label{bare:fermion:edge}
&\mathcal{S}_{edge}^0=\frac1{4\pi}\int\text{d}t\text{d}x\big(\partial_t\phi_1\partial_x\phi_1-\partial_t\phi_2\partial_x\phi_2\\
&\notag-\sum_{I,J}V_{I,J}\partial_x\phi_I\partial_x\phi_J\big)
\end{eqnarray}
on the edge where $V_{I,J}$ is a positive definite constant matrix, as discussed in section \ref{K+HIGGS}. The Kac-Moody algebra satisfied by fields $\{\phi_1,\phi_2\}$ on the edge is
\begin{eqnarray}
&\notag[\partial_x\phi_1(x),\partial_y\phi_1(y)]=-[\partial_x\phi_2(x),\partial_y\phi_2(y)]=\\
&2\pi\imth\partial_x\delta(x-y),~~~[\partial_x\phi_1(x),\partial_y\phi_2(y)]=0.\label{KM algebra:fermion non-chiral SRE}
\end{eqnarray}
In the absence of any symmetry, a set of independent Higgs term satisfying (\ref{condition:local}) and (\ref{condition:commute}) on the edge can be chosen as either $\{C_l\cos(l\phi_1+l\phi_2+\alpha_l),~l\in\mbz\}$ or $\{C_l\cos(l\phi_1-l\phi_2+\alpha_l),~l\in\mbz\}$. All degrees of freedom on the edge will be gapped once ``bosonic" variable $\phi_1+\phi_2$ (or $\phi_2-\phi_1$) is localized at a classical value by the Higgs terms. Again the two bosonic variables $\phi_1+\phi_2$ and $\phi_1-\phi_2$ cannot be localized simultaneously, according to Heisenberg uncertainty relation implied by Kac-Moody algebra (\ref{KM algebra:fermion non-chiral SRE}).

There is an intrinsic difference between fermions and bosons: \ie only bosonic quasiparticles can ``condense" in a bosonic/fermionic system described by a local Hamiltonian. This means in a bosonic system, all quasiparticles are bosons and should transform trivially under identity element $\bse=\prod_a\bsg_a^{n_a}$ of group $G$ as shown in (\ref{condition:algebra:bosonic SRE}). In a fermionic system on the other hand, any bosonic quasiparticle consists of an even number of fermions and is always invariant if every fermion creation (annihilation) operator obtains a minus sign. Therefore under symmetry transformation $\{W^\bse,\delta\phi^\bse\}$ corresponding to identity element $\bse$ of the same symmetry group $G$ in any local fermionic system, only those Higgs terms  satisfying condition (\ref{boson})-(\ref{boson_mutual}) should transform trivially. This can be generalized to a more universal situation where anyonic quasiparticles are present ($|\det{\bf K}|>1$): under identity element of group $G$ only local operators (\ie Higgs terms $\cos({\bf l}^T\phi+\alpha_{\bf l})$ satisfying (\ref{boson})-(\ref{boson_mutual}) which condense bosonic quasiparticles) should transform trivially.  This means with the same symmetry group $G=\{g_a\}$, the symmetry transformations $\{W^{g_a}_b,\delta\phi^{g_a}_b\}$ of a bosonic SRE state form a faithful representation of group $G$, while symmetry transformations $\{W^{g_a}_f,\delta\phi^{g_a}_f\}$ of a fermionic SRE state (or more generally a gapped Abelian phase containing fermionic and anyonic quasiparticles) form a \emph{projective representation}\cite{Wen2012} of group $G=\{g_a\}$. And for these systems the identity element $\bse$ in group compatibility conditions (\ref{algebra}) and (\ref{condition:algebra}) doesn't always correspond to a trivial transformation on the fermionic (anyonic) quasiparticles.

In a fermionic non-chiral SRE phase with ${\bf K}=\sigma_z$ here, it is easy to verify that such a bosonic quasiparticle is labeled by any vector ${\bf l}=\begin{pmatrix}l_1\\l_2\end{pmatrix}$ satisfying
\begin{eqnarray}
l_1=l_2\mod2\label{boson:fermion non-chiral SRE}
\end{eqnarray}
\emph{Locality} requires that only Higgs terms $\cos({\bf l}^T\phi+\alpha_{\bf l})$ satisfying (\ref{boson:fermion non-chiral SRE}) can be added to bare action (\ref{bare:fermion:bulk}) and (\ref{bare:fermion:edge}), i.e. fermions are not allowed to condense. Hence in the absence of any symmetry, all these Higgs terms $\cos({\bf l}^T\phi+\alpha_{\bf l})$ satisfying (\ref{boson:fermion non-chiral SRE}) are allowed and should be added to a fermionic non-chiral SRE state. As a result, the identity element $\bse$ (no symmetry)in a fermionic system is implemented by the following generic form of symmetry transformations
\begin{eqnarray}\label{sym trans:fermion}
&W^{\bse}=I_{2\times2},~~~\delta\phi^{\bse}=\eta_f\pi\begin{pmatrix}1\\1\end{pmatrix}.
\end{eqnarray}
where $\eta_f=0,1$. Notice that the above symmetry transformations are invariant under any gauge transformation (\ref{gauge transformation on symmetry}). When $\eta_f=1$ the fermionic operators $\sim\exp[\imth\phi_\alpha],~\alpha=1,2$ obtains a minus sign, corresponding to the \emph{fermion number parity} operation $\fnp=(-1)^{\hat{N}_f}$. When $\eta_f=0$ on the other hand, every fermion remains invariant and it corresponds to the actual identity element $\bse_f$ of the symmetry group $G_f$ (including $\fnp$) for the underlying fermions. If we incorporate the fermion number parity $\fnp$ into the symmetry group $G_f$, one easily notices that $Z_2^f=\{\bse_f,\fnp\}$ is always a normal subgroup of fermion symmetry group $G_f$, which means $\fnp$ is involutory ($\fnp^2=\bse_f$) and central in $G_f$\cite{Fidkowski2011}.

If we always incorporate fermion number parity $\fnp$ into fermion symmetry group $G_f$, then a fermionic system with symmetry $G_f$ is naturally related to a bosonic system with symmetry $G=G_f/Z_2^f$ (since $Z_2^f$ is a normal subgroup of $G_f$ the quotient group $G_f/Z_2^f$ can be defined). Physically this means if fermions with symmetry $G_f$ pair up to form Cooper pairs (which are bosons), these bosonic Cooper pairs have symmetry $G=G_f/Z_2^f$. Different SPT phases of bosonic Cooper pairs (where fermions are confined) with symmetry $G=G_f/Z_2^f$ are necessarily different fermionic SRE phases with symmetry $G_f$. Hence the different classes of fermionic SRE phases with symmetry $G_f$ must contain all different bosonic SRE phases with symmetry $G=G_f/Z_2^f$ as a subset\cite{Gu2012}. To be more precise, the group $H_b(G_f/Z_2^f)$ formed by different bosonic (non-chiral) SRE phases with symmetry $G=G_f/Z_2^f$ is always a subgroup of $H_f(G_f)$, the group formed by different fermionic (non-chiral) SRE phases with symmetry $G_f$.

Before discussing specific examples of non-chiral topological phases, we point out that SRE {\em chiral} phases of fermions are readily obtained (\eg integer quantum Hall states) so we skip their discussion. It should also be noted that our formalism currently is restricted to topological phases in which the gapless edge, when present, has integer central charge (\eg $c=1$ in many cases). So chiral and non-chiral Majorana modes (\eg of a $p_x+ip_y$ superconductor, or the D=2+1 class DIII topological superconductor\cite{Schnyder2008,Kitaev2009}, with a pair of counter-propagating Majorana modes) are not captured in our formulation.

\subsection{$G_f/Z_2^f=\{\bse\}\Rightarrow G_f=Z_2^f$ symmetry: $\mbz_1$ class}

If we choose $G=G_f/Z_2^f=\{\bse\}$ is the trivial group, then the fermion symmetry group is $G_f=Z_2^f$. The generator of $Z_2^f$ \ie fermion number parity operator is
\begin{equation}
\fnp\equiv(-1)^{\hat{N}_f}
\end{equation}
 where $\hat{N}_f$ denotes the total fermion number. 
~The existence of $Z_2^f$ symmetry is a basic requirement for any fermionic system described by a \emph{local} Hamiltonian. Simply speaking $\fnp$ guarantees that one single fermion cannot condense like the bosons: only a bosonic conglomerate containing an even number of fermions can condense and obtain a non-vanishing expectation value. A general form for such a bosonic quasiparticle in a fermionic system is labeled by a integer vector ${\bf l}$ satisfying condition (\ref{boson}).

This means $Z_2^f$ is more like a constraint for fermionic system due to locality, rather than a true "symmetry". As discussed earlier it is implemented by nothing but the non-trivial ($n_f=1$) realization for $\bse$ in $G=G_f/Z_2^f$:
\begin{eqnarray}\label{sym trans:z2f}
&W^\fnp=I_{2\times2},~~~\delta\phi^\fnp=\pi\begin{pmatrix}1\\1\end{pmatrix}.
\end{eqnarray}
It guarantees that in the absence of any symmetry, all Higgs terms $\cos({\bf l}^T\phi+\alpha_{\bf l})$ satisfying (\ref{boson:fermion non-chiral SRE}) can be added to a fermionic non-chiral SRE state and these are the only terms that can be added. Notice that (\ref{sym trans:z2f}) is invariant under any gauge transformations (\ref{gauge transformation on symmetry}) for fermions with ${\bf K}=\sigma_z$. Apparently we have $\fnp^2=\bse_f$ \ie the fermion number parity acting twice would yield the identity operation for fermions.

The $Z_2^f$ symmetry-allowed Higgs terms are all terms associated with bosonic quasiparticles (\ref{boson:fermion non-chiral SRE})
\begin{eqnarray}\label{higgs:z2f}
&\mathcal{S}^1_{edge}=\sum_{\{l_1=l_2\mod 2\}}C_{\bf l}\int\text{d}x\text{d}t\cos({\bf l}^T\phi+\alpha_{\bf l}).
\end{eqnarray}
Apparently the bosonic variable $\phi_1+\phi_2$ (or $\phi_1-\phi_2$) can be localized at a classical value and the edge will be gapped. Physically $\phi_1+\phi_2$ corresponds to the pairing between right mover and left mover, while $\phi_1-\phi_2$ is backscattering between right and left movers. They are both allowed in the absence of any symmetry. This describes the (trivial) $\mbz_1$ class of non-chiral fermionic phase with $Z_2^f$ symmetry.

Since fermion number parity $\fnp$ is always realized by (\ref{sym trans:fermion}), in the following we'll not specifically mention this symmetry but only requires $Z_2^f$ to be a normal subgroup of the full symmetry group $G_f$ of fermions. And we use $\bse$ to denote the identity element in the ``bosonic" symmetry group $G=G_f/Z_2^f$. Therefore in the fermion system $\bse$ can be either $\bse_f$ (all fermion operators keep invariant) or $\fnp$ (all fermion operators change sign).

\subsection{$G_f/Z_2^f=Z_2^T\Rightarrow G_f=Z_2^T\times Z_2^f$ symmetry: $\mbz_1$ class}

In the presence of time reversal $Z_2^T$ symmetry with generator $\bst$, the algebraic structure of full symmetry group $G_f/Z_2^T=Z_2^T$ is given by
\begin{eqnarray}\label{symmetry:z2f:z2t}
\bst^2=\bse.
\end{eqnarray}
Here in our notation $\bse$ can be either $\bse_f$, the identity element for fermions or $\fnp$, the fermion number parity operation. From (\ref{condition:algebra}) this leads to the following constraint on symmetry transformations $\{W^\bst,\delta\phi^\bst\}$ are
\begin{eqnarray}
&\label{constraint:fermion:z2t:W} (W^\bst)^2=I_{2\times2},~~~(W^\bst)^T{\bf K}W^\bst=-{\bf K};\\
&\label{constraint:fermion:z2t:dphi}(I_{2\times2}-W^\bst)\delta\phi^\bst=\eta_{\bst}\pi\begin{pmatrix}1\\1\end{pmatrix}\mod2\pi,\\
&\notag(I_{2\times2}-W^\bst)(\delta\phi^\bst+\delta\phi^\fnp)\\
&=\eta_{\bst\fnp}\pi\begin{pmatrix}1\\1\end{pmatrix}\mod2\pi.\label{constraint:z2f:z2t:dphi}
\end{eqnarray}
where $\eta_\bst,\eta_{\bst\fnp}=0,1.$ Notice that fermion number parity symmetry $\fnp$ is always implemented by (\ref{sym trans:z2f}), independent of the gauge choice. Here with ${\bf K}=\sigma_z$ the gauge inequivalent solutions to (\ref{constraint:fermion:z2t:W}) is $W^\bst=\sigma_x$. Then solving (\ref{constraint:fermion:z2t:dphi}) and (\ref{constraint:z2f:z2t:dphi}) we get $\eta_\bst=\eta_{\bst\fnp}$ and $\delta\phi^\bst=\begin{pmatrix}0\\ \eta_\bst\pi\end{pmatrix}$. Therefore the inequivalent symmetry transformations for $Z_2^T\times Z_2^f$ group is (\ref{sym trans:z2f})
\begin{eqnarray}
&W^\bst=\sigma_x,~~\delta\phi^\bst= \eta_\bst\begin{pmatrix}0\\\pi\end{pmatrix},~~\eta_\bst=0,1.\label{sym trans:z2f:z2t}
\end{eqnarray}
And the symmetry-allowed Higgs terms are
\begin{eqnarray}
&\notag\mathcal{S}^1_{edge}=\sum_{\{l_1=l_2\mod 2\}}C_{\bf l}\int\text{d}x\text{d}t\Big[\cos({\bf l}^T\phi+\alpha_{\bf l})\\
&+\cos(-{\bf l}^T\sigma_x\phi+l_2\eta_{\bst}\pi+\alpha_{\bf l})\Big].\label{higgs:z2f:z2t}
\end{eqnarray}
where $\eta_{\bst}=0,1$. It turns out the above Higgs term always describes the same trivial SPT phase no matter $\eta_\bst=0$ or $1$: \eg variable $\phi_1+\phi_2$ can be always localized at an expectation value $\langle\phi_1+\phi_2\rangle=\eta_2\pi/2$ by the Higgs terms, and the gapless edge states will be destroyed without breaking any symmetry. Therefore (\ref{bare:fermion:edge}) together with (\ref{higgs:z2f:z2t}) describes the (trivial) $\mbz_1$ class of fermionic non-chiral SPT with $Z_2^T\times Z_2^f$ symmetry, no matter $\bst^2=\bse_f$ ($\eta_\bst=\eta_{\bst\fnp}=0$) or $\bst^2=\fnp$ ($\eta_\bst=\eta_{\bst\fnp}=1$). Note, the $Z_2$ classification of free fermions in class DIII with these symmetries is missed by this classification. The reason of course is that we are not able to describe Majorana modes (the class DIII topological superconductor has a counter-propagating pair of Majorana modes with central charge $c=1/2$) within the $K$-matrix formulation.


\subsection{$G_f/Z_2^f=U(1)\rtimes Z_2^T$ symmetry}

By labeling the $U(1)$ group elements as $U_\theta$, the algebraic structure of $G=U(1)\rtimes Z_2^T$ group is given by
\begin{eqnarray}\label{symmetry:fermion:u1:z2t:semi}
&\bst^2=\bst U_\theta\bst U_\theta=U_{(\theta=0\mod2\pi)}=\bse.
\end{eqnarray}
in addition to (\ref{symmetry:U(1)}). Again here in our notation $\bse$ can be either identity element $\bse_f$ for fermions or fermion number parity $\fnp$. Again a general form of symmetry transformation for $U_\theta$ is given by
\begin{eqnarray}\label{sym trans:fermion:u1}
&W^{U_\theta}=I_{2\times2},\delta\phi^{U_\theta}=\theta{\bf t}
\end{eqnarray}
and the we have the following constraints for the symmetry transformations
\begin{eqnarray}\label{constraint:fermion:u1:dphi}
&2\pi{\bf t}=\eta_{U(1)}\pi\begin{pmatrix}1\\1\end{pmatrix}\mod2\pi,\\
&(I_{2\times2}-W^\bst)(\theta{\bf t}+\delta\phi^\bst)=\eta\pi\begin{pmatrix}1\\1\end{pmatrix}\mod2\pi.\label{constraint:fermion:u1:z2t}
\end{eqnarray}
in addition to (\ref{constraint:fermion:z2t:W}-\ref{constraint:fermion:z2t:dphi}). The last line in (\ref{symmetry:fermion:u1:z2t:semi}) are automatically satisfied. The gauge inequivalent solutions to these constraint equations are
\begin{eqnarray}\label{sym trans:fermion:u1:z2t_semid}
&{\bf t}=(t+\frac{\eta_{U(1)}}2)\begin{pmatrix}1\\1\end{pmatrix},~~~t\in\mbz;\\
&W^\bst=\sigma_x,~~~\delta\phi^\bst=\begin{pmatrix}0\\ \eta\pi\end{pmatrix}\mod2\pi.
\end{eqnarray}
where $\eta_{U(1)},~\eta=0,1$. Notice that the fermion parity generator $\fnp$ is always a subgroup of $U(1)$ since $t+\eta_{U(1)}/2\neq0$. Hence $Z_2^f$ is always a subgroup of $U(1)$ group associated with fermion number conservation. If $\eta=1$ we have $\bst^2=\fnp$, while $\eta=0$ corresponds to $\bst^2=1$.

\subsubsection{$G_f=U(1)\rtimes Z_2^T$ with $\bst^2=1$:~$\mbz_2$ classes}

When $\eta=0$ the algebra of symmetry group $G_f=U(1)\rtimes Z_2^T$ is
\begin{eqnarray}\label{symmetry:fermion:u1:z2t:semi:T^2=1}
&\bst^2=\bst U_\theta\bst U_\theta=U_{(\theta=0\mod2\pi)}=\bse_f.
\end{eqnarray}
where $\bse_f$ is the identity element of symmetry group $G_f$ for fermions. The symmetry-allowed Higgs terms associated with symmetry transformations (\ref{sym trans:fermion:u1}) and (\ref{sym trans:fermion:u1:z2t_semid}) are
\begin{eqnarray}
&\notag\mathcal{S}^1_{edge}=\sum_{l\in\mbz}C_{l}\int\text{d}x\text{d}t~\cos\Big[l(\phi_1-\phi_2)+\alpha_{l}\Big].
\end{eqnarray}
for $\eta=0$. Hence in $\eta=0$ case there is only one trivial phase, since independent bosonic variable $\phi_1-\phi_2$ can be localized at a classical value by the Higgs terms and the edge will be gapped without breaking any symmetry (under $\bst$ we have $\phi_1-\phi_2\rightarrow\phi_1-\phi_2-\eta\pi$). Meanwhile notice that for a bosonic system with $G=U(1)\rtimes Z_2^T$ symmetry ($\bst^2$=1) there are $\mbz_2$ classes of different phases. Hence the nontrivial bosonic SPT phase of Cooper pairs (fermions are confined) protected by $U(1)\rtimes Z_2^T$ symmetry form a nontrivial SPT phase of fermions with $G_f=U(1)\rtimes Z_2^T$ symmetry. As a result there are $\mbz_2$ classes of different fermionic (non-chiral) SRE phases in the presence of $U(1)\rtimes Z_2^T$ symmetry with $\bst^2=1$. The $\mbz_2$ classification comes purely from the bosonic SPT phases (bosonic QSH insulator) of Cooper pairs in the molecule limit where fermions are confined.

\subsubsection{$G_f=U(1)\rtimes Z_2^T$ with $\bst^2=\fnp$:~$\mbz_2$ classes}

When $\eta=1$ the algebra of symmetry group $G_f=U(1)\rtimes Z_2^T$ is
\begin{eqnarray}\label{symmetry:fermion:u1:z2t:semi:T^2=1}
&\bst^2=\bst U_\theta\bst U_\theta=U_{(\theta=0\mod2\pi)}=\fnp.
\end{eqnarray}
And the symmetry-allowed Higgs terms on the edge are
\begin{eqnarray}
&\notag\mathcal{S}^1_{edge}=\sum_{l\in\mbz}C_{l}\int\text{d}x\text{d}t~\cos\Big[2l(\phi_1-\phi_2)+\alpha_{l}\Big].
\end{eqnarray}
for $\eta=1$. This corresponds to the nontrivial SPT phase, whose edge cannot be gapped without breaking the symmetry. We use $[\eta]$ with $\eta=1$ and $\eta=1$ to label these two phases. Now let's examine the group structure $\{\Psi_{U(1)\rtimes Z_2^T\times Z_2^f}\}$ formed by these states.

The trivial state is labeled by identity element $\bse_{U(1)\rtimes Z_2^T}$. If we put two $[\eta=1]$ states together, we can gap out the edge states without breaking the symmetry so $[1]\oplus[1]=\bse_{U(1)\rtimes Z_2^T}$. They form a $\mbz_2$ group. As a result, $[\eta=1]$ and $[1]\oplus[1]$ label the $\mbz_2$ classes of fermionic non-chiral SRE phases in the presence of $U(1)\rtimes Z_2^T$ symmetry with $\bst^2=\fnp$. When $U(1)$ symmetry corresponds to fermion charge conservation, these two different phases are nothing but the trivial band insulator and $Z_2$ topological band insulator (quantum spin Hall insulator) of fermions in $2+1$-D\cite{Kane2005a,Kane2005,Bernevig2006}.

Naively the bosonic SPT phases of Cooper pairs (where fermions are confined) with $G=U(1)\rtimes Z_2^T$ symmetry gives rise to another $\mbz_2$ classification. However it turns out that bosonic quantum spin Hall insulator (QSHI) of Cooper pairs becomes trivial in the presence of electrons \footnote{We thank Chong Wang and T. Senthil for pointing this out to us. C. Wang (unpublished)}. To be specific, assume that the fermions form bosonic molecules of charge-$2e$ (Cooper pairs) which preserve time reversal symmetry, and these charge-$2e$ molecules form a bosonic QSHI in 2d. Its edge state is described by two chiral bosons $\phi_{1,2}$ with $2\times2$ matrix ${\bf K}=\sigma_x$ in Eq. (6). They transform under $U(1)$ charge rotation $U_\theta$ and time reversal $\bst$ as
\bea
\vec\phi\equiv\bpm\phi_1\\ \phi_2\epm\overset{U_\theta}\longrightarrow\vec\phi+\theta\bpm2\\0\epm,~~~~~\vec\phi\overset{\bst}\longrightarrow\bpm-\phi_1\\ \phi_2+\pi\epm.
\eea
Now let's couple two layers of (spin-$1/2$) electronic QSHI to this QSHI of bosonic charge-$2e$ molecules. The gapless edge states of these electronic QSHIs are described by chiral bosons $\phi_{L/R,a}$ with matrix ${\bf K}=\sigma_z$ in Eq. (6), where $a=u/d$ is the layer index. Under symmetry operations they transform as
\bea
\bpm\phi_{L,a}\\ \phi_{R,a}\epm\overset{U_\theta}\longrightarrow\bpm\phi_{L,a}+\theta\\ \phi_{R,a}+\theta\epm,~~~~\bpm\phi_{L,a}\\ \phi_{R,a}\epm\overset{\bst}\longrightarrow\bpm-\phi_{R,a}\\-\phi_{L,a}+\pi\epm.
\eea
All excitations on the edge can be fully gapped by the following ``backscattering'' terms without breaking any symmetry
\bea
&\notag\mathcal{H}_{\text{edge}}^1=C_1\sin(\phi_1-\phi_{L,u}-\phi_{R,d})\\
&\notag+C_1\sin(\phi_1-\phi_{R,u}-\phi_{L,d})\\
&+C_2\cos(\phi_2+\phi_{L.u}-\phi_{R,u}+\alpha_2).
\eea
where $C_{1,2}$ and $\alpha_2$ are all real constants. Since two layers of electronic QSHI can be continuously tuned into a trivial insulator without closing the bulk energy gap, the QSHI of bosonic charge-$2e$ Cooper pairs is a trivial insulator in the presence of spin-$1/2$ electrons (with $\bst^2=\fnp$).

Hence in total there are just $\mbz_2$ classes of different fermionic SPT phases with $U(1)\rtimes Z_2^T$ symmetry ($\bst^2=\fnp$).

\subsection{$G_f/Z_2^f=U(1)\times Z_2^T$ symmetry: $\mbz_1$ class}

The algebraic structure of $G=G_f/Z_2^f=U(1)\times Z_2^T$ group is given by
\begin{eqnarray}\label{symmetry:fermion:u1:z2t:direct}
&\bst^2=\bst U_{-\theta}\bst U_\theta=U_{(\theta=0\mod2\pi)}=\bse.\\
\end{eqnarray}
in addition to (\ref{symmetry:U(1)}). The associated constraints (\ref{condition:algebra}) for symmetry transformations (\ref{sym trans:fermion:u1}) and $\{W^\bst,\delta\phi^\bst\}$ are (\ref{constraint:fermion:z2t:W}-\ref{constraint:fermion:z2t:dphi}), (\ref{constraint:fermion:u1:dphi}) and
\begin{eqnarray}
&(I_{2\times2}-W^\bst)\delta\phi^\bst+\theta(I_{2\times2}+W^\bst){\bf t}=\eta\pi\begin{pmatrix}1\\1\end{pmatrix}\mod2\pi.\notag
\end{eqnarray}
Again the last line of (\ref{symmetry:fermion:u1:z2t:direct}) is automatically satisfied. The gauge inequivalent solutions to these constraint equations are
\begin{eqnarray}\label{sym trans:fermion:u1:z2t:direct}
&{\bf t}=(t+\frac{\eta_{U(1)}}2)\begin{pmatrix}1\\-1\end{pmatrix},~~~t\in\mbz;\\
&W^\bst=\sigma_x,~~~\delta\phi^\bst=\begin{pmatrix}0\\ \eta\pi\end{pmatrix}\mod2\pi.
\end{eqnarray}
The associated symmetry-allowed Higgs terms are
\begin{eqnarray}
&\notag\mathcal{S}^1_{edge}=\sum_{l\in\mbz}C_{l}\int\text{d}x\text{d}t~\cos\Big[l(\phi_1+\phi_2)+\alpha_{l}\Big].
\end{eqnarray}
for $\eta=0$ and
\begin{eqnarray}
&\notag\mathcal{S}^1_{edge}=\sum_{l\in\mbz}\int\text{d}x\text{d}t\Big(C_l\sin\big[(2l+1)(\phi_1+\phi_2)\big]\\
&\notag+D_l\cos\big[2l(\phi_1+\phi_2)\big]\Big).
\end{eqnarray}
for $\eta=1$. In both cases variable $\phi_1+\phi_2$ can be localized at a classical value by the Higgs term, so the gapless edge states will be destroyed without breaking any symmetry (under $\bst$ we have $\phi_1+\phi_2\rightarrow\eta\pi-\phi_1-\phi_2$). So there is a (trivial) $\mbz_1$ class of fermionic non-chiral SRE phase in the presence of $U(1)\times Z_2^T\times Z_2^f$ symmetry. Note that $\bst^2=\bse_f$ if $\eta=0$ and $\bst^2=\fnp$ if $\eta=1$. Since there is no bosonic SPT phases with $U(1)\rtimes Z_2^T=G_f/Z_2^T$ symmetry, there are no new fermionic SPT phases coming from bosonic SPT phases of Cooper pairs.

\subsection{$G_f/Z_2^f=Z_2$ symmetry}

The generator $\bsg$ of $Z_2$ symmetry satisfies the following algebra
\begin{eqnarray}\label{symmetry:z2}
&\bsg^2=\bse.
\end{eqnarray}
Here $\bse$ stands for either identity element $\bse_f$ for fermions or fermion number parity $\fnp$. This algebraic constraints (\ref{condition:algebra}) for symmetry transformations $\{W^\bsg,\delta\phi^\bsg\}$ are
\begin{eqnarray}\label{constraint:fermion:z2:W}
&(W^\bsg)^2=I_{2\times2},~~~(W^\bsg)^T{\bf K}W^\bsg={\bf K},\\
&(I_{2\times2}+W^\bsg)\delta\phi^\bsg=\eta\pi\begin{pmatrix}1\\1\end{pmatrix}\mod2\pi.\label{constraint:fermion:z2:dphi}
\end{eqnarray}
where $\eta=0,1$. The gauge inequivalent solutions of (\ref{constraint:fermion:z2:W}) are $W^\bsg=\pm I_{2\times2},~\pm\sigma_z$. In the following we analyze those cases with $W^\bsg=\pm I_{2\times2}$ and the discussions about cases with $W^\bsg=\pm\sigma_z$ is put in the Appendix. It's not clear to us now whether the transformation laws with $W^\bsg=\pm\sigma_z$ can be realized in a microscopic model, therefore we didn't include these cases in the minimal set of different fermion SRE phases with $G_f$ symmetry.

(\rmnum{1}) For $W^\bsg=-I_{2\times2}$ the gauge inequivalent solution to (\ref{constraint:fermion:z2:dphi}) is $\delta\phi^\bsg=\begin{pmatrix}0\\0\end{pmatrix}$ and $\eta=0$. A set of independent symmetry-allowed Higgs terms can be either (\ref{higgs_fermion_z2_A_1st}) or (\ref{higgs_fermion_z2_A_2nd}) with $\alpha_l\equiv0,~\forall~l\in\mbz$. Hence it corresponds to the trivial phase, whose edge can be gapped without breaking any symmetry.

(\rmnum{2}) For $W^\bsg=I_{2\times2}$ the inequivalent solutions to (\ref{constraint:fermion:z2:dphi}) are
\begin{eqnarray}
&\delta\phi^\bsg=\pi\begin{pmatrix}t_1\\t_2\end{pmatrix}+{\frac\pi2}\eta\begin{pmatrix}1\\1\end{pmatrix},~~~t_1,t_2,\eta=0,1.
\end{eqnarray}
The symmetry allowed Higgs terms are those $\cos({\bf l}^T\phi+\alpha_{\bf l})$ terms satisfying
\begin{eqnarray}
l_1t_1+l_2t_2+\frac{l_1+l_2}{2}\eta=0\mod2
\end{eqnarray}
and the condition (\ref{boson:fermion non-chiral SRE}) for local operators. It's straightforward to verify that when $t_1+t_2+\eta=0\mod2$ a set of independent symmetry-allowed Higgs terms satisfying (\ref{condition:local}) is
\begin{eqnarray}\label{higgs_fermion_z2_A_1st}
&\mathcal{S}^1_{edge}=\sum_{l\in\mbz}C_l\int\text{d}x\text{d}t\cos\big[l(\phi_1+\phi_2)+\alpha_l\big]
\end{eqnarray}
and the edge states can be gapped without breaking the $Z_2$ symmetry (under $\bsg$ we have $\phi_1+\phi_2\rightarrow\phi_1+\phi_2+(t_1+t_2+\eta)\pi$), if variable $\phi_1+\phi_2$ is localized at any classical value. Similarly when $t_1-t_2=0\mod2$ a set of independent symmetry-allowed Higgs terms satisfying (\ref{condition:local}) is
\begin{eqnarray}\label{higgs_fermion_z2_A_2nd}
&\mathcal{S}^1_{edge}=\sum_{l\in\mbz}D_l\int\text{d}x\text{d}t\cos\big[l(\phi_1-\phi_2)+\beta_l\big]
\end{eqnarray}
and the edge states will be gapped without breaking the $Z_2$ symmetry (under $\bsg$ we have $\phi_1-\phi_2\rightarrow\phi_1-\phi_2+(t_1-t_2)\pi$), if variable $\phi_1-\phi_2$ is localized at any value. They all correspond to the trivial phase. Notice that when $\eta=0$ we have $\bsg^2=\bse_f$ while $\bsg^2=\fnp$ if $\eta=1$. This corresponds to the following two different symmetry groups:

\subsubsection{$G_f=Z_2\times Z_2^f$ symmetry:~~~$\mbz_4$ classes}

{\bf Intrinsic fermion phases}

This means $\eta=0$ and $t_1-t_2=1\mod2$. The algebra of symmetry group $G_f$ is
\begin{eqnarray}
\bsg^2=\bse_f.
\end{eqnarray}
When $[\eta,t_1,t_2]=[0,0,1]$ or $[0,1,0]$ a set of independent symmetry-allowed Higgs terms satisfying (\ref{condition:local}) can be chosen to be either
\begin{eqnarray}
\mathcal{S}^1_{edge}=\sum_{l\in\mbz}C_l\int\text{d}x\text{d}t\cos\big[2l(\phi_1+\phi_2)+\alpha_l\big]
\end{eqnarray}
or
\begin{eqnarray}
\mathcal{S}^1_{edge}=\sum_{l\in\mbz}D_l\int\text{d}x\text{d}t\cos\big[2l(\phi_1-\phi_2)+\beta_l\big]
\end{eqnarray}
They correspond to two nontrivial SPT phases, where the edge cannot be gapped without spontaneously breaking the $Z_2$ symmetry. Let's label these two states as $[\eta,t_1,t_2]=[0,0,1]$ and $[0,1,0]$. Notice that when we put a $[0,0,1]$ edge with variables $\{\phi_1,\phi_2\}$ together with a a $[0,1,0]$ edge with variables $\{\phi_1^\prime,\phi_2^\prime\}$, the edge can always be gapped by condensing \eg independent bosonic variables $\{\phi_1+\phi_2^\prime,\phi_1^\prime-\phi_2\}$ and no symmetry will be broken. Hence $[0,0,1]\oplus[0,1,0]=\bse_{Z_2\times Z_2^f}$ and we have $[0,0,1]=[0,1,0]^{-1}$. On the other hand, if we put four $[0,1,0]$ states with edge variables $\{\phi^a_R,\phi^a_L,~a=1,2,3,4\}$ together, the edge can be gapped without breaking the symmetry, by localizing the following independent bosonic variables:
\begin{eqnarray}
&\notag\phi_R^1+\phi_R^2+\phi_L^3+\phi_L^4,\\
&\notag\phi_R^3+\phi_R^4+\phi_L^1+\phi_L^2,\\
&\notag\phi_R^1+\phi_R^3+\phi_L^1+\phi_L^4,\\
&\notag\phi_R^1+\phi_R^4+\phi_L^1+\phi_L^3.
\end{eqnarray}
As a results we have $[0,1,0]^4=\bse_{Z_2\times Z_2^f}$ and hence $[0,1,0]^3=[0,0,1]$. Therefore all different fermionic phases form a $\mbz_4$ group.

To summarize, with $Z_2$ symmetry transformation $W^\bsg=\pm I_{2\times2}$, there are $\mbz_4$ classes of different fermionic non-chiral SRE phases in the presence of $Z_2\times Z_2^f$ symmetry. The nontrivial SPT phases such as $[0,1,0]$ can be realized by non-interacting fermions, as shown by the coupled wire construction in section \ref{COUPLED WIRE CONSTRUCTION:BOSON SPT}.

{\bf Interacting fermionic SPT phases from the bosonic SPT phase with $Z_2$ symmetry}

In the previous discussion of bosonic SPT phases, a $\mbz_2$ classification was found for bosons with $Z_2$ symmetry. Here in a fermionic system with $Z_2\times Z_2^f$ symmetry, we can always let the fermions combine to form bosonic Cooper pairs which can serve as the fundamental bosons, which then form the nontrivial bosonic SPT phase discussed in section \ref{BOSONIC SPT}. Notice that the fermion parity $Z_2^f$ symmetry can never be broken and have no effect on the Cooper pairs at all. Do these interacting bosonic SPT phases lead to an extra $\mbz_2$ classification for fermions with $Z_2\times Z_2^f$ symmetry, in the presence of deconfined fermions in the low-energy sector? If so these non-trivial SPT phases cannot be obtained from perturbing non-interacting fermions. However, we show now that the bosonic SPT phase with $Z_2$ symmetry is contained within the fermion classification discussed previously when there are gapless fermions on the edge. And it is a $Z_2$ subgroup of the $Z_4$ classes that were found. Thus they can be obtained from adding perturbation to a non-interacting fermion Hamiltonian.

Consider one bosonic $Z_2$-symmetric SPT state with edge variables $\{\phi_1,\phi_2\}$, whose symmetry transformations are $\phi_a\rightarrow\phi_a+\pi,~a=1,2$ under $Z_2$ generator $\bsg$. When this state is put together with two fermion $Z_2\times Z_2^f$-symmetric SPT states $[0,1,0]\oplus[0,1,0]$ with edge variables $\{\phi_R,\phi_L\}$ and $\{\phi_R^\prime,\phi_L^\prime\}$, its edge can be gapped out by simultaneously localizing the following bosonic fields on the edge:
\begin{eqnarray}
&\notag\phi_R+\phi_L+\phi_1,\\
&\notag-\phi_R^\prime+\phi_L+\phi_2,\\
&\label{higgs:fermion:z2}\phi_R-\phi_L^\prime-\phi_2.
\end{eqnarray}
Notice that under $Z_2$ generator $\bsg$ the edge variables $\{\phi_R,\phi_L\}$ transform as $\phi_R\rightarrow\phi_R+\pi$,~$\phi_L\rightarrow\phi_L$ and the same for $\{\phi_R^\prime,\phi_L^\prime\}$. Notices that the inverse of the above bosonic SPT phase is itself, hence we have shown that $[0,1,0]^2\equiv[0,1,0]\oplus[0,1,0]$, \ie the state obtained by putting two fermion SPT phases $[0,1,0]$ together is nothing but the bosonic $Z_2$-symmetric SPT phase. Therefore we conclude that bosonic SPT phase with $Z_2$ symmetry is contained within the fermionic classification $Z_4$.\\

\subsubsection{$G_f=Z_4$ symmetry:~~~$\mbz_2$ classes}

Also note that when $\eta=1$ we have the algebra
\begin{eqnarray}
\bsg^2=\fnp.
\end{eqnarray}
for symmetry group $G_f=Z_4$. Note that $Z_2^f$ is a normal subgroup of $Z_4$. Since all phases with $\eta=1$ are trivial, they do not give rise to nontrivial (intrinsic) fermionic SPT phases with $Z_4$ symmetry. However as discussed before, the bosonic SPT phase of Cooper pairs with symmetry $G_f/Z_2^f=Z_2$ (when fermions are confined) always automatically lead to one interacting fermionic SPT phase protected by symmetry $G_f=Z_4$. Hence all different SRE fermionic phases with symmetry group $G_f=Z_4$ have at least a $\mbz_2$ classification. These are physically related to charge-$4e$ superconductors in two dimensions protected by electron charge conservation modulo $4$. The nontrivial fermionic SPT phase protected by $G_f=Z_4$ symmetry cannot be obtained by perturbing non-interacting fermions, and therefore they are intrinsic interaction-driven fermion SPT phases (charge-$4e$ superconductors in this case) with $Z_4$ symmetry.\\

In summary, there are at least $\mbz_4$ classes of different fermionic non-chiral SRE phases in the presence of $Z_2\times Z_2^f$ symmetry. Bosonic SPT phases with symmetry $Z_2$ do not add any new phases. 

On the other hand, in a fermion system with $Z_4$ symmetry ($\bsg^2=\fnp$ or $\eta=1$) there are $\mbz_2$ classes of different fermionic SRE phases. This corresponds to two different classes of charge-$4e$ superconductors in 2+1-D.

{\bf Discussion of Results:} The fermionic topological phases protected by $Z_2\times Z_2^f$ symmetry form a $\mbz_4$ group. In comparison, super-cohomology theory\cite{Gu2012} obtains the same number of phases but with group structure $\mbz_2^2$. An advantage of the present formalism is that we can see how these phases connect to the bosonic SPT phases with the same symmetry, and verify they do not add any new phases.

\subsection{$G_f/Z_2^f=Z_2\times Z_2^T$ symmetry}

The algebraic structure of $Z_2\times Z_2^T$ group is
\begin{eqnarray}\label{symmetry:z2:z2t}
&\bsg^2=\bst^2=\bsg\bst\bsg\bst=\bse.
\end{eqnarray}
where $\bsg$ is the $Z_2$ generator and time reversal operation $\bst$ is the $Z_2^T$ generator. In our notation $\bse$ can be either identity element $\bse_f$ for fermions or fermion number parity $\fnp$. The associate constraint (\ref{condition:algebra}) are given by
\begin{eqnarray}
&\notag(W^\bsg)^2=(W^\bst)^2=(W^\bsg W^\bst)^2=I_{2\times2},\\
&\label{constraint:fermion:z2:z2t:W}(W^\bsg)^T{\bf K}W^\bsg=-(W^\bst)^T{\bf K}W^\bst={\bf K},\\
&\notag(1+W^\bsg)\delta\phi^\bsg=\eta_\bsg\pi\begin{pmatrix}1\\1\end{pmatrix},\\
&\notag(1-W^\bst)\delta\phi^\bst=\eta_\bst\pi\begin{pmatrix}1\\1\end{pmatrix},\\
&\label{constraint:fermion:z2:z2t:dphi}(1-W^\bst W^\bsg)(\delta\phi^\bst-W^\bst\delta\phi^\bsg)=\eta\pi\begin{pmatrix}1\\1\end{pmatrix}.
\end{eqnarray}
where $\eta,\eta_\bst,\eta_\bsg=0,1$. We can always choose a gauge so that $W^\bst=\sigma_x$ and from (\ref{constraint:fermion:z2:z2t:W}) $W^\bsg=\pm I_{2\times2}$. We haven't found a microscopic realization of symmetry transformation $W^\bsg=-I_{2\times2}$ so far, therefore we put the discussions of $W^\bsg=-I_{2\times2}$ case to the Appendix. Here we'll focus on symmetry transformation $W^\bsg_{2\times2}$ case.


For $W^\bsg=I_{2\times2}$ the inequivalent solutions to (\ref{constraint:fermion:z2:z2t:dphi}) are
\begin{eqnarray}\label{sym trans:fermion:z2:z2t}
&\delta\phi^\bsg=(\alpha+\frac{\eta_\bsg}2)\pi\begin{pmatrix}1\\1\end{pmatrix}+\pi\begin{pmatrix}\eta-\eta_\bst\\0\end{pmatrix},\\
&\delta\phi^\bst={\eta_\bst}\pi\begin{pmatrix}0\\1\end{pmatrix},~~~\alpha,\eta,\eta_\bst,\eta_\bsg=0,1.\notag
\end{eqnarray}
If $\eta_\bsg+\eta-\eta_\bst=0$ the variable $\phi_1+\phi_2$ can be localized without breaking any symmetry. If $\eta=\eta_\bst=0$ the variable $\phi_1-\phi_2$ can be localized without breaking any symmetry.

Note that when $\eta_\bsg=0$ we have $\bsg^2=\bse_f$ and hence the symmetry group is $G_f=(Z_2\rtimes Z_2^T)\times Z_2^f$ with $\bst^2=\bse_f$ if $\eta_\bst=0$ or $\bst^2=\fnp$ if $\eta_\bst=1$. When $\eta_\bsg=1$ on the other hand we have $\bsg^2=\fnp$, and symmetry group for fermions is $G_f=Z_4\rtimes Z_2^T$ or $G_f=Z_4\times Z_2^T$.

\subsubsection{$G_f=(Z_2\rtimes Z_2^T)\times Z_2^f$ symmetry:~~$\mbz_4$ classes}

There are 4 nontrivial SPT phases with $\eta_\bsg=0$: they have $\eta-\eta_\bst=1\mod2$ and $\alpha=0,1$. Let's label a state with symmetry transformations (\ref{sym trans:fermion:z2:z2t}) as $[\eta_\bsg,\eta_\bst,\eta,\alpha]$. We already showed $[1,0,0,\alpha]=[1,\eta+1,\eta,\alpha]=[0,\eta,\eta,\alpha]=\bse_{(Z_2\rtimes Z_2^T)\times Z_2^f}$. When a $[0,\eta+1,\eta,\alpha]$ state with edge variables $\{\phi_{1},\phi_{2}\}$ is put together with a $[0,\eta^\prime+1,\eta^\prime,\alpha]^{-1}$ state with edge variables $\{\phi_{1}^\prime,\phi_{2}^\prime\}$, its edge cannot be gapped without breaking the symmetry by localizing independent bosonic fields $\{\phi_1+\phi_1^\prime,\phi_2+\phi_2^\prime\}$. Therefore we know $[0,1,0,\alpha]=[0,0,1,\alpha]$ is the same nontrivial SPT phase. In the case $\eta_\bsg=0=\eta_\bst$ and $\eta=1$ the algebra of $(Z_2\rtimes Z_2^T)\times Z_2^f$ group is
\begin{eqnarray}
\bsg^2=\bst^2=\bse_f,~~~\bsg\bst\bsg^{-1}\bst^{-1}=\fnp.
\end{eqnarray}
As discussed in \Ref{Qi2012} this is the same as $\eta_\bsg=0=\eta$ and $\eta_\bst=1$, since one can always redefine the anti-unitary time reversal as $\bst^\prime\equiv\bsg\bst$. Just as discussed earlier for fermionic SPT phases with $Z_2\times Z_2^T$ symmetry, it's easy to verify that $[0,1,0,0]=[0,1,0,1]^{-1}$ and when four $[0,1,0,0]$ states are put together their edges can be gapped without breaking any symmetry \ie $[0,1,0,0]^4=\bse_{(Z_2\rtimes Z_2^T)\times Z_2^f}$. As a result $[0,1,0,0]^n,~n=1,2,3$ are the only three nontrivial SPT phase, whose edge cannot be gapped without breaking the symmetry. Hence all different phases with $W^\bsg=I_{2\times2}$ form a $\mbz_4$ group for $(Z_2\rtimes Z_2^T)\times Z_2^f$ symmetry ($\eta_\bsg=0$).

For the same reason mentioned earlier for fermions with $Z_2\times Z_2^f$ symmetry, here in the presence of $(Z_2\rtimes Z_2^T)\times Z_2^f$ symmetry, we can obtain interacting SPT phases from bosonic SPT phases of fermion pairs with $Z_2\times Z_2^T$ symmetry. Note there are $\mbz_2^2$ classes of bosonic non-chiral SRE phases with with $Z_2\times Z_2^T$ symmetry. Do these lead to an extra $\mbz_2^2$ group structure for fermions with this symmetry?  As before, we now show that this is {\em not} the case. These phases are already accounted for within the fermionic classification when fermions are present on the edge. 

Again let us consider a bosonic $Z_2\times Z_2^T$-symmetric SPT state $[1,n_2,n]_b$ with edge variables $\{\phi_1,\phi_2\}$, which transforms as
\begin{eqnarray}
&\notag\bsg:~~~\begin{pmatrix}\phi_1\\ \phi_2\end{pmatrix}\rightarrow\begin{pmatrix}\phi_1+\pi\\ \phi_2+n\pi\end{pmatrix}\\
&\notag\bst:~~~\begin{pmatrix}\phi_1\\ \phi_2\end{pmatrix}\rightarrow\begin{pmatrix}-\phi_1\\ \phi_2+n_2\pi\end{pmatrix}
\end{eqnarray}
under $Z_2$ generator $\bsg$ and time reversal $\bst$. When this state is put together with two fermion $Z_2\times Z_2^f$-symmetric SPT states $[0,1,0,0]_f\oplus[0,1,0,0]_f$ with edge variables $\{\phi_R,\phi_L\}$ and $\{\phi_R^\prime,\phi_L^\prime\}$, its edge can be gapped out by simultaneously localizing the bosonic fields (\ref{higgs:fermion:z2}) on the edge if $n=1$. This means
\begin{eqnarray}
&\notag[1,n_2,1]_b\oplus[0,1,0,0]_f\oplus[0,1,0,0]_f=e_{(Z_2\rtimes Z_2^T)\times Z_2^f}.
\end{eqnarray}
Similarly one can show that
\begin{eqnarray}
&\notag[1,n_2,0]_b\oplus[0,1,0,0]_f\oplus[0,1,0,1]_f=e_{(Z_2\rtimes Z_2^T)\times Z_2^f}\\
&\notag=[0,1,0,0]_f\oplus[0,1,0,1]_f.
\end{eqnarray}
again by localizing the same bosonic variables (\ref{higgs:fermion:z2}). Hence we have shown that $\notag[1,n_2,1]_b=[0,1,0,0]_f\oplus[0,1,0,0]_f$ and $\notag[1,n_2,0]_b=[0,1,0,0]_f\oplus[0,1,0,1]_f=e_{(Z_2\rtimes Z_2^T)\times Z_2^f}$. So in the presence of fermions, all the bosonic SPT phases protected by $(Z_2\rtimes Z_2^T)\times Z_2^f$ can be obtained by perturbing non-interacting fermions (constructed by putting several non-interacting $[0,1,0,0]_f$ together and add interactions). Therefore we conclude that bosonic SPT phase with $Z_2\times Z_2^T$ symmetry can all be obtained from perturbing non-interacting fermions in the presence of $Z_2\times Z_2^T$ symmetry. Hence in total there are $\mbz_4$ classes of different fermionic (non-chiral) SRE phases with $(Z_2\rtimes Z_2^T)\times Z_2^f$ symmetry: all can be obtained from perturbing free fermions.

{\bf Discussion of Results:} While this particular symmetry class cannot be discussed within group super-cohomology theory\cite{Gu2012}, recent work \cite{Qi2012,Yao2012,Ryu2012} have approached this problem from another angle, by starting with non-interacting fermions (which have a $\mbz$ classification with this symmetry) and then turning on interactions. They find a $\mbz_8$ classification that survives interactions. Odd integer members of this series have an odd number of pair of Majorana modes at the edge, that move in opposite directions. Although apparently quire different, these results are consistent with ours due to the following. Since we are unable to deal with unpaired Majorana modes, only the even members of the series are captured (hence $\mbz_4$ classes here). An advantage though is that this classification of topological phases which are stable to interaction emerges directly from the formalism, without the need to begin from free fermions. 

\subsubsection{$G_f=Z_4\rtimes Z_2^T$ with $\bst^2=\fnp$:~~~$\mbz_2^2$ classes}

On the other hand, when $\eta_\bsg=1$ we have $\bsg^2=\fnp=(-1)^{\hat{N}_f}$ and hence the corresponding symmetry group is actually $Z_4\times Z_2^T$, where $Z_2^f$ is a subgroup of $Z_4$. In this case the nontrivial SPT phases have $\eta_\bsg=1$ and $\eta=\eta_\bst=1$,~$\alpha=0,1$. The algebraic structure of the symmetry group $G_f$ is given by
\begin{eqnarray}
\bsg^2=\bst^2=\bsg\bst\bsg\bst=\fnp.
\end{eqnarray}
It's easy to check that $\bst\bsg=\bsg^{-1}\bst$ and hence the symmetry group is actually $Z_4\rtimes Z_2^T$ with $\bst^2=\fnp$. We still label the phases with symmetry transformations (\ref{sym trans:fermion:z2:z2t}) as $[\eta_\bsg,\eta_\bst,\eta,\alpha]$. When a $[1,1,1,0]$ state with edge variables $\{\phi_{1},\phi_{2}\}$ is put together with a $[1,1,1,1]^{-1}$ state with edge variables $\{\phi_{1}^\prime,\phi_{2}^\prime\}$, its edge cannot be gapped without breaking the symmetry by localizing independent bosonic fields $\{\phi_1+\phi_1^\prime,\phi_2+\phi_2^\prime\}$. Therefore $[1,1,1,0]=[1,1,1,1]$ is the same nontrivial SPT phase. Now let's put two $[1,1,1,0]$ states together with edge variables $\{\phi_L,\phi_R\}$ and $\{\phi_L^\prime,\phi_R^\prime\}$, it's easy to see the edge states will be all gapped out by simultaneously localizing the following bosonic variables: $\{\phi_R-\phi_L^\prime,\phi_R^\prime-\phi_L\}$. Hence we have $[1,1,1,0]^2=\bse_{Z_4\rtimes Z_2^T}$ and these different intrinsic fermionic SRE phases with $Z_4\times Z_2^T$ form a $\mbz_2$ group. The nontrivial fermionic SPT phase can be obtained from free fermion band structures, just like $G_f=(Z_2\rtimes Z_2^T)\times Z_2^f$ case. In fact its edge states are similar to those of quantum spin Hall insulators.

In the case of a fermion system with $Z_4\rtimes Z_2^T$ symmetry, the corresponding bosonic system of Cooper pairs have $Z_2\times Z_2^T$ symmetry and has a $\mbz_2^2$ topological classification. For fermions in the presence of $Z_4\rtimes Z_2^T$ symmetry, these bosonic SPT phases of strongly-bound Cooper pairs (with $Z_2\times Z_2^T$ symmetry and a $\mbz_2^2$ classification) may potentially bring in new phases, in addition to the the nontrivial fermionic SPT phase $[1,1,1,0]$ (with a $\mbz_2$ group structure). Again we use $[1,n_2,n]_b$ to label the bosonic $Z_2\times Z_2^T$-SPT phases where $n_2,n=0,1$. One can show that
\bea
\notag[1,n_2,0]_b\oplus[1,1,1,0]_f\oplus[1,1,1,0]_f=e_{Z_4\rtimes Z_2^T}
\eea
and hence bosonic $Z_2\times Z_2^T$-SPT phase $[1,1,0]_b=$ becomes trivial in the presence of fermions with $\bst^2=\fnp$. On the other hand bosonic SPT phases $[1,n_b,1]_b$ cannot be obtained by perturbing free fermions, and gives rise to an extra $\mbz_2$ classification.

Hence to summarize, all different fermionic SRE phases with $Z_4\rtimes Z_2^T$ symmetry have a $\mbz_2^2$ classification, where one $\mbz_2$ structure comes from bosonic SPT phases $[1,n_2,1]_b$ of Cooper pairs, and the other $\mbz_2$ associated with fermion state $[1,1,1,0]$ are intrinsic properties of fermionic systems. These $\mbz_2^2$ classes of phases physically correspond to different charge-$4e$ superconductors with time reversal symmetry $\bst^2=\fnp$. Recently, the possibility of realizing charge-$4e$ superconductivity (four fermion condensates), in imbalanced cold atomic gases\cite{Radzihovsky} (which break time reversal symmetry) and also in certain cuprate superconductors\cite{BergKivelson} (which preserve time reversal) has been discussed. While these phases were non-topological, the prospects for realizing topological phases with these symmetries remains to be seen.

\subsubsection{$G_f=Z_4\times Z_2^T$ symmetry:~~~$\mbz_2^2$ classes}

In this case we have $\eta_\bsg=1$ and $\eta_\bst-\eta=1$ and therefore
\begin{eqnarray}
&\bsg^2=\fnp,~~~\bsg\bst=\bst\bsg,\\
&\bst^2=\bse_f~\text{or}~\fnp.\notag
\end{eqnarray}

In this case there are no intrinsic fermionic SPT phases with $G_f=Z_4\times Z_2^T$ symmetry. However bosonic SPT phases of Cooper pairs with $G_f/Z_2^f=Z_2\times Z_2^T$ symmetry leads to $\mbz_2^2$ classes of different fermion SRE phases. Hence there are at least $\mbz_2^2$ classes of different fermion non-chiral SRE phases with $G_f=Z_4\times Z_2^T$ symmetry. Physically they correspond to time reversal symmetric topological superconductors with $Z_4$ spin rotation symmetry along certain axis.


\section{Coupled wire construction of bosonic and fermionic SPT phases}\label{COUPLED WIRE CONSTRUCTION:BOSON SPT}

In the previous sections we showed how to classify different SPT phases in the ${K}$ matrix + Higgs formulation. The edge structure of SPT phases is explicit in this formulation: \eg the edge of a bosonic SPT phase is characterized by bare action (\ref{bare:boson:edge}), Kac-Moody algebra (\ref{KM algebra:boson non-chiral SRE}), as well as symmetry transformation rules (\ref{unitary}) for unitary symmetry $g$ and (\ref{antiunitary}) for anti-unitary symmetry $h$ on the bosonic variable $\{\phi_a,~a=1,2\}$. However, a more microscopic construction of these $2+1$-D SPT phases is still lacking in this formulation. In this section we present a microscopic construction of these SPT phases in the anisotropic (quasi-1D) limit, from an array of coupled one-dimensional quantum wires. This approach has been used to construct Abelian\cite{Kane2002} and non-Abelian FQH states\cite{Teo2011}. We first give a short introduction to the coupled wire construction, and then use the this method to explicitly construct bosonic SPT phases in the presence of symmetry group $G=U(1)$ and $G=U(1)\rtimes Z_2^T$, as well as fermionic SPT phases with symmetry group $Z_2\times Z_2^F$. Generalizations to other symmetry groups are straightforward.

\begin{figure}
 \includegraphics[width=0.45\textwidth]{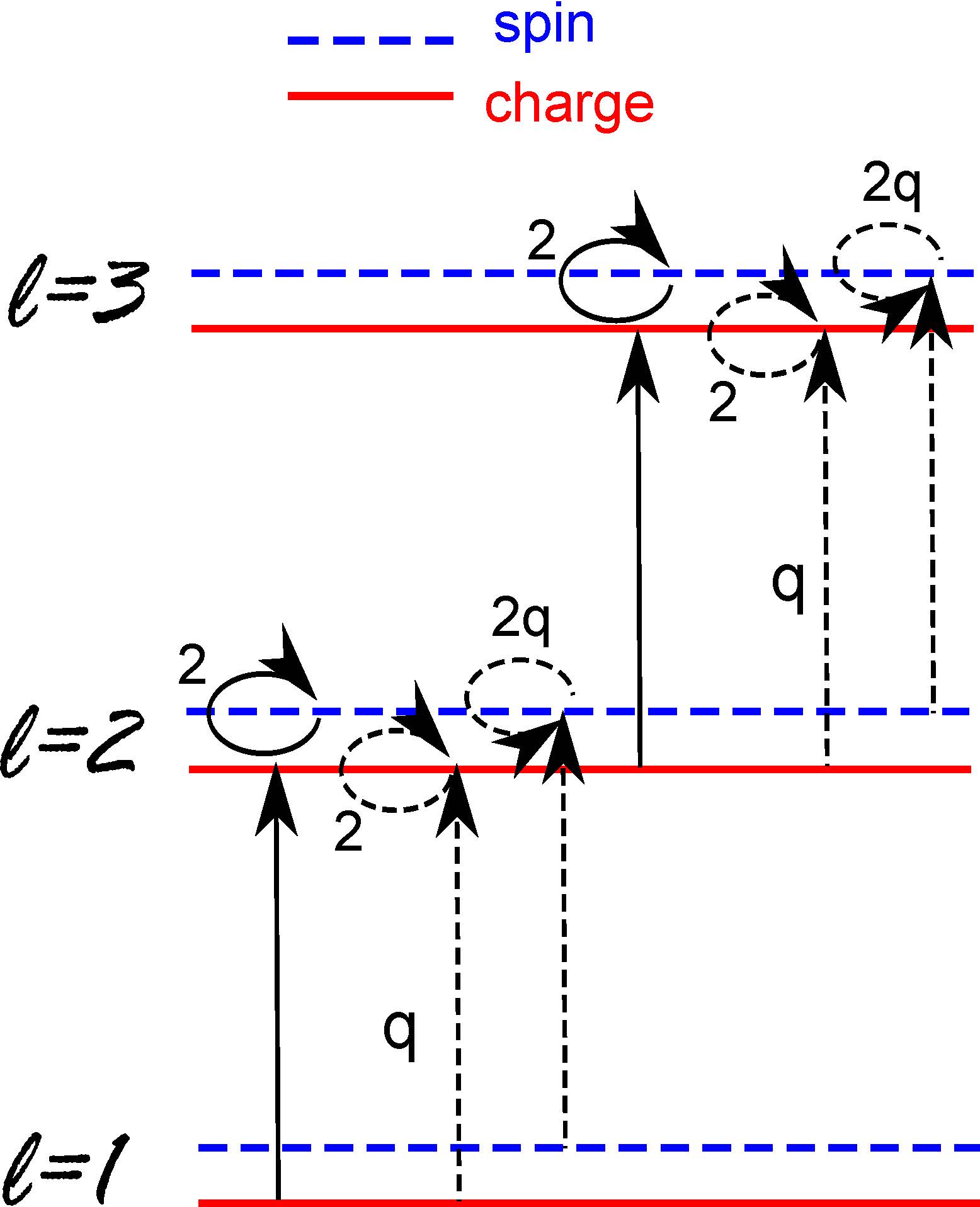}
\caption{Schematic illustration of interwire coupling terms which stabilize the bosonic SPT phases protected by $U(1)$ symmetry, with Hall conductance $\sigma_{xy}=2q$. Solid horizontal lines stand for quantum wires of charged bosons (each carries unit $U(1)$ charge) while dashed horizontal lines represent quantum wires composed of neutral (say spin) degrees  of freedom. Dashed and solid arrows illustrate the two interwire coupling terms in (\ref{interwire:U(1)}) that gap the bulk, but leave behind nontrivial edge states.}\label{fig:u1 coupled wire}
\end{figure}

\subsection{Coupled wire construction in a nutshell}

Consider an array of uncoupled identical one-dimensional quantum wires, each wire being described by a non-chiral Luttinger liquid. The bosonic fields associated with the Luttinger liquid in the $l$-th wire ($1\leq l\leq N_w$,~$N_w$ being the total number of quantum wires) are $\{\theta_l(x),\varphi_l(x)\}$ satisfying the following commutation relation
\begin{eqnarray}\label{luttinger liquid:commutation relation}
\big[\theta_m(x),\varphi_l(y)\big]=\imth\frac{\pi}{2}\text{Sign}(x-y)\delta_{m,l}.
\end{eqnarray}
$\varphi(x)$ is a bosonic phase field, while $\theta(x)$ describes the density fluctuations in the Luttinger liquid. The long-wavelength density fluctuations on $l$-th wire is given by
\begin{eqnarray}
\rho_l(x)-\bar\rho_l=\partial_x\theta_l(x)/\pi.
\end{eqnarray}
where $\bar\rho_l$ is the average particle (boson or fermion) density in the $l$-th wire. In terms of these two variables, the Luttinger liquid Hamiltonian for decoupled quantum wires is given by
\begin{eqnarray}\label{luttinger liquid:bare hamiltonian}
H_{LL}=\sum_l\frac{v_l}{2\pi}\int\text{d}x\Big[\frac1{g_l}(\partial_x\theta_l)^2+g_l(\partial\varphi_l)^2\Big].
\end{eqnarray}
where $g_l>0$ is the Luttinger parameter for the $l$-th wire.

The idea of coupled wire construction is to add inter-wire and intra-wire interactions between electrons as well as tunneling between wires. For example the forward scattering terms between different wires is written as
\begin{eqnarray}\label{luttinger liquid:forward scattering}
H_{FC}=\sum_{k\neq l}\int\text{d}x~(\partial\varphi_k,\partial\theta_k){\bf M}_{k,l}\begin{pmatrix}\partial\varphi_l\\ \partial\theta_l\end{pmatrix}
\end{eqnarray}
where ${\bf M}_{j,k}$ are all $2\times2$ matrices describing the forward scattering interactions between the $j$-th and $k$-th wire. Other inter-channel scattering terms in general have the following form:
\begin{eqnarray}
&\notag\mathcal{O}^{\{m_l,n_l\}}(x)\sim\cos\Big[\imth\sum_{l}\big(m_l\theta_l(x)+n_l\varphi_l(x)\big)+\alpha_{\{m_l,n_l\}}\Big].\\\
&\label{luttinger liquid:interwire scattering}
\end{eqnarray}
where $\alpha_{\{m_l,n_l\}}$ are real constants and $\{m_l,n_l\}$ all take integer values. For a bosonic system in the absence of any symmetry (or associated conservation laws), the above interwire scattering term must satisfy the following condition:
\begin{eqnarray}\label{coupled wire:constraint:boson}
m_l=0\mod2,~~~\forall~1\leq l\leq N_w.
\end{eqnarray}
This is because the boson density fluctuations are mainly contributed by density waves at vector $q_n\sim2\pi\bar\rho_ln,~n\in\mbz$ and the density fluctuations at $q_n$ is given by
\begin{eqnarray}\notag
&\rho_n^l(x)\propto e^{\imth n\big(2\pi\bar\rho_lx+2\theta_l(x)\big)},~~~n\in\mbz,~1\leq l\leq N_w.
\end{eqnarray}
for the $l$-th quantum wire. For a fermionic system on the other hand, in the absence of any symmetry the constraint on interwire scattering term (\ref{luttinger liquid:interwire scattering}) is
\begin{eqnarray}\label{coupled wire:constraint:fermion}
m_l=n_l\mod2,~~~\forall~1\leq l\leq N_w.
\end{eqnarray}
This is because inter channel scattering terms must be composed of single fermion operators: \ie left mover $\psi^R_{l}\sim\exp[\imth(\varphi_l-\theta_l-\pi\bar\rho_lx)]$ and right mover $\psi^R_{l}\sim\exp[\imth(\varphi_l+\theta_l+\pi\bar\rho_lx)]$. The presence of symmetry group $G$ will lead to further constraints on inter-wire coupling terms (\ref{luttinger liquid:interwire scattering}) and forward scattering terms (\ref{luttinger liquid:forward scattering}): symmetry-allowed scattering terms must transform trivially under any symmetry operation. The bare Luttinger liquid Hamiltonian together with symmetry-allowed forward scattering (\ref{luttinger liquid:forward scattering}) and inter channel scattering (\ref{luttinger liquid:interwire scattering}) forms the generic Hamiltonian for a coupled wire construction:
\begin{eqnarray}\label{coupled wire:hamiltonian}
&\notag H=H_{LL}+H_{FC}+\sum_{\{m_l,n_l\}}\int\text{d}xC_{\{m_l,n_l\}}\mathcal{O}^{\{m_l,n_l\}}(x).\\
\end{eqnarray}

In the coupled wire construction (\ref{coupled wire:hamiltonian}), one can properly choose a set of symmetry-allowed interwire scattering terms $\{\mathcal{O}^{\{m_l,n_l\}}(x)\}$ in (\ref{coupled wire:hamiltonian}), so that
\begin{eqnarray}\label{coupled wire:constraint:commute}
\sum_lm_ln_l=\sum_l(m_ln^\prime_l+m_l^\prime n_l)=\sum_lm_l^\prime n_l^\prime=0.
\end{eqnarray}
for any two terms $\mathcal{O}^{\{m_l,n_l\}}(x)$ and $\mathcal{O}^{\{m_l^\prime,n_l^\prime\}}(x)$ in Hamiltonian (\ref{coupled wire:hamiltonian}). Therefore the set of bosonic variables $\{\sum_l(m_l\theta_l+n_l\varphi_l)\}$ can be simultaneously localized at classical values  by minimizing the interwire scattering terms $\{\mathcal{O}^{\{m_l,n_l\}}(x)\}$. When chosen properly all degree of freedom in the bulk will be gapped by these interwire scattering terms, and the only low-energy degrees of freedom left free are on the edge. To be specific the variable $p_1\theta_1+q_1\varphi_1$ on the left edge $l=1$ will remain gapless if
\begin{eqnarray}\label{coupled wire:constraint:edge}
p_1n_1+q_1m_1=0,~~~\forall~\{m_l,n_l\}~\text{in}~(\ref{coupled wire:hamiltonian}).
\end{eqnarray}
Note that one can always tune the forward scattering terms (\ref{luttinger liquid:forward scattering}) so that interwire coupling terms $\{\mathcal{O}^{\{m_l,n_l\}}(x)\}$ are relevant in the renormalization group sense. Then one expects the coupled wire system will be driven into strong coupling phase of interwire scattering $\{\mathcal{O}^{\{m_l,n_l\}}(x)\}$, and hence all bosonic variables $\{\sum_l(m_l\theta_l+n_l\varphi_l)\}$ will be localized simultaneously at classical values.

One of the simplest example is the Laughlin state\cite{Laughlin1983} of spinless fermions, \ie FQH state at filling fraction $\nu=1/m,~m=$odd integer. The interwire scattering terms whose strong coupling phase corresponding to the Laughlin state are\cite{Kane2002}
\begin{eqnarray}
H_{1/m}=\sum_{l=1}^{N_w-1}\int\text{d}xC_{l}\cos\Big[\varphi_l-\varphi_{l+1}-m(\theta_l+\theta_{l+1})\Big].\notag
\end{eqnarray}
Its gapless variable on the left edge is $\phi^L_{1/m}=\varphi_1+m\theta_1$ which satisfies the following Kac-Moody algebra\cite{Wen1990}
\begin{eqnarray}
[\partial_x\phi_{1/m}^L(x),\partial_y\phi_{1/m}^L(y)]=2\pi m\imth\partial_x\delta(x-y).\notag
\end{eqnarray}
It's easy to verify that $\phi_{1/m}^L$ satisfies condition (\ref{coupled wire:constraint:edge}) and is the only gapless degree of freedom on the edge.

\subsection{Bosonic SPT phases with $U(1)$ symmetry}

As discussed in section \ref{BOSONIC SPT}, in the presence of $U(1)$ symmetry there are $\mbz$ (integer group) different classes of bosonic non-chiral SRE states, which are labeled by their $U(1)$ charge vector ${\bf t}=(1,q)^T$. The bosonic variables $\{\phi_1,\phi_2\}$ on its edge satisfies Kac-Moody algebra (\ref{KM algebra:boson non-chiral SRE}). Under group element $U_{\Delta\phi}$ of symmetry group $U(1)$ they transform as
\begin{eqnarray}\label{sym trans:boson:U(1)}
&U_{\Delta\phi}:~~~\begin{pmatrix}\phi_1\\ \phi_2\end{pmatrix}\rightarrow\begin{pmatrix}\phi_1\\ \phi_2\end{pmatrix}+\Delta\phi\begin{pmatrix}1\\q\end{pmatrix},\\
&\notag~~~~~\Delta\phi\in[0,2\pi).
\end{eqnarray}
The nontrivial SPT phases correspond to different integers $q\neq0$, whose edge cannot be gapped out without breaking the $U(1)$ symmetry. Here we present an explicit construction of these SPT phases with $U(1)$ symmetry in the coupled wire approach.

We start from an array of quantum wires ($1\leq l\leq N_w$) where each wire is composed of two chains: a chain of charged bosons (each boson carries a unit of $U(1)$ electric charge) and a spin chain. Each chain forms a $c=1$ Luttinger liquid described by bosonic fields: $\{\varphi^s_l(x),\theta^s_l(x)\}$ for the spin chain, $\{\varphi^c_l(x),\theta^c_l(x)\}$ for the chain of charged bosons in the $l$-th wire. These bosonic fields satisfies the commutation relation (\ref{luttinger liquid:commutation relation}):
\begin{eqnarray}\label{spin+charge:commutation relation}
\big[\theta_m^\alpha(x),\varphi^\beta_l(y)\big]=\imth\frac{\pi}{2}\text{Sign}(x-y)\delta_{m,l}\delta_{\alpha,\beta}.
\end{eqnarray}
where $\alpha,\beta=c/s$ denotes charge/spin degree of freedom and $1\leq m,l\leq N_w$ are the wire index. $\varphi_l^c$ are phase fields of charged bosons while $2\partial_x\theta_l^c$ describes charged boson density fluctuations. For the spin chain $\partial_x\theta_l^s(x)\sim S^z_l(x)$ and $\exp[\imth\varphi_l^s(x)]\sim S^+_l(x)$. Without interwire scattering terms, the bare Hamiltonian density of the system takes the form (\ref{luttinger liquid:bare hamiltonian}) of Luttinger liquids
\begin{eqnarray}\label{spin+charge:bare hamiltonian}
\mathcal{H}_{LL}=\sum_{l=1}^{N_w}\sum_{\alpha=c/s}\frac{v_l^\alpha}{2\pi}\Big[\frac1{g_{l}^\alpha}(\partial_x\theta_l^\alpha)^2+g_l^\alpha(\partial\varphi_l^\alpha)^2\Big].
\end{eqnarray}
The $U(1)$ symmetry associated with $\{\varphi^c_l(x),\theta^c_l(x)\}$ boson charge conservation leads to the following symmetry transformations for the bosonic fields:
\begin{eqnarray}\label{spin+charge:U(1) transf}
&\varphi_l^c(x)\rightarrow\hat{U}_{\Delta\phi}\varphi_l^c(x)\hat{U}_{\Delta\phi}^{-1}=\varphi_l^c(x)+\Delta\phi,\\
&\notag\hat{U}_{\Delta\phi}\equiv e^{\imth\Delta\phi\int\text{d}x\sum_l2\partial_x\theta_l^c(x)},~~0\leq\Delta\phi<2\pi.
\end{eqnarray}
The other fields $\theta_l^c,~\varphi_l^s,~\theta_l^s$ are invariant under the above $U(1)$ charge rotation $\hat{U}_{\Delta\phi}$.

In the presence of the above $U(1)$ symmetry associated with boson charge conservation, the different phases labeled by charge vector ${\bf t}=(1,q)^T$ are stablized by the following inter-wire coupling terms:
\begin{eqnarray}
&\notag\mathcal{H}^1_{(1,q)}=\sum_{l=1}^{N_w-1}\Big[C_l\cos(\varphi^c_l-\varphi^c_{l+1}-2\theta_l^s+\lambda_l)+D_l\cdot\\
&\notag\cos\big(\varphi_l^s-\varphi_{l+1}^s+q(\varphi_l^c-\varphi_{l+1}^c)-2(\theta_{l+1}^c-q\theta_{l+1}^s)+\lambda_l^\prime\big)\Big].\\
\label{interwire:U(1)}
\end{eqnarray}
where $C_l,D_l,\lambda_l,\lambda_l^\prime$ are real constants. A pictorial illustration of the above interwire scattering terms is given in FIG. \ref{fig:u1 coupled wire}. Clearly the above interwire scattering terms all satisfy constraint (\ref{coupled wire:constraint:boson}) for bosonic systems, and they are also invariant under $U(1)$ rotation (\ref{spin+charge:U(1) transf}).

As argued in \Ref{Kane2002,Teo2011} one can always choose proper forward scattering terms (\ref{luttinger liquid:forward scattering}) to make the above interwire coupling terms become relevant and drive the system into their strong coupling phase. Notice that the arguments of the above $\cos$ terms commute with each other, so they can be localized at certain classical values simultaneously. It's straightforward to show that all bosonic fields in the bulk with $2\leq l\leq N_w-1$ are gapped while the gapless edge states on the left edge $l=1$ are described by variables $\{\phi_1^1(x),\phi_1^2(x)\}$ defined as
\begin{eqnarray}\label{coupled wire:U(1):variable:left edge}
&\phi_l^1\equiv\varphi_l^c,\\
&\notag\phi_l^2\equiv\varphi_l^s+q\varphi_l^c+2(\theta_l^c-q\theta_l^s).
\end{eqnarray}
They transform exactly like $\{\phi_1,\phi_2\}$ in (\ref{sym trans:boson:U(1)}) under charge $U(1)$ symmetry (\ref{spin+charge:U(1) transf}). Besides they also obtain the Kac-Moody algebra (\ref{KM algebra:boson non-chiral SRE}) for a bosonic non-chiral SRE system. As a result the strong coupling phase of interwire couplings (\ref{interwire:U(1)}) is nothing but the bosonic SPT phases labeled by charge vector ${\bf t}=(1,q)^T$ with charge $U(1)$ symmetry.

Now let us elaborate on why the interwire coupling (\ref{interwire:U(1)}) can gap out everything in the bulk and leave variables (\ref{coupled wire:U(1):variable:left edge}) on the edge. In addition to variables $\{\phi_l^1(x),\phi_l^2(x)\}$ in (\ref{coupled wire:U(1):variable:left edge}) one can define another pairs of variables $\{\tilde\phi_l^1(x),\tilde\phi_l^2(x)\}$ as
\begin{eqnarray}\label{coupled wire:U(1):variable:right edge}
&\tilde\phi_l^1\equiv\varphi_l^c-2\theta_l^s,\\
&\notag\tilde\phi_l^2\equiv\varphi_l^s+q\varphi_l^c.
\end{eqnarray}
They also satisfy Kac-Moody algebra (\ref{KM algebra:boson non-chiral SRE}) except for an extra minus sign for all commutators. Notice that the two pairs of variables $\{\tilde\phi_l^1(x),\tilde\phi_l^2(x)\}$ and $\{\tilde\phi_l^1(x),\tilde\phi_l^2(x)\}$ commute with each other. They are just a linear combination of the original charge and spin variables $\{\varphi_l^c,\theta_l^c,\varphi_l^s,\theta_l^s\}$. The interwire scattering terms (\ref{interwire:U(1)}) can be written as
\begin{eqnarray}
&\notag\mathcal{H}^1_{(1,q)}=\sum_{l=1}^{N_w-1}\Big[C_l\cos(\tilde\phi_l^1-\phi_{l+1}^1+\lambda_l)\\
&\notag+D_l\cos\big(\tilde\phi_l^2-\phi_{l+1}^2+\lambda_l^\prime\big)\Big].
\end{eqnarray}
Hence everything in the bulk \ie $\{\tilde\phi_l^{1,2}(x),\phi_l^{1,2}(x),~2\leq l\leq N_w-1\}$ are all gapped since they don't commute with at least one of interwire scattering terms in (\ref{interwire:U(1)}). For the $l=1$ wire on the left edge, variables $\{\tilde\phi_1^1(x),\tilde\phi_1^2(x)\}$ are gapped for the same reason while variables $\{\phi_l^1(x),\phi_l^2(x)\}$ are left gapless. For the $l=N_w$ wire on the right edge, things happen in the opposite way: variables $\{\phi_{N_w}^1(x),\phi_{N_w}^2(x)\}$ are gapped while variables $\{\tilde\phi_{N_w}^1(x),\tilde\phi_{N_w}^2(x)\}$ remain gapless.

\subsection{Bosonic SPT phase with $U(1)\rtimes Z_2^T$ symmetry}

As discussed in section \ref{BOSONIC SPT}, in the presence of $U(1)\rtimes Z_2^T$ symmetry there are $\mbz_2$ different classes of non-chiral bosonic SRE phases in $2+1$-D. Among them there is only one nontrivial SPT phase, whose edge cannot be gapped without breaking the $U(1)\rtimes Z_2^T$ symmetry. Its gapless edge is described by bosonic fields $\{\phi_1,\phi_2\}$ which satisfies the Kac-Moody algebra (\ref{KM algebra:boson non-chiral SRE}). Under charge $U(1)$ rotation the two bosonic variables transform as (\ref{sym trans:boson:U(1)}) while under time reversal $\bst$ they transform as
\begin{eqnarray}\label{sym trans:boson:Z2T}
&\bst:~~~\begin{pmatrix}\phi_1\\ \phi_2\end{pmatrix}\rightarrow\begin{pmatrix}-\phi_1\\ \phi_2+\pi\end{pmatrix}
\end{eqnarray}
for the nontrivial SPT phase.

Such a SPT phase is nothing but the strong coupling phase of interwire scattering terms (\ref{interwire:U(1)}) with $q=0$. Its gapless bosonic fields on the left edge $l=1$ are
\begin{eqnarray}\label{coupled wire:U(1)rtimes Z2T:variable:left edge}
&\phi_l^1\equiv\varphi_1^c,\\
&\notag\phi_1^2\equiv\varphi_1^s+2\theta_l^c.
\end{eqnarray}
while on the right edge $l=N_w$ the gapless boson fields are
\begin{eqnarray}\label{coupled wire:U(1)rtimes Z2T:variable:right edge}
&\tilde\phi_{N_w}^1\equiv\varphi_{N_w}^c-2\theta_{N_w}^s,\\
&\notag\tilde\phi_{N_w}^2\equiv\varphi_{N_w}^s.
\end{eqnarray}
Under time reversal $\bst$ the original boson fields $\{\varphi_l^c,\theta_l^c,\varphi_l^s,\theta_l^s\}$ naturally transform as
\begin{eqnarray}\notag
&\bst\varphi_l^c\bst^{-1}=-\varphi_l^c,~~~&\bst\theta_l^c\bst^{-1}=\theta_l^c,\\
&\bst\varphi_l^s\bst^{-1}=\varphi_l^s+\pi,~~~&\bst\theta_l^s\bst^{-1}=-\theta_l^s.\notag
\end{eqnarray}
since all components of the spin $S^z\sim\partial_x\theta^s,~S^+\sim\exp[\imth\varphi^s]$ change sign under time reversal. It's easy to verify that interwire scattering terms (\ref{interwire:U(1)}) with $q=0$ is invariant under time reversal as long as we choose $\lambda_l=0$. Hence these interwire couplings are allowed by symmetry. It's also straightforward to show that the pair of bosonic fields, \ie both $\{\phi_l^1(x),\phi_l^2(x),~l=1\}$ and $\{\tilde\phi_{l}^1(x),\tilde\phi_{l}^2(x),~l=N_w\}$ transform in the same way as (\ref{sym trans:boson:U(1)}) and (\ref{sym trans:boson:Z2T}) under $U(1)\rtimes Z_2^T$ symmetry. Hence the strong coupling phase of interwire couplings (\ref{interwire:U(1)}) with $q=0$ indeed corresponds to the nontrivial bosonic SPT phase in the presence of $U(1)\rtimes Z_2^T$ symmetry.
\subsection{Fermionic SPT phases with $Z_2\times Z_2^f$ symmetry}

Here we show that fermionic SPT phase $[\eta=0,t_1=1,t_2=0]$ with $W^\bsg=I_{2\times2}$ discussed in section \ref{FERMION SPT} can be constructed in the coupled wire approach. Its coupled wire construction also indicates this SPT phase responsible for the $\mbz_4$ group structure can be obtained from non-interacting fermion band structures. The edge variables $\{\phi_1,\phi_2\}$ satisfies Kac-Moody algebra (\ref{KM algebra:fermion non-chiral SRE}) and transform in the following way under $Z_2$ symmetry $\bsg$:
\begin{eqnarray}\label{sym trans:fermion:z2}
\bsg:~~~\begin{pmatrix}\phi_1\\ \phi_2\end{pmatrix}\rightarrow\begin{pmatrix}\phi_1+\pi\\ \phi_2\end{pmatrix}.
\end{eqnarray}
and under fermion parity $Z_2^f=\{\bse,\fnp\}$:
\begin{eqnarray}\label{sym trans:fermion:z2f}
\fnp:~~~\begin{pmatrix}\phi_1\\ \phi_2\end{pmatrix}\rightarrow\begin{pmatrix}\phi_1+\pi\\ \phi_2+\pi\end{pmatrix}.
\end{eqnarray}

Consider right now each quantum wire contains electrons of both spins \ie two left movers $\psi^L_{\uparrow/\downarrow}$ and two right movers $\psi^R_{\uparrow/\downarrow}$:
\begin{eqnarray}
&\notag\psi^R_{l,\uparrow/\downarrow}\sim\exp\Big[\imth(\varphi_{l,\uparrow/\downarrow}+\theta_{l,\uparrow/\downarrow}+k_{\uparrow/\downarrow}x)\Big],\\
&\notag\psi^L_{l,\uparrow/\downarrow}\sim\exp\Big[\imth(\varphi_{l,\uparrow/\downarrow}-\theta_{l,\uparrow/\downarrow}-k_{\uparrow/\downarrow}x)\Big].
\end{eqnarray}
where bosonic fields $\varphi_{l,\sigma}$ and $\theta_{l,\sigma}$ satisfy commutation relation (\ref{luttinger liquid:commutation relation}). Let's assume here $Z_2$ symmetry $\bsg$ is the fermion number parity of spin-$\uparrow$ fermions, which is naturally realized by
\begin{eqnarray}
\bsg:~~~\varphi_{l,\uparrow}(x)\rightarrow\varphi_{l,\uparrow}(x)+\pi.
\end{eqnarray}
while $\theta_{l,\uparrow},\varphi_{l,\downarrow},\theta_{l,\downarrow}$ remain invariant under $\bsg$. On the other hand under total fermion parity $\fnp$ we have
\begin{eqnarray}
\fnp:~~~\varphi_{l,\uparrow/\downarrow}(x)\rightarrow\varphi_{l,\uparrow/\downarrow}(x)+\pi.
\end{eqnarray}
where $\theta_{l,\uparrow},\theta_{l,\downarrow}$ remain invariant under $\fnp.$

By defining the following variables
\begin{eqnarray}
\phi_l^R=\varphi_{l,\uparrow}+\theta_{l,\uparrow},~~~\phi_l^L=\varphi_{l,\downarrow}-\theta_{l,\downarrow}.
\end{eqnarray}
and
\begin{eqnarray}
\tilde\phi_l^L=\varphi_{l,\uparrow}-\theta_{l,\uparrow},~~~\tilde\phi_l^R=\varphi_{l,\downarrow}+\theta_{l,\downarrow}.
\end{eqnarray}
It's easy to verify that both $\{\phi_l^R,\phi_l^L\}$ and $\{\tilde\phi_l^R,\tilde\phi_l^L\}$ satisfy Kac-Moody algebra (\ref{KM algebra:fermion non-chiral SRE}) and the symmetry transformations (\ref{sym trans:fermion:z2}) and (\ref{sym trans:fermion:z2f}). The two pairs of variables commute with each other. The two variables $\{\phi_l^R,\phi_l^L\}$ are nothing but the right mover for spin-$\uparrow$ and left mover for spin-$\downarrow$. Clearly the following single-fermion tunneling terms
\begin{eqnarray}
&\notag\mathcal{H}^1_{Z_2\times Z_2^f}=\sum_{l=1}^{N_w-1}A_l{\psi^L_{l,\uparrow}}^\dagger\psi^R_{l+1,\downarrow}+B_l{\psi^R_{l,\downarrow}}^\dagger\psi^L_{l+1,\uparrow}+~h.c.\\
&\notag=\sum_{l=1}^{N_w-1}C_l\cos(\tilde\phi_l^L-\phi_{l+1}^R+\lambda_l)+D_l\cos(\tilde\phi_l^R-\phi_{l+1}^L+\tilde\lambda_l).
\end{eqnarray}
will gap out everything in the bulk in its strong coupling phase. The gapless variables are $\{\phi_1^R,\phi_1^L\}$ on the left edge $l=1$: they do transform as (\ref{sym trans:fermion:z2}) and (\ref{sym trans:fermion:z2f}). On the right edge $l=N_w$ the gapless variables are $\{\tilde\phi_{N_w}^L,\tilde\phi_{N_w}^R\}$. Since the interwire scattering term which stablizes this SPT phase is just a single-electron tunneling term, we expect such a fermionic SPT phase should be realized in a non-interacting band structure with symmetry $Z_2\times Z_2^f$.

\section{Concluding remarks}

We have discussed an algebraic method to systematically classify interacting topological phases in two dimensions in the absence of topological order. The key development is a general formalism for incorporating symmetry transformations into the $\bf K$ matrix formalism. Various examples of interacting boson and fermion topological phases that are well defined in the presence of disorder were presented. The method provides both long wavelength information of these phases (bulk effective field theory and edge theory) as well as suggests microscopic realizations in model systems (such as quasi 1D realizations presented here). It also opens the door to study various transitions out of these phases - \eg topology or symmetry changing transitions driven by disorder and interactions. Future work will focus on extending these results to symmetry protected distinctions between topologically ordered states, and extending this formalism to $d=1$ and $d=3$. It remains to be seen if a more general formalism that subsumes the present one can be devised, which, for example, can handle unpaired Majorana edge modes. A deeper understanding of the remarkable connection between this formalism and the Borel group cohomology/super-cohomology is also needed. Perhaps the most important question is to determine how topological phases of the interacting variety can be obtained in an experimentally realistic setting. We leave these questions for future work.

While completing this manuscript \Ref{Levin2012a} appeared which studies the specific case of $G=U(1)\rtimes Z_2^T$ (topological insulators) using a K matrix approach. Our results agree in the areas where they overlap.

\acknowledgements

We are indebted to Zheng-Cheng Gu for numerous discussions and Xiao-Gang Wen, Michael Levin and Xie Chen for feedback on the manuscript. A.V. was supported by NSF DMR 0645691 and Y.L. by Office of BES, Materials Sciences Division of the U.S. DOE under contract No. DE-AC02-05CH11231. We thank Chong Wang and T. Senthil for pointing out that bosonic SPTs with $U(1)\rtimes Z_2^T$ symmetry are destabilized by the presence of electrons.

\appendix

\section{A short note on $GL(N,\mbz)$}\label{app:gl(n,z)}

$GL(N,\mbz)$ is the group of all $N\times N$ unimodular matrices. All $GL(N,\mbz)$ matrices can be generated by the following basic transformations ($i\neq j$):
\begin{eqnarray}
&\notag T^{(i,j)}_{a,b}=\delta_{a,b}+\delta_{a,i}\delta_{b,j},\\
&\notag S^{(i,j)}_{a,b}=\delta_{a,b}(1-\delta_{a,i})(1-\delta_{a,j})+\delta_{a,j}\delta_{b,i}-\delta_{a,i}\delta_{b,j},\\
& D_{a,b}=\delta_{a,b}-2\delta_{a,N}\delta_{b,N}.\label{gl(n,z):generator}
\end{eqnarray}
$T^{(i,j)}{\bf K}$ will add the $j$-th row of matrix ${\bf K}$ to the $i$-th row of ${\bf K}$, while $S^{(i,j)}{\bf K}$ will exchange the $i$-th and $j$-th row of ${\bf K}$ with a factor of $-1$ multiplied on the $i$-th row. $D{\bf K}$ will just multiply the $N$-th row of ${\bf K}$ by a factor of $-1$. ${\bf K}T^{(i,j)},~{\bf K}S^{(i,j)}$ and ${\bf K}D$ correspond to similar operations to columns (instead of rows). A subgroup of $GL(N,\mbz)$ with determinant $+1$ is called $SL(N,\mbz)$ and it's generated by $\{T^{(i,j)},~S^{(i,j)}\}$.

As a simple example when $N=2$, group $GL(2,\mbz)$ is generated by the following basic transformations:
\begin{eqnarray}\label{gl(2,z):generator}
&S=\begin{pmatrix}0&-1\\1&0\end{pmatrix},~~T=\begin{pmatrix}1&1\\0&1\end{pmatrix},~~D=\begin{pmatrix}1&0\\0&-1\end{pmatrix}.
\end{eqnarray}
The following results will be useful
\begin{eqnarray}
&\notag T^n=\begin{pmatrix}1&n\\0&1\end{pmatrix},~(-STS)^n=\begin{pmatrix}1&0\\-n&1\end{pmatrix},~n\in\mathbb{Z}.
\end{eqnarray}

\section{A theorem on $2\times2~{\bf K}$ matrices with $\det{\bf K}=-n^2$}\label{app:K2x2}

In general a gauge transformation which relabels the quasiparticles in the ${\bf K}$ matrix formulation (\ref{cs action}) is implemented by a $GL(N,\mbz)$ matrix ${\bf W}$ as shown in (\ref{GL(N,Z) guage field}) and (\ref{gl(n,z)}). Therefore in the absence of any symmetry, any two ${\bf K}$ matrices related by (\ref{gl(n,z)}) are equivalent to each other \ie
\begin{eqnarray}\label{gl(n,z):equivalency}
{\bf K}\simeq{\bf M}^T{\bf K}{\bf M},~~~\forall~{\bf M}\in GL(n,\mbz).
\end{eqnarray}
We use $\simeq$ to denote the equivalency. In the presence of $U(1)$ symmetry with a charge vector ${\bf t}$ the equivalency requires
\begin{eqnarray}\label{gl(n,z):equivalency:charge sector}
\{{\bf K},{\bf t}\}\simeq\{{\bf M}^T{\bf K}{\bf M},{\bf M}^T{\bf t}\},~~~\forall~{\bf M}\in GL(n,\mbz).
\end{eqnarray}

In the absence of any symmetry, here we prove the following theorem: \emph{any $2\times2~{\bf K}$ matrix with determinant $\det{\bf K}=-n^2$ can be transformed into the standard form}
\begin{eqnarray}
&\notag\begin{pmatrix}0&n\\n&a\end{pmatrix},~~~0\leq a\leq2n-1.\label{K2x2:standard form:general}
\end{eqnarray}
In the special case of SRE phases with $n=1$,
\begin{eqnarray}
&\notag\begin{pmatrix}0&n\\n&2a\end{pmatrix},~~~0\leq a\leq n-1~\emph{for bosons},\\
&\begin{pmatrix}0&n\\n&2a+1\end{pmatrix},~~~0\leq a\leq n-1~\emph{for fermions}.\label{K2x2:standard form}
\end{eqnarray}

A generic $2\times2~{\bf K}$ matrix with determinant $-n^2$ can be written as
\begin{eqnarray}\label{K2x2:generic form}
&{\bf K}_{2\times2}=\begin{pmatrix}a&n+k\\n+k&b\end{pmatrix},\\
&\notag ab=k(2n+k),~a,b,k\in\mbz.
\end{eqnarray}
Apparently for a bosonic system $a,b$ are both even integers and $k$ is also an even integer. For a fermionic system there are two possibilities: $a,b,k$ are all odd integers; $k=0$ or $-2n$ and one of $a,b$ equals zero while the other being an odd integer.

Notice that under $GL(2,\mbz)$ transformations $\sigma_x$ and $\imth\sigma_y$ ($\sigma_\alpha,~\alpha=x,y,z$ are Pauli matrices) we have
\begin{eqnarray}\label{gl(2,z):Sx,iSy}
\begin{pmatrix}a&b\\b&c\end{pmatrix}\simeq\begin{pmatrix}c&b\\b&a\end{pmatrix}\simeq\begin{pmatrix}c&-b\\-b&a\end{pmatrix}.
\end{eqnarray}
If $a=0$ or $b=0$ we have $k=0$ or $k=-2n$ \ie $n+k=\pm n$ in (\ref{K2x2:generic form}). Using relations (\ref{gl(2,z):Sx,iSy}) and generator $T$ in (\ref{gl(2,z):generator}) one can show
\begin{eqnarray}
&\notag{\bf K}\simeq\begin{pmatrix}0&n\\n&x\end{pmatrix}\simeq T\begin{pmatrix}0&n\\n&x\end{pmatrix}T^T=\begin{pmatrix}0&n\\n&x+2n\end{pmatrix}
\end{eqnarray}
and (\ref{K2x2:standard form}) can be easily verified.

If $ab\neq0$, without loss of generality we can assume that $|a|\leq|b|$ and therefore $|a|\leq\max(|k|,|2n+k|)=\max(|{\bf K}_{1,2}-n|,|{\bf K}_{1,2}+n|)$. We use the following strategy: if $|k|\leq|2n+k|$ choose ${\bf M}=\begin{bmatrix}1&-\text{sign}\big(a(2n+k)\big)\\0&1\end{bmatrix}$ in (\ref{gl(n,z):equivalency}) so that $|{\bf K}_{1,2}+n|\rightarrow|{\bf K}_{1,2}+n|-|a|$; if $|k|>|2n+k|$ choose ${\bf M}=\begin{bmatrix}1&-\text{sign}\big(ak\big)\\0&1\end{bmatrix}$ in (\ref{gl(n,z):equivalency}) so that $|{\bf K}_{1,2}-n|\rightarrow|{\bf K}_{1,2}-n|-|a|$. The value of $\max(|{\bf K}_{1,2}-n|,|{\bf K}_{1,2}+n|)$ will decrease monotonically when this procedure is repeated and finally one will end up with a $2\times2~{\bf K}$ matrix whose off-diagonal elements are $\pm n$. This means $ab=0$ in (\ref{K2x2:generic form}). Therefore theorem (\ref{K2x2:standard form}) is proved.

\section{A theorem on bosonic  $2\times2~{\bf K}$ matrices with $\det{\bf K}=-1$ and charge vector ${\bf t}$}\label{app:K2x2:boson:U(1)}

In this section we prove the following theorem (we use $(a,b)$ to denote the greatest common divisor of two integers $a$ and $b$): \emph{for a $2+1$-D bosonic system any ${\bf K}$ with $\det{\bf K}=-1$ and a charge vector ${\bf t}=\begin{pmatrix}t_1\\t_2\end{pmatrix}$ with $(t_1,t_2)=1$ is equivalent to ${\bf K}=\begin{pmatrix}0&1\\1&0\end{pmatrix}$ and ${\bf t}=\begin{pmatrix}1\\-l\end{pmatrix},~l\in\mbz$ by a $GL(2,\mbz)$ gauge transformation, \ie}
\begin{eqnarray}
&\notag\{{\bf K},{\bf t}=\begin{pmatrix}t_1\\t_2\end{pmatrix}\}\simeq\{\begin{pmatrix}0&1\\1&2l\end{pmatrix},\begin{pmatrix}1\\0\end{pmatrix}\}\simeq\{\begin{pmatrix}0&1\\1&0\end{pmatrix},\begin{pmatrix}1\\-l\end{pmatrix}\},\\
&\text{if}~(t_1,t_2)=1,~~~l\in\mbz.\label{K2x2:U(1):standard form}
\end{eqnarray}

First we notice that according to the Euclidean division algorithm on integers $\mathbb{Z}$, for any pairs of integers \eg $t_1$ and $t_2$ here, there is always a list $[q_1,q_2,\cdots,q_{n+1}]$ and $[r_1,r_2,\cdots,r_n]$ such that (let's assume $|t_1|\geq|t_2|$ without loss of generality)
\begin{eqnarray}
&\notag t_1=q_1t_2+r_1,\\
&\notag t_2=q_2r_1+r_2,\\
&\notag r_1=q_3r_2+r_3,\\
&\notag\cdots,\\
&\notag r_{n-2}=q_n r_{n-1}+r_n,\\
&\notag r_{n-1}=q_{n+1}r_n.
\end{eqnarray}
where $r_n=(t_1,t_2)$ is the \emph{greatest common divisor} of $t_1$ and $t_2$, and $1\leq|r_{m+1}|\leq|r_m|,~\forall m$. Therefore one can always find two integers $u_1$ and $u_2$ such that
\begin{eqnarray}
(t_1,t_2)=r_n=r_{n-2}-q_n r_{n-1}=\cdots=t_1u_2-t_2u_1.
\end{eqnarray}
As a result for $(t_1,t_2)=1$ we have
\begin{eqnarray}
\begin{pmatrix}t_1\\t_2\end{pmatrix}={\bf M_0}^T\begin{pmatrix}1\\0\end{pmatrix},~~~{\bf M_0}^T\equiv\begin{pmatrix}t_1&u_1\\t_2&u_2\end{pmatrix}\in GL(2,\mbz).
\end{eqnarray}
and hence
\begin{eqnarray}\label{gl(2,z):charge vector}
&\{{\bf K},{\bf t}\}\simeq\{{\bf K}^\prime\equiv({\bf M_0}^{-1})^T{\bf K}{\bf M_0}^{-1},\begin{pmatrix}1\\0\end{pmatrix}\}.
\end{eqnarray}
as long as $(t_1,t_2)=1$. In the following we prove that an arbitrary $2\times2~{\bf K}$ matrix with $\det{\bf K}=-1$ for a bosonic system is equivalent to the standard form $\begin{pmatrix}0&1\\1&2l\end{pmatrix}$ by a gauge transformation which keeps the charge vector $\begin{pmatrix}1\\0\end{pmatrix}$ invariant. To prove this we need to enlarge the Hilbert space by introducing a $4\times4$ matrix $\tilde{\bf K}={\bf K}_{2\times2}\oplus\sigma_x$ and associated charge vector $\tilde{\bf t}\equiv(1,0,0,0)^T$. This describes a direct product of the original $2\times2$ bosonic ${\bf K}$ matrix with a $U(1)$ charge conservation and another trivial $2\times2$ bosonic ${\bf K}$ matrix without any symmetry. A generic form for ${\bf K}$ is $\begin{pmatrix}2a&2k+1\\2k+1&2b\end{pmatrix}$ satisfying $ab=k(k+1)$ ($\det{\bf K}=-1$). One can prove that
\begin{eqnarray}
&\notag\{\tilde{\bf K}=\begin{pmatrix}2a&2k+1&0&0\\2k+1&2b&0&0\\0&0&0&1\\0&0&1&0\end{pmatrix},\begin{pmatrix}1\\0\\0\\0\end{pmatrix}\}\\
&\simeq\{\begin{pmatrix}0&1&0&0\\1&2l&0&0\\0&0&0&1\\0&0&1&0\end{pmatrix},\begin{pmatrix}1\\0\\0\\0\end{pmatrix}\}.\label{4x4:equivalency}
\end{eqnarray}
by the following $GL(4,\mbz)$ transformations:
\begin{eqnarray}
&\notag{\bf M_1}^T\begin{pmatrix}2a&2k+1&0&0\\2k+1&2b&0&0\\0&0&0&1\\0&0&1&0\end{pmatrix}{\bf M_1}=\\
&\notag\begin{pmatrix}0&1&0&0\\1&2b&2(ab-k)&-2b\\0&2(ab-k)&2a(ab-2k)&2(k-ab)+1\\0&-2b&2(k-ab)+1&2b\end{pmatrix}\simeq\\
&\notag\begin{pmatrix}0&1&0&0\\1&2b&a^\prime&b^\prime\\0&a^\prime&0&1\\0&b^\prime&1&0\end{pmatrix}=({\bf M_2}^T)^{-1}\begin{pmatrix}0&1&0&0\\1&2(b-a^\prime b^\prime)&0&0\\0&0&0&1\\0&0&1&0\end{pmatrix}{\bf M_2}^{-1},\\
&\notag a,b,k,a^\prime,b^\prime\in\mbz.
\end{eqnarray}
where we defined
\begin{eqnarray}
&\notag{\bf M_1}\equiv\begin{pmatrix}1& 0& 0& 0\\0&1& a& -1\\1&0& 1& 0\\-a& -2 k& -2 k a& 2 k + 1\end{pmatrix},\\
&\notag{\bf M_2}\equiv\begin{pmatrix}1&0&0&0\\0&1&0&0\\0&-b^\prime&1&0\\0&-a^\prime&0&1\end{pmatrix}.
\end{eqnarray}
We have used the theorem (\ref{K2x2:standard form}) for a $n=1$ bosonic system since one can easily verify that $\det{\begin{bmatrix}2a(ab-2k)&2(k-ab)+1\\2(k-ab)+1&2b\end{bmatrix}}=-1$. Notice that the charge vector ${\bf t}=(1,0,0,0)^T$ remains invariant under these $GL(4,\mbz)$ transformations. Combing relations (\ref{gl(2,z):charge vector}) and (\ref{4x4:equivalency}) we have proved theorem (\ref{K2x2:U(1):standard form}).

\section{Faithful vs. unfaithful representations of the symmetry group}\label{Appendix:faithful}

{\em Transformations $\{W^{g_a},\delta\phi_I^{g_a}\}$ form a faithful or unfaithful representation of symmetry group $G$}: By solving the constraint equations (\ref{algebra}) and choosing a proper gauge in (\ref{gauge transformation on symmetry}), one can obtain a set of transformation rules $\{W^{g_a},\delta\phi_I^{g_a}\}$ as the solution.  Apparently the transformations $\{W^{g_a},\delta\phi_I^{g_a}\}$ always form a representation of the symmetry group $G$ of the system (or the Hamiltonian of the system) in the sense that
\begin{eqnarray}\label{representation}
&\forall~g_1,g_2\in G:~~~W^{g_1\cdot g_2}=\eta_1\eta_2W^{g_2}W^{g_1},\\
&\notag \delta\phi^{g_1\cdot g_2}=\delta\phi^{g_2}+\eta_2W^{g_2}\delta\phi^{g_1}.
\end{eqnarray}
where $\eta_1=\pm1$ if $g_1$ is a unitary (anti-unitary) symmetry and $\eta_2$ is associated with $g_2$. This representation of group $G$ is \emph{faithful} if and only if \emph{the identity element $\bse$ is the only symmetry element under which all bosonic quasiparticle fields $\{\sum_Il_I\phi_I\}$ on the edge (or $\{\prod_Ib_I^{l_I}\}$ in the bulk) transform trivially}. In other words, under any element $g\neq\bse$ of symmetry group $G$ at least one bosonic quasiparticles ${\bf l}^T\phi$ satisfying (\ref{boson}) will transform nontrivially.

In contrast to faithful representations, an \emph{unfaithful} representation $\{W^{g_a},\delta\phi_I^{g_a}\}$ of symmetry group $G$ means there exists a \emph{nontrivial subgroup} $G_\psi$ of $G$, so that \emph{under any symmetry element $g\in G_\psi$, all bosonic quasiparticle fields $\{\sum_Il_I\phi_I\}$ on the edge (or $\{\prod_Ib_I^{l_I}\}$ in the bulk) transform trivially}. This means for a phase described by ${\bf K}$ matrix and symmetry transformations $\{W^{g_a},\delta\phi_I^{g_a}\}$, its edge states can be gapped by condensing the bosonic quasiparticles without breaking the subgroup $G_\psi$ of symmetry group $G$ since under any symmetry $g\in G_\psi$ all bosonic quasiparticles are left invariant. As a result when the edge is gapped, the symmetry group $G$ of the Hamiltonian breaks down to its subgroup, the ground state symmetry group $G_\psi$. As a result the symmetry breaking phases can be naturally incorporated in the ${\bf K}$ matrix+Higgs formulation.

\section{Other Bosonic SPT Phases:}
\label{Appendix:BosonicSPT}

\subsection{$U(1)\times Z_2^T$ symmetry: $\mbz_1$ class}
\label{Appendix:U(1)xZ_2^T}

In contrast to the $U(1)\rtimes Z_2^T$ symmetry discussed in the previous section, here we study a direct product of $U(1)$ and time reversal $Z_2^T$ symmetry. This can be realized by time reversal and $U(1)$ spin rotational symmetry in an integer spin system. The algebraic relations for the $U(1)\times Z_2^T$ group are
\begin{eqnarray}\label{symmetry:U(1):T}
&\bst^2=\bst U_{-\theta}\bst U_\theta=\bse.
\end{eqnarray}
 The corresponding constraints (\ref{condition:algebra:bosonic SRE}) for symmetry transformations $\{W^\bst,\delta\phi^\bst\}$ and $\{W^{U_\theta}=I_{2\times2},\delta\phi^{U_\theta}=\theta{\bf t}\}$ are
\begin{eqnarray}
&\notag(I_{2\times2}-W^\bst){\bf\delta\phi}^\bst+(I_{2\times2}+W^\bst)\theta{\bf t}\\
&\label{u1xT:constaint:dphi}=\begin{pmatrix}0\\0\end{pmatrix}\mod2\pi,~~\forall~\theta.
\end{eqnarray}
and (\ref{z2t:constraint:W}-\ref{z2t:constraint:dphi}). The gauge inequivalent solutions to these constraint equations lead to
\begin{eqnarray}
&W^{U_\theta}=I_{2\times2},~~~\delta\phi^{U_\theta}=\theta\begin{pmatrix}0\\1\end{pmatrix}\\
&W^\bst=\sigma_z,~~~\delta\phi^\bst=\begin{pmatrix}0\\n\pi\end{pmatrix},~~~n=0,1.
\end{eqnarray}
For both $n=0,1$ corresponding symmetry-allowed Higgs terms are
\begin{eqnarray}\label{higgs:u1xT}
&\mathcal{S}^1_{edge}=\sum_{l\in\mbz}C_{l}\int\text{d}x\text{d}t\cos(l\phi_1)
\end{eqnarray}
Apparently there is only one ($\mbz_1$ class) trivial phase $\bse_{U(1)\times Z_2^T}$ with $U(1)\times Z_2^T$ symmetry, whose edge states can be gapped without breaking the symmetry.

\subsection{$Z_N\times Z_2^T$ symmetry}
\label{Appendix:Z_NxZ_2^T}

The algebraic structure for $Z_N\times Z_2^T$ group is given by
\begin{eqnarray}\label{symmetry:zn:z2t:direct}
&\bsg^N=\bst^2=\bst\bsg^{-1}\bst\bsg=\bse.
\end{eqnarray}
where $\bsg$ is the $Z_N$ symmetry generator and $\bst$ is time reversal. The associated constraint equations for symmetry transformations are
\begin{eqnarray}\label{constraint:zn:z2t_direct:W}
&W^\bsg W^\bst (W^\bsg)^{-1} W^\bst=I_{2\times2},\\
&\label{constraint:zn:z2t_direct:dphi}(I_{2\times2}+W^\bsg W^\bst(W^\bsg)^{-1})\delta\phi^\bsg+\\
&W^\bsg(1-W^\bst(W^\bsg)^{-1})\delta\phi^\bst=\begin{pmatrix}0\\0\end{pmatrix}\mod2\pi.\notag
\end{eqnarray}
in addition to (\ref{z2t:constraint:W}-\ref{z2t:constraint:dphi}) and (\ref{constraint:zn:W}-\ref{constraint:zn:dphi}).

\subsubsection{$N=$~odd integer: $\mbz_1$ classes}

The gauge inequivalent solutions to the above constraint equations are (\ref{sym trans:z2t}) and
\begin{eqnarray}\label{sym trans:zn:odd:z2t_direct:+I}
W^\bsg=I_{2\times2},~~~\delta\phi^\bsg=\frac{2\pi k}{N}\begin{pmatrix}0\\1\end{pmatrix},~k\in\mbz.
\end{eqnarray}
It's easy to verify that a set of independent Higgs terms are $\int\text{d}x\text{d}t\sum_lC_l\cos(l\phi_1)$. Hence the variable $\phi_1$ can be localized at value $\langle\phi_1\rangle=0$ without breaking any symmetry. So they all correspond to the trivial phase. There is only one trivial phase with $Z_N\times Z_2^T$ symmetry for $N=$~odd.

\subsubsection{$N=$~even integer: Minimal set: $\mbz_2^2$ classes}

Solving (\ref{constraint:zn:W}) and (\ref{constraint:zn:z2t_direct:W}) we have $W^\bst=\sigma_z$ and $W^\bsg=\pm I_{2\times2}$.

(\rmnum{1}) For $W^\bsg=I_{2\times2}$ the gauge inequivalent solutions are
\begin{eqnarray}\label{sym trans:zn:even:z2t_direct:+I}
&\delta\phi^\bst=n_2\pi\begin{pmatrix}0\\1\end{pmatrix},~~\delta\phi^\bsg=\pi\begin{pmatrix}n_1\\2k/N\end{pmatrix},\\
&0\leq k\leq N-1,~n_1,n_2=0,1.\notag
\end{eqnarray}
If $n_1=0$ the variable $\phi_1$ can be localized at $\langle\phi_1\rangle=0$ by Higgs term $-\cos\phi_1$ without breaking the symmetry and it corresponds to the trivial phase. If $n_1=1$, $n_2=k=0$ corresponds to the trivial phase again since $\phi_2$ can be localized. Notice that when $n_1=1$ we require $(k,N/2)=1$ so that transformations (\ref{sym trans:zn:even:z2t_direct:+I}) form a faithful representation of symmetry group $G=Z_N\times Z_2^T,~N=$even. Let's label a state with the above transformations (\ref{sym trans:zn:even:z2t_direct:+I}) as $[k,n_1,n_2]$ and we have $[0,1,0]=[k,0,n_2]=\bse_{Z_N\times Z_2^T}$. In the following we analyze the group structure formed by states $[k,n_1,n_2]$.

Now let's put together a state $[k,n_1,n_2]$ with edge variable $\{\phi_1,\phi_2\}$ is put together with a state $[k^\prime,n_1^\prime,n_2^\prime]^{-1}$ with edge variable $\{\phi_1^\prime,\phi_2^\prime\}$, we can choose the following independent bosonic variables $\{k^\prime\phi_2-k\phi_2^\prime,k\phi_1-k^\prime\phi_1^\prime\}$ and gap all the edge states if $(k,k^\prime)=1$. The associated Higgs terms will preserve the $Z_{N}\times Z_2^T$ symmetry if $kn_1-k^\prime n^\prime_1=0\mod2$ and $k^\prime n_2-kn^\prime_2=0\mod2$. As a result $[k,n_1,n_2]\oplus[k^\prime,n_1^\prime,n_2^\prime]^{-1}=\bse_{Z_{N}\rtimes Z_2^T}$ or equivalently
\begin{eqnarray}\notag
&\text{For}~(k,k^\prime)=1:~~~[k,n_1,n_2]=[k^\prime,n_1^\prime,n_2^\prime],\\
&\notag\text{if}~kn_1-k^\prime n^\prime_1=0\mod2,~~k^\prime n_2-kn^\prime_2=0\mod2.
\end{eqnarray}
Therefore we have $[2k+1,n_1,n_2]=[1,n_1,n_2]$ by choosing $k^\prime=1$. On the other hand, if $k=$even we again have $N/2=$odd since $(k,N/2)=1$ for a faithful representation of symmetry group $Z_N\times Z_2^T$. We can localize bosonic variable $\{\frac{N}{2}\phi_2-\phi_2^\prime,~\frac{N}{2}\phi_1^\prime-\phi_1\}$ without breaking any symmetry if we choose $k^\prime=0$. Hence we also have $[2k,n_1,n_2]=[0,n_1,n_2]$. These relations result in only three nontrivial SPT phases: $[1,1,0]$, $[1,1,1]$ and $[0,1,1]$.

Similarly by putting together a state $[k,1,n_2]$ with edge variable $\{\phi_1,\phi_2\}$ is put together with a state $[k^\prime,1,n_2^\prime]$ with edge variable $\{\phi_1^\prime,\phi_2^\prime\}$, we can always localize bosonic variable $\phi_1-\phi_1^\prime$ and gap out part of the edge. What is left on the edge are described by variables $\{\tilde\phi_1=\phi_1,\tilde\phi_2=\phi_2+\phi_2^\prime\}$. They obey Kac-Moody algebra (\ref{KM algebra:boson non-chiral SRE}) and transform as a $[k+k^\prime,1,n_2+n_2^\prime]$ state. Hence we've shown that
\begin{eqnarray}
&\notag[k,1,n_2]\oplus[k^\prime,1,n_2^\prime]\\
&=[k+k^\prime\mod2,1,n_2+n_2^\prime\mod2].
\end{eqnarray}
Since $k,n_2$ are both $\mbz_2$ integers, so clearly all different 4 states $[k,1,n_2]$ form a $\mbz_2^2$ group. Consequently there are $3$ nontrivial SPT phases labeled by $n_1=1$ and $[k,n_2]=[0,1],~[1,0]$ or $[1,1]$ in (\ref{sym trans:z2t}) and (\ref{sym trans:zn:even:z2t_direct:+I}).

(\rmnum{2}) For $W^\bsg=-I_{2\times2}$ we can always choose a gauge so that inequivalent solutions to constraints are
\begin{eqnarray}
\delta\phi^\bsg=\begin{pmatrix}0\\0\end{pmatrix},~~~\delta\phi^\bst=\pi\begin{pmatrix}n_1\\n_2\end{pmatrix}~~n_1,n_2=0,1.
\end{eqnarray}
However the above symmetry transformations $\{W^\bsg=-I_{2\times2},\delta\phi^\bsg=\begin{pmatrix}0\\0\end{pmatrix}\}$ do not correspond to a faithful representation of $Z_N\times Z_2^T$ group for $N=$even, unless $N=2$. And it is not clear how the states with symmetry transformations $W^\bsg=-I$ can be realized in a physical bosonic system. Therefore we won't include the states with symmetry transformations $W^\bsg=-I$ in the minimal set of topological phases with $Z_N\times Z_2^T$ symmetry, as discussed earlier for $W^\bsg=-I$ phases with $Z_N\rtimes Z_2^T$ symmetry.

%
%

In summary, there are $\mbz_2^2$ classes of different bosonic non-chiral SRE phase in the presence of symmetry group $Z_N\times Z_2^T,~N=$even. When $N$=odd there are no nontrivial SPT phases.

\subsection{$U(1)\times Z_2$ symmetry: $\mbz\times\mbz_2^2$ classes}

Denoting the group elements of $U(1)$ by $U_\theta,~0\leq\theta<2\pi$ and generator of $Z_2$ by $\bsg$, the group $U(1)\times Z_2$ has the following algebraic structure:
\begin{eqnarray}
\bsg^2=U_\theta\bsg U_{-\theta}\bsg=\bse.
\end{eqnarray}
and (\ref{symmetry:U(1)}). The associated constraints for symmetry transformations are
\begin{eqnarray}\label{constraint:z2:u1:W}
&(W^\bsg)^2=I_{2\times2},~~~(I_{2\times2}+W^\bsg)\delta\phi^\bsg=\begin{pmatrix}0\\0\end{pmatrix},\\
&(I_{2\times2}+W^\bsg)\delta\phi^\bsg+\theta(I_{2\times2}-W^\bsg){\bf t}=\begin{pmatrix}0\\0\end{pmatrix}.\label{constraint:z2:u1:dphi}
\end{eqnarray}
where we have $W^{U_\theta}=I_{2\times2}$ $\delta\phi^{U_\theta}=\theta{\bf t}$ with $t_1,t_2\in\mbz$ and $(t_1,t_2)=1$ for $U(1)$ symmetry. Solving (\ref{constraint:z2:u1:W}) we have $W^\bsg=\pm I_{2\times2}$ or $\pm\sigma_x$.

\subsubsection{$\mbz\times\mbz_2^2$ classes with $W^\bsg=\pm I_{2\times2}$}

(\rmnum{1}) For $W^\bsg=I_{2\times2}$, as guaranteed by theorem (\ref{K2x2:U(1):standard form}) we can always transform the ``charge vector" ${\bf t}$ into a standard form ${\bf t}=\begin{pmatrix}1\\q\end{pmatrix},~q\in\mbz$. And the inequivalent ``faithful" symmetry transformations satisfying constraints (\ref{constraint:z2:u1:W}-\ref{constraint:z2:u1:dphi}) are
\begin{eqnarray}
\delta\phi^\bsg=\begin{pmatrix}n_1\\n_2\end{pmatrix}\pi,~~~\delta\phi^{U_\theta}=\theta\begin{pmatrix}1\\q\end{pmatrix},~~n_1,n_2=0,1,~q\in\mbz.
\end{eqnarray}
Let's label a state with the above transformation rules as $[q,n_1,n_2]$. Similar as earlier discussions for other symmetries, by putting two states $[q,n_1,n_2]$ and $[q^\prime,n_1^\prime,n_2^\prime]$ together, one can show the following multiplication rule:
\begin{eqnarray}
&\notag[q,n_1,n_2]\oplus[q^\prime,n_1^\prime,n_2^\prime]\\
&=[q+q^\prime,n_1+n_1^\prime\mod2,n_2+n_2^\prime\mod2]
\end{eqnarray}
Hence there are $\mbz\times(\mbz_2)^2$ classes of different phases labeled by integer $q$ and $\mbz_2$ integers $n_1,n_2$. The trivial phase corresponds to $q=n_1=n_2=0$.

(\rmnum{2}) For $W^\bsg=-I_{2\times2}$, one can always choose a gauge so that $\delta\phi^\bsg=\begin{pmatrix}0\\0\end{pmatrix}$ and by solving (\ref{constraint:z2:u1:dphi}) we get ${\bf t}=\begin{pmatrix}0\\0\end{pmatrix}$. It's easy to verify this corresponds to the trivial phase.

\subsubsection{Other solutions to (\ref{constraint:z2:u1:W}-\ref{constraint:z2:u1:dphi}) with $W^\bsg=\pm\sigma_x$}

(\rmnum{3}) For $W^\bsg=\sigma_x$, the inequivalent solutions to constraint equations are
\begin{eqnarray}
\delta\phi^\bsg=n\pi\begin{pmatrix}1\\1\end{pmatrix},~~\delta\phi^{U_\theta}=\theta\begin{pmatrix}1\\1\end{pmatrix},~~n=0,1.
\end{eqnarray}
These two nontrivial SPT phases are labeled as $[\sigma_x,n]$ with $n=0,1$. Their physical realization and group structure are not clear.

(\rmnum{4}) For $W^\bsg=-\sigma_x$, the inequivalent solutions to constraint equations are
\begin{eqnarray}
\delta\phi^\bsg=n\pi\begin{pmatrix}1\\-1\end{pmatrix},~~\delta\phi^{U_\theta}=\theta\begin{pmatrix}1\\-1\end{pmatrix},~~n=0,1.
\end{eqnarray}
These two nontrivial SPT phases are labeled as $[-\sigma_x,n]$ with $n=0,1$. It's easy to show that
\begin{eqnarray}
[\sigma_x,n]^{-1}=[-\sigma_x,n],~~~n=0,1.
\end{eqnarray}
Their physical realization and group structure are not clear either as with the discussion in Section \ref{unphysical} and therefore we do not include these phases.

To summarize, there are $\mbz\times(\mbz_2)^2$ classes of different phases with $W^\bst=\pm I_{2\times2}$ for symmetry group $U(1)\times Z_2$.

Besides, there are 4 extra possible nontrivial SPT phases with $W^\bsg=\pm\sigma_x$ for $U(1)\times Z_2$ symmetry in a bosonic non-chiral SRE system.



\section{Other solutions to (\ref{constraint:fermion:z2:dphi}) for fermion SPT phases with $G_f/Z_2^f=Z_2$ symmetry}

\subsubsection{$W^\bsg=\pm\sigma_z$:~~~$\mbz^2$ classes}

(\rmnum{3}) For $W^\bsg=\sigma_z$ the inequivalent solutions to (\ref{constraint:fermion:z2:dphi}) are $\eta=0$ and $\delta\phi^\bsg=\begin{pmatrix}n\pi\\0\end{pmatrix}$ where $n=0,1$. Since under symmetry $\bsg$ we have $\phi_1\pm\phi_2\rightarrow\phi_1\mp\phi_2+n\pi$, the variables cannot be localized without breaking the $Z_2$ symmetry. These two different nontrivial SPT phases are labeled as $[\sigma_z,n]$ where $n=0,1$. In the following we identify their group structure.

First notice that when a $[\sigma_z,0]$ state with edge variables $\{\phi_{1},\phi_{2}\}$ is put together with a $[\sigma_z,1]^{-1}$ state with edge variables $\{\phi_{1}^\prime,\phi_{2}^\prime\}$, its edge cannot be gapped without breaking the symmetry, suggesting $[\sigma_z,0]\neq[\sigma_z,1]$.
~Then let's consider $N$ copies of $[\sigma_z,n]$ states put together and their edge variables are $\{\phi_1^a,\phi_2^a,~~~1\leq a\leq N\}$ . A generic bosonic variable that can be localized on the edge is written as $\sum_{a=1}^N(A_a\phi_1^a+B_a\phi_2^a),~A_a,B_a\in\mbz$ satisfying $\sum_aA_a^2-B_a^2=0$ due to condition (\ref{condition:local}). Under $Z_2$ symmetry generator $\bsg$ this bosonic variable becomes $\sum_{a=1}^N(A_a\phi_1^a-B_a\phi_2^a)$. In order for the two bosonic variables to be localized simultaneously (\ie they are independent bosons) they have to satisfy (\ref{condition:commute}) and hence $\sum_aA_a^2+B_a^2=0$. This leads to $A_a=B_a=0$ and hence no bosonic variable on the edge can be localized without breaking the symmetry. Hence whenever we add an extra $[\sigma_z,0]$ state into the system, there is one more $c=1$ gapless state on the edge. Hence all the different states $\{[\sigma_z,0]^{M}\oplus[\sigma,1]^{N},~M,N\in\mbz\}$ form the $\mbz^2$ group.

(\rmnum{4}) For $W^\bsg=-\sigma_z$ the inequivalent solutions to (\ref{constraint:fermion:z2:dphi}) are $\eta=0$ and $\delta\phi^\bsg=\begin{pmatrix}0\\n\pi\end{pmatrix}$ where $n=0,1$. We label these states by $[-\sigma_z,n]$ and it's straightforward to show that $[\sigma_z,n]^{-1}=[-\sigma_z,n]$.

To summarize, with $Z_2$ symmetry transformation $W^\bsg=\pm\sigma_z$, there are $\mbz^2$ classes of different fermionic non-chiral SRE phases in the presence of $Z_2\times Z_2^f$ symmetry. It is presently unclear to us if these transformation laws can be realized in a physical system of fermions. We have not found a microscopic realization, hence we do not include it in the minimal set of topological phases with this symmetry.

\section{Other solutions to $ (\ref{constraint:fermion:z2:z2t:dphi})$ for fermion SPT phases with $G_f/Z_2^f=Z_2\times Z_2^T$ symmetry}

\subsubsection{$W^\bsg=-I_{2\times2}$:~$\mbz_2$ classes}

(\rmnum{2}) If $W^\bsg=-I_{2\times2}$ the gauge inequivalent solutions to (\ref{constraint:fermion:z2:z2t:dphi}) are $\eta_\bsg=0$ and
\begin{eqnarray}
&\delta\phi^\bsg=\begin{pmatrix}0\\0\end{pmatrix},~~\delta\phi^\bst=(\frac{\eta}2+n)\pi\begin{pmatrix}1\\1\end{pmatrix}+\frac{\eta_\bst\pi}2\begin{pmatrix}1\\-1\end{pmatrix},\notag\\
&n,\eta,\eta_\bst=0,1.\label{sym trans:fermion:z2:z2t:-I}
\end{eqnarray}
If $\eta_\bst=0$ we can destroy the gapless edge states without breaking the symmetry, by localizing the bosonic variable $\phi_1-\phi_2$ at a classical value. If $\eta=0$, we can destroy the gapless edge states without breaking the symmetry, by localizing a differnt bosonic variable $\phi_1+\phi_2$ at a classical value. Hence only $\eta=\eta_\bst=1$ and $n=0,1$ correspond to non-trivial SPT phases, with $\delta\phi^\bst=(\pi,0)^T$ or $(0,\pi)^T$ ($n=0$ or $1$). Let's label the states with symmetry transformations (\ref{sym trans:fermion:z2:z2t:-I}) by $[\eta,\eta_\bst,n]$. It's easy to verify that $[1,1,0]\oplus[1,1,0]=[1,1,0]\oplus[1,1,1]=\bse_{Z_2\times Z_2^T\times Z_2^f}$, since the edge from two phases put together can be gapped by condensing independent bosons $\{\phi_1+\phi_2^\prime,\phi_2+\phi_1^\prime\}$. Since $[0,0,n]=[0,1,n]=[1,0,n]=\bse_{Z_2\times Z_2^T\times Z_2^f}$ we see that different phases form a $\mbz_2$ group. The only nontrivial SPT phase is $[1,1,0]=[1,1,1]$. The microscopic realization of this particular SPT phase is not clear, (we have not found a realization in the coupled wire approach) and as discussed previously, we omit it from the minimal set of topological phases.




\end{document}